\documentclass[prd,10pt,aps,nofootinbib, floatfix, notitlepage,longbibliography, superscriptaddress,preprintnumbers]{revtex4-2}

\usepackage{placeins}
\usepackage{amsmath,amsfonts,amssymb}
\usepackage{mathrsfs}
\usepackage{graphicx}
\usepackage[english]{babel} 
\usepackage{ulem}
\usepackage{url} 
\usepackage{color}
\usepackage{bbold}
\usepackage{adjustbox}
\usepackage[utf8]{inputenc} 
\usepackage[colorlinks=true,citecolor=blue,urlcolor=blue]{hyperref}

\newcommand{\beq}{\begin{equation}}
\newcommand{\eeq}{\end{equation}}
\newcommand{\bea}{\begin{eqnarray}}
\newcommand{\eea}{\end{eqnarray}}
\newcommand{\bef}{\begin{figure}}
\newcommand{\eef}{\end{figure}}

\newcommand{\cH}{\mathcal{H}}
\newcommand{\NUV}{\Delta N^{\text{UV}}_{\text{eff}}}
\newcommand{\NIR}{\Delta N^{\text{IR}}_{\text{eff}}}
\newcommand{\rg}{r_g}
\newcommand{\zt}{z_t}
\newcommand{\GamSDR}{\Gamma}
\newcommand{\GamOSDR}{\Gamma_0}
\newcommand{\fdm}{f_{\textrm{DM}}}
\newcommand{\cs}{c_s^2}

\newcommand{\SDR}{SDR}
\newcommand{\SDRIDM}{IDM}
\newcommand{\sdr}{sdr}
\newcommand{\idm}{idm}

\newcommand{\thetaSDR}
{\theta_{\textrm{\sdr}}}
\newcommand{\deltaSDR}{\delta_{\textrm{\sdr}}}
\newcommand{\thetaIDM}{\theta_{\textrm{\idm}}}

\newcommand{\thetaSDRdot}{\dot{\theta}_{\textrm{\sdr}}}
\newcommand{\thetaIDMdot}{\dot{\theta}_{\textrm{\idm}}}
\newcommand{\rhoSDR}{\rho_{\textrm{\sdr}}}
\newcommand{\rhoIDM}{\rho_{\textrm{\idm}}}
\newcommand{\rhoDM}{\rho_{\textrm{DM,tot}}}

\newcommand{\Dnl}{\mathcal{D}_{\textrm{nl}}}
\newcommand{\Dbase}{\mathcal{D}_{\textrm{b}}}
\newcommand{\Dh}{\mathcal{D}_{H}}
\newcommand{\DS}{\mathcal{D}_{S}}
\newcommand{\DHS}{\mathcal{D}_{HS}}
\newcommand{\DEFT}{\mathcal{D}_{\textrm{b}}^{EFT}}
\newcommand{\DEFTS}{\mathcal{D}_{S}^{EFT}}
\newcommand{\DEFTH}{\mathcal{D}_{H}^{EFT}}
\newcommand{\DEFTHS}{\mathcal{D}_{HS}^{EFT}}

\newcommand{\QDMAP}{Q_\textrm{DMAP}}
\newcommand{\AIC}{\Delta \textrm{AIC}}
\newcommand{\MCMC}{\textrm{\tiny{MC}}}

\definecolor{green2}{RGB}{34,139,34}
\begin{document}

\title{Dark Sectors with Mass Thresholds Face Cosmological Datasets}

\author{Itamar~J.~Allali}
\email{itamar.allali@tufts.edu}
\affiliation{Institute of Cosmology, Department of Physics and Astronomy, Tufts University, Medford, MA 02155, USA
\looseness=-1}
\author{Fabrizio Rompineve}
\email{fabrizio.rompineve@cern.ch}
    \affiliation{CERN, Theoretical Physics Department, 1211 Geneva 23, Switzerland
    \looseness=-1}
	\affiliation{Departament de F\'isica, Universitat Aut\`onoma de Barcelona, 08193 Bellaterra, Barcelona, Spain
		\looseness=-1}
    \affiliation{Institut de F\'isica d’Altes Energies (IFAE) and The Barcelona Institute of Science and Technology (BIST),
    Campus UAB, 08193 Bellaterra (Barcelona), Spain
		\looseness=-1}
  \author{Mark P.~Hertzberg}
\email{mark.hertzberg@tufts.edu}
\affiliation{Institute of Cosmology, Department of Physics and Astronomy, Tufts University, Medford, MA 02155, USA
\looseness=-1}

\date{\today}

\begin{abstract}
\noindent Interacting dark sectors may undergo changes in the number of their relativistic species during the early universe, due to a mass threshold $m$ (similar to changes in the Standard Model bath), and in doing so affect the cosmic history. When such changes occur close to recombination, i.e., for $m\sim (0.1-10)~\text{eV}$, the stringent bound on the effective number of neutrino species, $N_{\text{eff}}$, can be relaxed and the value of the Hubble expansion rate $H_0$ inferred from cosmic microwave background (CMB) observations raised. We search for such sectors (with and without mass thresholds) in the latest cosmological datasets, including the full-shape (FS) of BOSS DR12 galaxy power spectrum. We perform a detailed analysis, accounting for the choice of prior boundaries and additionally exploring the possible effects of dark sector interactions with (a fraction of) the dark matter. We find $\Delta N_{\text{eff}}\leq 0.55\, (0.46)$ at 95\% C.L. with (without) a mass threshold. While a significantly larger Hubble rate is achieved in this scenario, $H_0=69.01^{+0.66}_{-1.1}$, the overall fit to CMB+FS data does not provide a compelling advantage over the $\Lambda$CDM model. Furthermore, we find that dark matter interactions with the dark sector do not significantly improve the (matter fluctuations) $S_8$ tension with respect to the $\Lambda$CDM model. Our work provides model-independent constraints on (decoupled) dark sectors with mass thresholds around the eV scale.
\end{abstract}

\preprint{CERN-TH-2023-084}
\maketitle


\section{Introduction}

Observations of primordial abundances and of the cosmic microwave background (CMB) reveal that the early Universe at $\text{eV}\lesssim T\lesssim \text{MeV}$ is dominated by a hot bath of photons and three neutrinos, while yet to be discovered beyond the Standard Model relativistic species, or {\it dark radiation} (DR), can only contribute a small fraction. More precisely, the current bound from CMB observations, commonly expressed in terms of the effective number of neutrino species $\Delta N_{\text{eff}}\equiv \rho_{\text{DR}}/\rho_{\nu,1}$, where $\rho_{\nu,1}$ is the energy density of a single neutrino species, is $\Delta N_{\text{eff}}\leq 0.28$~\cite{Planck:2018vyg} ($95\%$ C.L., Planck 2018 +  baryon acoustic oscillations, at the epoch of recombination). CMB and Large Scale Structure (LSS) observations also similarly constrain scenarios where the would-be DR has a mass around and above the eV scale, as in this case it behaves as ``hot" dark matter (DM) and suppresses structure formation (see e.g.~\cite{Viel:2005qj, Osato:2016ixc, Xu:2021rwg}). While these are powerful constraints, they apply only to the simplest dark sectors, made either of massless (i.e. $m\ll \text{eV}$) or light but massive ($m\gtrsim \text{eV}$) relics. On the other hand, the only particle physics sector that we have detected, the Standard Model (SM), features both light and heavy degrees of freedom that interact with each other. It is thus important to assess whether a light dark sector that more closely resembles the SM (albeit with very different mass scales) can evade the constraints above.

Interestingly, any model that succeeds in alleviating the $\Delta N_{\text{eff}}$ constraint may simultaneously prove promising to address the $\gtrsim 5\sigma$ tension between inferrals~\cite{Planck:2018vyg, Schoneberg:2019wmt, Philcox:2020vvt} and local measurements~\cite{Wong:2019kwg, Riess2022a,  Scolnic:2023mrv} (see however~\cite{Blum:2020mgu, Freedman:2021ahq} for alternative takes) of the Hubble expansion rate $H_0$ (barring underestimated systematics in any of the two types of measurements).
The addition of dark radiation is indeed arguably the simplest extension of the $\Lambda$CDM model that can result in a larger value of $H_0$~\cite{Planck:2018vyg} (see also~\cite{Vagnozzi:2019ezj}). However, the aforementioned constraint on $\Delta N_{\text{eff}}$ implies that the so-called ``Hubble tension" remains at $\simeq 4\sigma$ level in this seven-parameter cosmological model. The situation improves when the dark radiation is non-free streaming, as occurs in the presence of sizable self-interactions, because the phase shift of the CMB high-$\ell$ and BAO peaks (see e.g.~\cite{Baumann:2015rya}) is absent in this case. The constraint is then relaxed to $\Delta N_{\text{eff}}\leq 0.46$ at $95\%$ C.L., and the Hubble tension further reduced to around $3.5\sigma$  (see Appendix~\ref{app:data} and~\cite{Blinov:2020hmc, Schoneberg:2021qvd}), although the fit to CMB+BAO data only is not improved with respect to $\Lambda$CDM.

In this work, we aim to test simple interacting dark sector models with mass thresholds $m\sim (0.1-10)~\text{eV}$ and relic light species against cosmological datasets, and to assess their impact on cosmological tensions (for previous work related to the Hubble tension see~\cite{Aloni:2021eaq, Joseph:2022jsf, Buen-Abad:2022kgf}, and~\cite{Escudero:2019gvw, Sandner:2023ptm} for different scenarios featuring interactions with SM neutrinos). As equilibrium is maintained by interactions in the dark sector, entropy is transferred from species with mass scale $m$ to the remaining light degrees of freedom. As a consequence, the temperature $T_{d}$ of the dark sector temporarily scales slower than that of neutrinos, i.e. $T_{d}\propto g_{*,s}^{-1/3} a^{-1}$, $g_{*,s}$ being the (temperature dependent) number of relativistic degrees of freedom in entropy in the dark sector, and therefore $\Delta N_{\text{eff}}$ increases rapidly around the mass threshold. This is completely analogous to the familiar case of photons around the electron mass scale.

Since the mass threshold of interest is close to the temperature of recombination, the high-$\ell$ and low-$\ell$ CMB modes can be affected differently \cite{Aloni:2021eaq}; in particular, while both modes experience the usual background effect of dark radiation, the high-$\ell$ modes ``see" a smaller value of $\Delta N_{\text{eff}}$, corresponding to the early (pre-threshold) abundance, than the low-$\ell$ modes (for an analysis of such models comparing constraints from different ends of the multipole spectrum, see~\cite{Schoneberg:2022grr}). A specific phase shift is then induced mostly of the high-$\ell$ peaks, which in practice allows for a larger (post-threshold, but still pre-recombination) value of $\Delta N_{\text{eff}}$ and may thus lead to a more decisive alleviation of the $H_0$ tension. Importantly, below the mass threshold, the abundance of massive particles is rapidly depleted via annihilations, and therefore the usual hot dark matter bound from CMB and LSS is evaded.

From a particle physics perspective, these models can be straightforwardly implemented by assuming some particles in the dark sector to have a small mass $m$ (this is also technically natural if the particles are fermions as in~\cite{Buen-Abad:2022kgf}), or alternatively by considering a phase transition analogous to the electroweak or the QCD ones (in the latter case again avoiding naturalness issues in the dark sector). This offers a key advantage over other popular scenarios to raise $H_0$, such as early dark energy~\cite{Poulin:2018cxd, Niedermann:2020dwg} (EDE); the crucial ingredient of EDE is a fluid which decays faster than radiation and which does not have an obviously natural particle physics realization. On the other hand, those latter models have been thoroughly tested against cosmological datasets, and in particular the effects of prior choices as well as the constraining role of the full shape of the BOSS DR12~\cite{Gil-Marin:2014sta, BOSS:2016wmc, BOSS:2016hvq, BOSS:2016psr} galaxy power spectrum extracted by means of the effective field theory of LSS \cite{Baumann_2012,Carrasco_2012,Hertzberg_2014} have been highlighted~\cite{DAmico:2019fhj, Ivanov:2019pdj,Hill:2020osr, Ivanov:2020ril, DAmico:2020ods, Niedermann:2020qbw, Smith:2020rxx, Schoneberg:2021qvd, Simon:2022adh}.

The main aim of this work is thus to fill this gap, by presenting a careful analysis of light dark sectors with mass thresholds, also referred to as {\it stepped dark radiation} (\SDR), including the aid of the latest cosmological datasets. Specifically, we present a Bayesian analysis, which accounts for effects of prior choices on dark sector parameters and includes galaxy clustering data. Our analysis applies model-independently to any interacting dark sector with a mass threshold around the eV scale, that is decoupled from the SM at the epochs probed by CMB observations. For comparison, we also provide results for the case without a mass threshold.

Beyond the simple \SDR\, model described above, we also analyze extensions that include interactions between dark matter and dark radiation~\cite{Joseph:2022jsf, Buen-Abad:2022kgf}. These are motivated by an additional, albeit much milder, discrepancy between CMB and late Universe measurements of cosmological parameters; this concerns the amplitude of matter fluctuations at late times, conventionally quantified by the parameter $S_8\equiv \sigma_8\sqrt{\Omega_m/0.3}$, where $\sigma_8$ is the amplitude of the matter power spectrum at redshift $z=0$ averaged over $8 \mbox{ Mpc}/h$ scales ($h\equiv H_0/(100\,\text{km/s/Mpc})$)  and $\Omega_m$ is the total matter relic abundance. Recent galaxy clustering~\cite{Philcox:2021kcw} and shear surveys, among them most importantly KiDS-1000~\cite{KiDS:2020suj, Heymans:2020gsg} and the Dark Energy Survey (DES)~\cite{DES:2021bvc, DES:2021vln, DES:2021wwk}, currently prefer a smaller value of $S_8$ than what is inferred by Planck CMB observations assuming the $\Lambda$CDM model, with the discrepancy around $3\sigma$. Models that add energy density around recombination to address the $H_0$ tension typically cause a shift of $S_8$ to larger values to keep the goodness of the fit to CMB data, and therefore exacerbate this so-called ``$S_8$ tension." For EDE-like models, simple extensions have already been proposed and tested with LSS data~\cite{Allali:2021azp}, which allow simultaneous alleviation of both tensions. Here we provide a similar analysis for the extended stepped dark radiation models of~\cite{Joseph:2022jsf, Buen-Abad:2022kgf}.

Our work is the first one to test SDR models (and simple interacting DR models without mass thresholds) with LSS data. Previous work~\cite{Schoneberg:2022grr} has investigated the effects of priors in these models on the Hubble tension and derived constraints using big bang nucleosynthesis (BBN) observations. In our work, we discuss different prior choices (to avoid possibly important volume effects) and assume that the dark radiation is produced after BBN, since this does not require new ingredients at the scales probed by the CMB, and can be accommodated with model building (for instance by the post-BBN decay of a massive particle, see e.g.~\cite{Hasenkamp:2012ii}). Furthermore, the extended SDR models that we consider differ importantly from that constrained in~\cite{Schoneberg:2022grr}, in that we include a rapid turn-off of DM-DR interactions below the mass threshold, as predicted by particle physics implementations of these scenarios~\cite{Joseph:2022jsf, Buen-Abad:2022kgf}. Refs.~\cite{Aloni:2021eaq, Joseph:2022jsf} presented results under restrictive prior choices, and keeping some dark sector parameters fixed in their cosmological analyses. In contrast and in order to at least partially account for ``look-elsewhere" effects, we allow for all parameters to vary, with more conservative prior choices. Finally,~\cite{Buen-Abad:2022kgf} proposed a particle physics extension of the SDR model that features a stronger turn-off of the DM-DR interaction rate than in~\cite{Joseph:2022jsf}, without however testing it against cosmological datasets. In our work, we test both proposals~\cite{Joseph:2022jsf, Buen-Abad:2022kgf}, and we also allow for the interacting DM fraction to vary. We also include terms in the SDR perturbation equations that were missed by~\cite{Aloni:2021eaq, Schoneberg:2021qvd, Joseph:2022jsf} and provide clarifications.

This paper is organized as follows: In Section~\ref{sec:SDR}, we describe the SDR models of interest and present the perturbation equations used in this work;
In Section~\ref{sec:data}, we outline the datasets, discuss methods for assessing cosmological tensions and present the results of our Bayesian analyses. Finally, in Section~\ref{sec:conclusions}, we provide the overall conclusions of our work.

\section{Dark Sector Model}
\label{sec:SDR}

\noindent The properties of an interacting  dark sector that undergoes a change in its number of relativistic degrees of freedom (referred to as the stepped dark radiation, or \SDR , model) can be described in terms of an effective fluid model with redshift-dependent equation of state parameter $w$ and sound speed of perturbations $c_s^2$:
\begin{equation}
\label{eq:fluidpars}
w(z)\equiv \frac{p(z)}{\rho(z)}, \quad c_s^2(z)\equiv \frac{dp(z)/dz}{d\rho(z)/dz},
\end{equation}
where $z$ is the redshift. Much before the epoch $z_t$ at which the change in number of degrees of freedom occurs, the fluid behaves simply as radiation, i.e. $w=c_s^2=1/3$. As the Universe approaches $z_t$, $w$ and $c_s^2$ temporarily decrease since a non-negligible fraction of the energy density in the fluid is initially made of massive species (which are pressureless and thus have $w=0$). Since those species remain in thermal equilibrium with the remaining light degrees of freedom, their abundance is rapidly depleted and the sector is again described by a simple radiation fluid. The redshift dependence of  $w$ and $c_s^2$ can then be simply determined starting from \eqref{eq:fluidpars}, see~\cite{Aloni:2021eaq} and Appendix~\ref{app:SDR} for details. Fundamentally, these features are due to the existence of a mass scale $m$, such that once $T\simeq m$ a certain number of states becomes nonrelativistic. 

\begin{figure}
\includegraphics[width=0.48\linewidth]{ 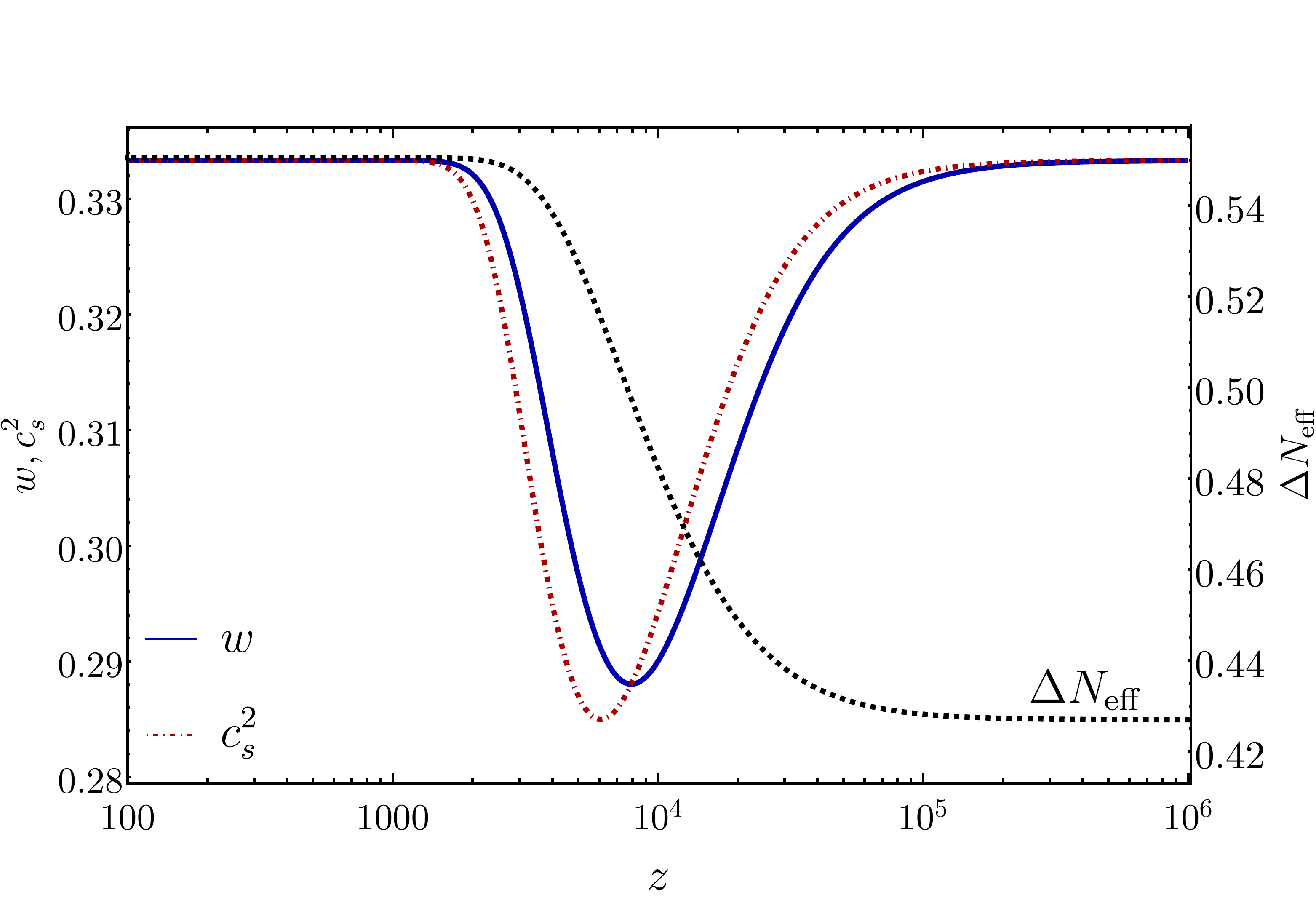}
\hspace{0.5cm}
\includegraphics[width=0.47\linewidth]{ 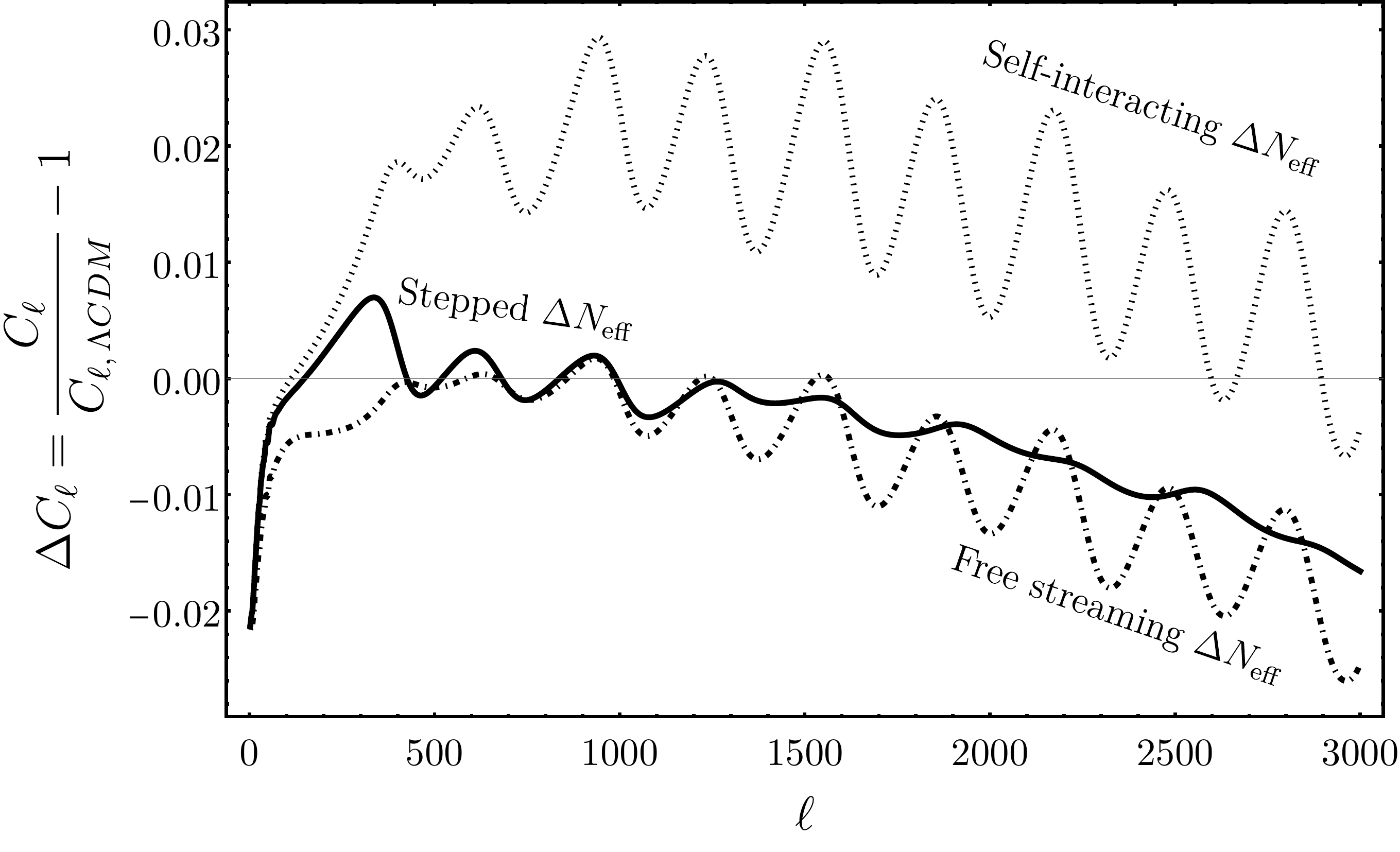}
  \caption{{\it Left:} dark sector fluid equation of state parameter and sound speed of perturbations as a function of redshift. The black dashed line shows the resulting behavior of the dark radiation abundance (see right vertical axis for its values). {\it Right:} the fractional change in multipole coefficients $C_\ell$ of the TT power spectrum in the self-interacting radiation models with (solid curve) and without (dotted curve) a mass threshold, and in the free-streaming radiation model (dot-dashed curve), all with respect to the $\Lambda$CDM model. $\Delta N^{\text{IR}}_{\text{eff}}=0.28$ has been used for the three dark radiation models, for which cosmological parameters have been fixed to the best-fit values of the stepped DR model obtained with Planck18+BAO+Pantheon (in particular, $H_0=69.34~\text{km/s/Mpc}$), while best-fit parameters for the $\Lambda$CDM model have been used for comparison ($H_0=67.9~\text{km/s/Mpc}$), see Table~\ref{tab:appdata1} for details.}
\label{fig:Nofz}
\end{figure}

It is customary to parameterize the energy density of relativistic species in terms of  $\Delta N_{\text{eff}}$. We are interested in a dark sector that decouples from neutrinos at sufficiently early times before recombination (if it has ever been in thermal contact at all). Therefore, the change in $g_*$ in the light sector induces a time-dependence in $\Delta N_{\text{eff}}$: indeed approximate entropy conservation in the dark sector implies that its temperature decreases temporarily more slowly than the temperature of neutrinos with cosmic expansion, i.e. $T_{d}\sim (g^d_{*,s})^{-1/3} a^{-1}$, where $a$ is the scale factor, whereas $T_\nu\sim a^{-1}$. Notice that $g^d_{*,s} (g^d_{*})$ is the temperature-dependent number of relativistic degrees of freedom in entropy (energy) in the dark sector, defined via the entropy density $s_{d}=(2\pi^2/45) g^d_{*,s}(T_d) T_{d}^3$ (or via the energy density). Therefore, $\Delta N_{\text{eff}}\propto g^d_{*}(T_d)(T_d/T_\nu)^4\propto g^d_{*}(T_d)(g^d_{*,s}(T_d))^{-4/3}$ increases as the dark sector undergoes a decrease in its number of relativistic species. Away from $z_t$, $g^d_{*}$ and $g^d_{*,s}$ are constant and equal in our case, since the dark sector species are all in thermal equilibrium, and thus one can define the relative change (we often drop the script $d$ in what follows and take all quantities to refer always to the dark sector, unless otherwise noted)
\begin{equation}
r_g\equiv \frac{g_{*}^{\text{UV}}-g_{*}^{\text{IR}}}{g_{*}^{\text{IR}}}=\left(\frac{\Delta N^{\text{IR}}_{\text{eff}}}{\Delta N^{\text{UV}}_{\text{eff}}}\right)^3-1,
\end{equation}
where we use superscript IR and UV for quantities evaluated at $z\ll z_t$ and $z\gg z_t$ respectively. Away from $z_t$, the dark sector temperature $T_d$ is related to the temperature of the SM bath by
\begin{equation}
T^{\text{IR,UV}}_d\sim 0.5\left(\frac{2}{g^{\text{IR,UV}}_*}\right)^{\frac{1}{4}}\left(\frac{\Delta N^{\text{IR,UV}}_{\text{eff}}}{0.3}\right)^{\frac{1}{4}} T_{\text{SM}}.
\label{TdTSM}
\end{equation}
For the values of $\Delta N^{\text{IR}}_{\text{eff}}$, $ g_{*}$ of interest, the dark sector is only slightly colder than the SM bath. Furthermore, the dark sector temperature today is related to the fundamental mass scale and the redshift $z_t$ by $m=T^0_d(1+z_t)$, or in terms of model parameters it is given by
\beq 
\label{eq:mass}
m \simeq 1.2 \,\mbox{eV}\,\left(\frac{1+\zt}{10^4}\right)\left(\frac{\Delta N_{\text{eff}}^{\text{IR}}}{0.3}\right)^{1/4}\left(\frac{2}{g_{*}^{\text{IR}}}\right)^{1/4}
\eeq 

The redshift dependence of $\Delta N_{\text{eff}}$ around $z_t$ can be determined by computing the evolution of $\rho$ (which we review in Appendix~\ref{app:SDR}) and is shown in Fig.~\ref{fig:Nofz}, together with $w$ and $c_s^2$ for example values of $\Delta N_{\text{eff}}^{\text{IR}}$ and $r_g$. Overall, the effective fluid of the \SDR\, model is thus characterized by three independent parameters, which can be chosen to be $\Delta N_{\text{eff}}^{\text{IR}}, r_g$ and $z_t$. 

As usual, the inclusion of fluid perturbations is crucial for cosmological analyses. We will assume that the dark sector bath is sufficiently strongly interacting that it effectively behaves as an ideal relativistic fluid rather than as free-streaming radiation (see e.g.~\cite{Blinov:2020hmc}). In such case, the perturbation equations read (in synchronous gauge)~\cite{Ma:1995ey}:
\beq \dot{\delta} = -(1+w)(\theta + \frac{\dot{h}}{2})-3\cH (c_s^2-w)\delta
\label{deltaeq}\eeq
\beq \dot{\theta} = -\cH (1-3w)\theta - \frac{\dot{w}}{1+w} \theta + \frac{c_s^2}{1+w}k^2 \delta - k^2 \sigma
\label{thetaeq}\eeq
where $\delta \equiv \delta \rho/\bar{\rho}, \theta= i k^j v_j$ are the density and velocity perturbations, respectively ($k^j$ is the wave mode and $v^j$ is the fluid velocity); overdots indicate conformal time derivatives; $h$ is the trace of the scalar metric perturbation; $\cH\equiv \dot{a}/{a}$ is the Hubble parameter in conformal time; and $\sigma$ is the shear perturbation of the fluid. One can readily confirm that in the limit of a perfect radiation fluid these equations take on the familiar form for radiation. 

Taking the dark radiation fluid to be shearless, we can see that in addition to the changing values of $w$ and $\cs$, there are ``new" terms in the perturbation equations for $\deltaSDR$ and $\thetaSDR$ which vanish for pure radiation. First, the term proportional to $(\cs-w)$ does not vanish near the step since the shifts of $w$ and $\cs$ are not lockstep, see Fig.~\ref{fig:Nofz}. Second, the term proportional to $(1-3w)$ does not vanish for all times. Finally, the term proportional to $\dot{w}$ is nontrivial during the step when $w$ evolves.

The impact of the dark sector model considered so far on CMB anisotropies is shown on the right side of Fig.~\ref{fig:Nofz} (solid blue curve). We plot the fractional change in the $C_l$'s for best-fit values of dark sector parameters (reported in Tab.~\ref{tab:WZdata}, for the baseline dataset, see also Table~\ref{tab:appdata1}), with respect to the $\Lambda$CDM model (with its own best-fit values of cosmological parameters). For comparison, we show also the simple (without mass threshold) free-streaming and self-interacting dark radiation models, both with $\Delta N_{\text{eff}}=\Delta N_{\text{eff}}^{\text{IR}}$ and with the same values of cosmological parameters as for the dark sector model. One can appreciate that the change with respect to $\Lambda$CDM is significantly smaller for the dark sector model than for free-streaming radiation at high-$\ell$. With respect to the simple self-interacting dark radiation model, the difference is $O(2-3\%)$ for $\ell\gtrsim 500$.

\subsection{Interactions with dark matter}

\noindent In addition to the model described so far, we will also be interested in extensions that allow for interactions between the dark radiation sector and (a fraction of) the dark matter. Such extended models are observationally motivated by the $S_8$ tension (see also~\cite{Buen-Abad:2015ova, Archidiacono:2019wdp}), since the growth of matter fluctuations is suppressed in the presence of interactions with other components.
In these scenarios, the background evolution of the dark radiation (\SDR) and interacting dark matter (\SDRIDM) fluids remains as above, while the perturbation equations of the two fluids, again in synchronous gauge, are modified as follows (see also Appendix~A4 of~\cite{Buen-Abad:2022kgf}):

\begin{align}
\label{thetasdr}
\thetaSDRdot &= -\cH (1-3w)\thetaSDR - \frac{\dot{w}}{1+w} \thetaSDR + \frac{\delta P /\delta \rho}{1+w}k^2 \deltaSDR -a\GamSDR \frac{\rhoIDM}{\rhoSDR(1+w)} (\thetaSDR-\thetaIDM)\\
\label{thetaidm}
\thetaIDMdot &= -\cH\thetaIDM  +a\GamSDR(\thetaSDR-\thetaIDM),
\end{align}
where $\Gamma$ is the (thermally averaged) dark matter-dark radiation momentum-transfer rate.
In general, only a fraction $\fdm\equiv \rhoIDM/\rhoDM\rvert_{z\gg z_\text{rec}}$ of the total dark matter may have been interacting with the dark radiation at early times, where $\rhoDM=\rhoIDM+\rho_{\text{cdm}}$ and $\rho_{\text{cdm}}$ is the standard noninteracting cold dark matter component (with the same velocity perturbation equation as \eqref{thetaidm}, except for the absence of the interaction term proportional to $\Gamma$). 

\begin{figure}
\includegraphics[width=0.6\linewidth]{ 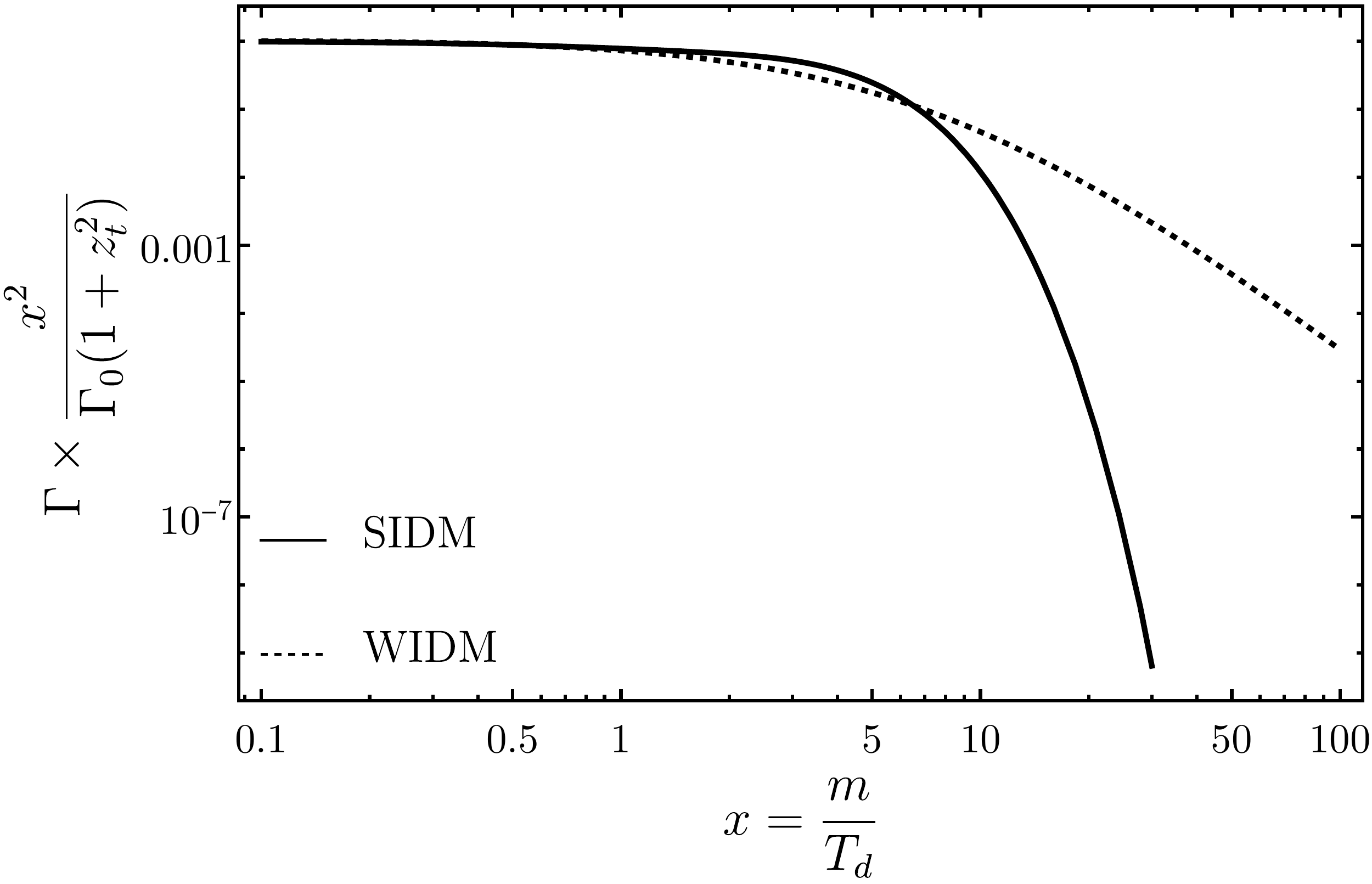}
  \caption{Dark matter-dark radiation interaction rates considered in this work, according to the models of~\cite{Joseph:2022jsf}, referred to as weakly interacting dark matter, or WIDM (dashed curve), and~\cite{Buen-Abad:2022kgf}, referred to as strongly interacting dark matter, or SIDM (solid curve).}
\label{fig:rates}
\end{figure}

Several possible types of dark matter-dark radiation interactions can be envisioned~\cite{Cyr-Racine:2015ihg} (see also~\cite{Lesgourgues:2015wza, Buen-Abad:2017gxg, Archidiacono:2019wdp} for cosmological studies), leading to different redshift dependence of the corresponding interaction rates. However, the models of interest for this work share a peculiar behavior: the rate $\Gamma$ rapidly decreases after a certain redshift relatively close to $z_t$. Therefore, no matter how big the interaction rate, the dark matter is entirely noninteracting shortly after this redshift. Such a behavior is due to the fact that dark matter-dark radiation interactions involve particles that become heavy around $z_t$ (thereby causing the change in the number of relativistic species). We focus here on two well-motivated scenarios, recently considered in~\cite{Joseph:2022jsf} and~\cite{Buen-Abad:2022kgf}. In both cases, $\Gamma\sim \alpha^2 T^2/m_{\text{\idm}}$ at $z\gg z_t$, where $m_{\text{\idm}}$ is the mass of the interacting dark matter component, taken to be much heavier than $\text{keV}$ so that the component is certainly cold at recombination, and $\alpha$ is the coupling strength. The difference between the two scenarios is in the strength of the interactions and their decrease after the redshift $z_t$. In particular, the DM-DR interactions can be:

\begin{itemize}
    \item{\textbf{Mediated by particles that become heavy around $z_t$}~\cite{Joseph:2022jsf}: in this case the most important interaction is scattering of the interacting dark matter component with the dark radiation species that remain light at $z_t$. A simple reference model has the interacting dark matter component being a fermion $\chi$, and the dark radiation made of a complex scalar field $\phi$ (with mass $m_\phi$ which becomes relevant at $z_t$) and a fermion $\psi$ (effectively massless at $z_t$). Both $\chi$ and $\psi$ are coupled to $\phi$ via Yukawa terms. At $z\gg z_t$, i.e. $T_d\gg m_\phi$, scatterings $\chi \psi\rightarrow \chi \psi$ mediated by $\phi$ in the t-channel give rise to $\Gamma\sim \alpha^2 T_d^2/m_{\chi}$. At $z\lesssim z_t$, i.e. $T_d\lesssim m_\phi$, the (four-fermion) scattering rate decreases more rapidly as:
    \begin{equation}
    \Gamma_{z\lesssim z_t}\sim \alpha^2\frac{T^2_d}{m_{\chi}}\left(\frac{T_d}{m_{\phi}}\right)^4.
    \end{equation}
    Below, we will mostly consider this scenario in the regime of weak interaction rates between the DM and the SDR, and we thus denote it as weakly interacting dark matter (WIDM) (here weak refers to small interacting rates, not to the electroweak scale).}
    \item{\textbf{Mediated by particles that remain light at $z_t$}~\cite{Buen-Abad:2022kgf} (see also~\cite{Chacko:2016kgg}): in this case the scattering process of interest involves particles that become heavy as external states. A simple reference model has the interacting dark matter component being a scalar $\chi$ charged under a dark $U(1)$ gauge sector, whose photon $A$ is the light component of the dark radiation sector. The latter also features charged fermions $\psi$, with typical mass $m_\psi$. At $T_d\gg m_\psi$, $\psi$-$\chi$ scatterings mediated by $A$ again lead to $\Gamma\sim \alpha T_d^2/m_\chi$. At $z\lesssim z_t$, the rate is however exponentially (Boltzmann) suppressed, because the $\psi$ population is non-relativistic. The rate is then
    \begin{equation}
    \Gamma_{z\lesssim z_t}\sim \alpha^2\frac{T^2_d}{m_{\chi}}e^{-\frac{m_\psi}{T_d}}.
    \end{equation}
     Below, we will mostly consider this scenario in the regime of large interaction rates between the DM and the SDR, and we thus denote it as strongly interacting dark matter (SIDM).}
    
\end{itemize}

\noindent In what follows, we encode the strength of interactions using the following parametrization for the momentum transfer rate:
\begin{equation}
\Gamma = \Gamma_0\left(\frac{1+z_t}{x}\right)^2 \left[1+b h_1(x)\right] h_2(x),
\label{gammaform}
\end{equation}
where $x\equiv m/T_d$ and $h_{1,2}$ are functions such that $(1+b h_1(x)) h_2(x)\rightarrow 1$ at early times before $z_t$, i.e. as $x\rightarrow 0$. The WIDM model of~\cite{Joseph:2022jsf} corresponds to setting $b=h_1(x)=0$ and $h_2(x)\sim x^{-4}$ at $x\gg 1$. The interactions thus introduce only one extra parameter $\Gamma_0$ (beyond $\fdm$). On the other hand, the SIDM model of~\cite{Buen-Abad:2022kgf} has $h_1(x)\sim x, h_2(x)\sim x^2 e^{-x}$ at $x\gg 1$, and thus introduces an additional parameter $b$ with respect to the previous model (see Appendix~\ref{app:SDRDM} for more details on the temperature dependence of $\GamSDR$). We plot the rates for the two scenarios of~\cite{Joseph:2022jsf} and~\cite{Buen-Abad:2022kgf} in Fig.~\ref{fig:rates}. Fundamentally, the parameter $\Gamma_0$ contains the combination $\alpha^2 m^2/m_{\text{idm}}$, where $\alpha$ is the coupling strength of the interaction and $m_{\text{idm}}$ is the mass of the interacting dark matter species. One can think of $\GamOSDR$ as the would-be momentum transfer rate today in the case of no mass threshold (since this corresponds to $x\propto m\rightarrow 0$); with the mass threshold, the interaction rate goes quickly to zero after $\zt$. In the SIDM scenario, the additional parameter $b$ is given by $1/\log[\pi/(g^\psi_{*}\alpha^3)]$, where $g^\psi_{*}$ corresponds to the degrees of freedom in the SDR component that becomes massive at $\zt$ (the fermion $\psi$ described above for SIDM); this term arises from the regularization of infrared divergences in scatterings involving massless gauge bosons.

We can determine the efficiency of interactions by comparing the momentum transfer rate $\GamSDR$ to the Hubble parameter $H$. At early times, $(1+z_t)/x \propto a$ (see also Appendix~\ref{app:SDR}), and since $(1+b h_1(x)) h_2(x)\rightarrow 1$, both $\GamSDR$ and $H$ go as $a^{-2}$. Therefore, their ratio is roughly a constant in the early universe: 
\beq\frac{\GamSDR}{H}\sim \frac{\GamOSDR}{10^{-6}\, \mbox{Mpc}^{-1}} \,\left(\frac{\Delta N_{\text{eff}}^{\text{IR}}}{0.3}\right)^{1/2} \left(\frac{2}{g_{*}^{\text{IR}}}\right)^{1/2}
\label{Gam0bound}
\eeq
where we have properly related the temperatures of the visible and dark sectors, and obtained a relationship that depends only loosely on $\Delta N_{\text{eff}}^{\text{IR}}$ and $g_{*}^{\text{IR}}$. Therefore $\GamOSDR$ also determines whether the DM and DR are tightly coupled at early times, which occurs for $\GamOSDR\gtrsim 10^{-6}~\text{Mpc}^{-1}$.

\begin{figure}
\includegraphics[width=0.48\linewidth]{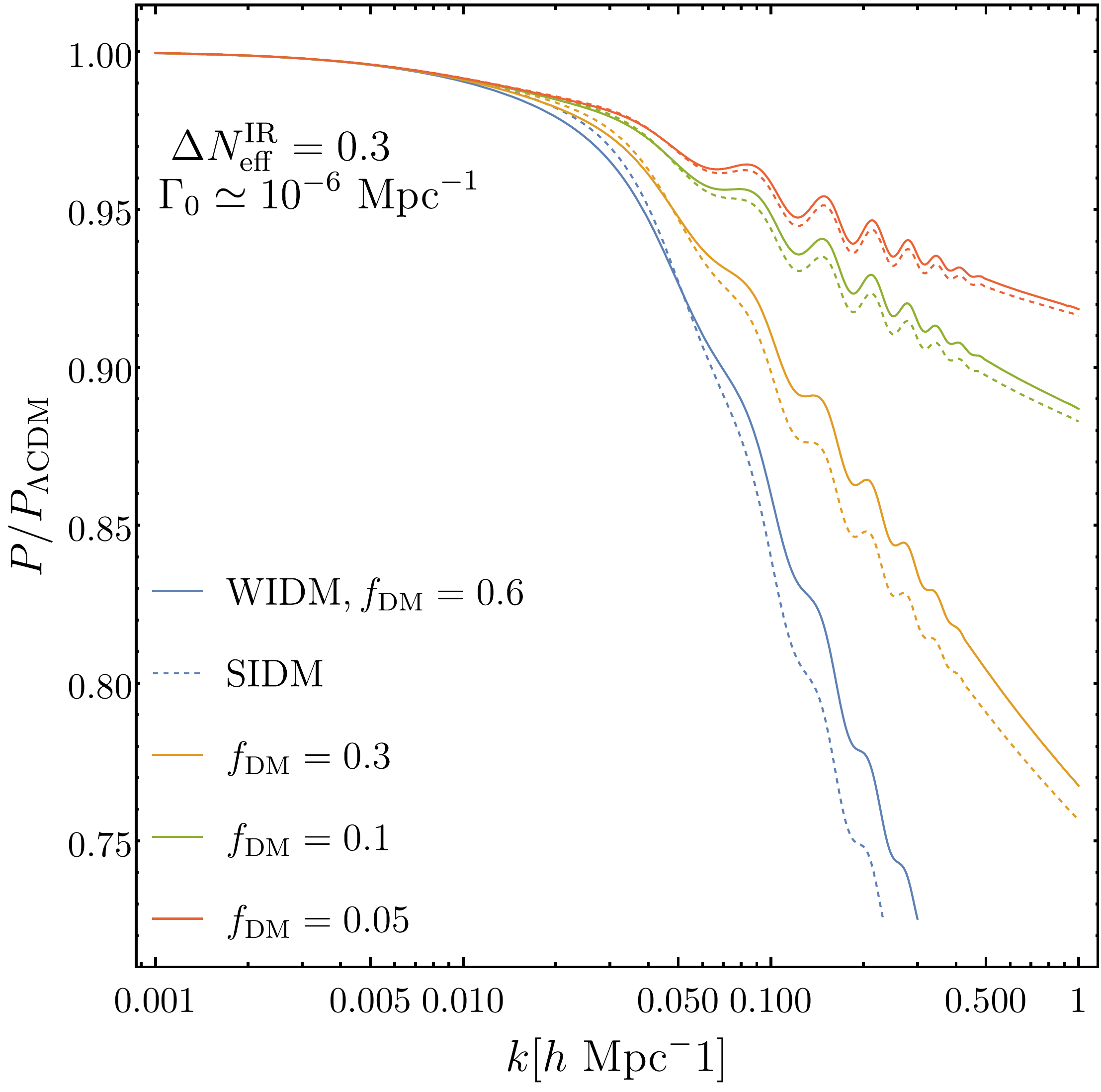}
\hfill
\includegraphics[width=0.48\linewidth]{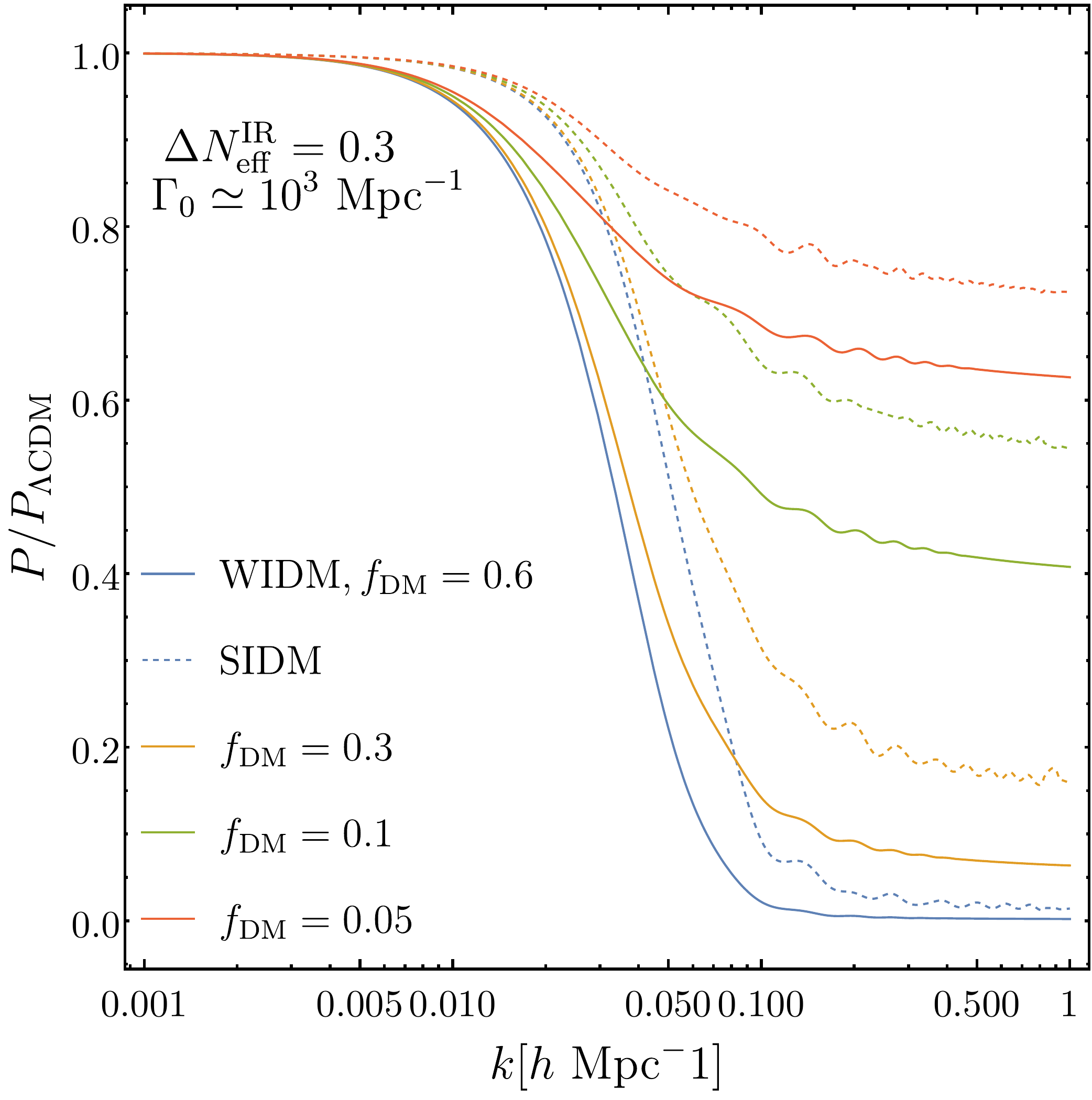}
\caption{Suppression of the matter power spectrum in models considered in this work, compared to the $\Lambda$CDM model. {\it Left:} $\Gamma_0=10^{-6}~\text{Mpc}^{-1}$. {\it Right:} $\Gamma_0=10^{3}~\text{Mpc}^{-1}$. For both plots, we have taken $\NIR=0.3, r_g=1.75$ and $\log_{10} z_t =3.8$. Solid (dashed) curves refer to the WIDM (SIDM) model, while the interacting DM fraction increases from the top curve to the bottom curve.}
\label{fig:suppression}
\end{figure}

The effects of dark matter-dark radiation interactions on the matter power spectrum in the WIDM and SIDM scenarios are shown in Fig.~\ref{fig:suppression}. 
We have considered two example cases of interaction strength: in the left panel, we show results for $\Gamma_0\simeq 10^{-6}~\text{Mpc}^{-1}\simeq 10^{-36}~\text{eV}\ll H_0$, corresponding to a would-be interaction rate that is slower than the Hubble rate today. Even at early times, this interaction rate is not very efficient according to \eqref{Gam0bound}.
In such case, large values of $f_\text{DM}$ are allowed, since the suppression effect is otherwise small. In the right panel, we show the suppression for a much larger rate $\Gamma_0\simeq 10^{3}~\text{Mpc}^{-1}\simeq 10^{-27}~\text{eV}$, such that dark matter-dark radiation interactions are efficient at early times
all the way until $\zt$. Small fractions of interacting dark matter are then enough to provide a strong suppression of the power spectrum. Notice that the WIDM and SIDM models give similar results for small values of the interaction rate, as expected since in this case neither is very efficient. This is different for the case of large interaction rates, where the decay of the rate with redshift becomes important. One can indeed appreciate that the SIDM model leads to a sharper (in $k$) suppression  than the WIDM model. Based on these results, we expect the SIDM and WIDM models to perform similarly (their background behavior is the same) for small $\Gamma_0$, and to possibly differ significantly only for large $\Gamma_0$ [as determined by \eqref{Gam0bound}].

With the addition of interactions, the dark sector models of interest for this work introduce a total of five or six parameters beyond $\Lambda$CDM depending on whether the WIDM or SIDM model for interactions is adopted. In the next section, we shall first consider a dark sector model without dark matter interactions, whose impact we assess only at a second stage.

Before moving on to the presentation of our results, an important comment is in order. Beyond the CMB and LSS spectra, it is well known that additional relativistic degrees of freedom affect big bang nucleosynthesis (BBN) as well. In the models of interest, BBN would then lead to a constraint on $\Delta N_{\text{eff}}^{\text{UV}}\leq 0.39$ ($95\%$ C.L., BBN+$Y_p$+$D$)~\cite{Fields:2019pfx} (see also~\cite{Planck:2018vyg} for discussion on uncertainties). However, the dark sector might be populated after BBN, so that the constraint above would not apply. For instance, one may consider a massive particle that decays into a light dark sector after BBN, while contributing a negligible fraction of the energy density at the epoch of BBN (see also~\cite{Aloni:2023tff} for a post-BBN dark sector model involving dark sector-neutrino interactions, or~\cite{Ghosh:2023ocl} for another scenario with relativistic degrees of freedom generated after BBN). Therefore, in order not to exclude such possibilities, we will not be imposing a BBN constraint on $\Delta N_{\text{eff}}^{\text{UV}}$ in our work (see instead~\cite{Schoneberg:2022grr} for a different perspective). Therefore, an additional layer of detail needs to be included in any viable particle physics model.

\section{Datasets and Results}
\label{sec:data}

We implemented the dark sector fluid above in the Boltzmann solver {\tt CLASS}~\cite{Lesgourgues:2011re, Blas:2011rf}. We review the details of the numerical implementation in Appendix~\ref{app:SDR} (see also Appendix~A of~\cite{Aloni:2021eaq}). For perturbations, we implemented~\eqref{deltaeq} and~\eqref{thetaeq} with $\sigma=0$. We consider first a model with only the stepped dark radiation (\SDR), which is characterized by nine parameters in total: six from the $\Lambda$CDM model, plus three from the dark radiation fluid. Then, we will consider the addition of interactions between the SDR and a fraction of the dark matter. In the weakly interacting dark matter (WIDM) scenario, there are two additional free parameters, for a total of eleven parameters. Meanwhile in the strongly interacting dark matter (SIDM) scenario, there are, in principle, three additional parameters, but for the sake of comparison we will fix one parameter ($b$, whose value anyway varies only logarithmically with fundamental parameters) such that this model also has eleven free parameters. We also fix the neutrino sector according to the standard choice of one massive (with mass $m_\nu=0.06~\text{eV}$) and two massless species.

We perform Bayesian searches using the Markov Chain Monte Carlo (MCMC) sampler {\tt MontePython}\footnote{\href{https://github.com/brinckmann/montepython_public}{\tt https://github.com/brinckmann/montepython\_public}}~\cite{Audren:2012wb, Brinckmann:2018cvx}. All our searches have Gelman-Rubin parameter $R-1<0.02$ (most of them $<0.01$), except for some results on the SIDM model, see below. To analyze and plot the posterior distributions of parameters, we make use of {\tt GetDist}\footnote{\href{https://getdist.readthedocs.io}{\tt https://getdist.readthedocs.io}}~\cite{Lewis:2019xzd}. 

We use the following datasets to test the dark sector model described in the previous section:

\begin{itemize}
\item{\textbf{Baseline dataset: P18+BAO+Pantheon}. Planck 2018 high-$\ell$ and low-$\ell$ TT, TE, EE and lensing data~\cite{Aghanim:2019ame}; BAO measurements from 6dFGS at $z = 0.106$~\cite{Beutler:2011hx}, SDSS MGS at $z = 0.15$~\cite{Ross:2014qpa} (BAO smallz), and CMASS and LOWZ galaxy samples of BOSS DR12 at $z = 0.38$, $0.51$, and $0.61$~\cite{Alam:2016hwk}; Pantheon Supernovae data sample~\cite{Pan-STARRS1:2017jku}. This is our baseline dataset. }
\item{+\textbf{FS}: the baseline dataset with the addition of the full-shape of the power spectrum of galaxies in the BOSS/SDDS sample, extracted by means of the EFTofLSS~\cite{DAmico:2019fhj, Ivanov:2019pdj, Colas:2019ret}. We use the publicly released {\tt PyBird}\footnote{\href{https://github.com/pierrexyz/pybird}{\tt https://github.com/pierrexyz/pybird}}~\cite{DAmico:2020kxu} code as a combined likelihood with BAO data from the same sample. We use the latest version of the likelihood, which accounts for a correction to the normalization of BOSS window functions, see~\cite{Simon:2022adh}.\footnote{Using {\tt CLASS-PT}~\cite{Chudaykin:2020aoj} rather than {\tt PyBird} has been shown to lead to milder constraints from LSS, see e.g.~\cite{Simon:2022adh}.}}
\item{$\mathbf{+S_8}$: any of the two datasets above with the addition of two measurements of the $S_8$ parameter from cosmic shear analyses of KiDS-1000, $S_8=0.759^{+0.024}_{-0.021}$~\cite{KiDS:2020suj} and DES-Y3, $S_8=0.772^{+0.018}_{-0.017}$~\cite{DES:2021bvc}. For computation of tension measures, we use the combined value of $S_8 = 0.767 \pm 0.014$, see below.}
\item{$\mathbf{+M_b}$: any of the datasets above with the addition of the latest measurement of the intrinsic SNIa magnitude $M_b=-19.253\pm0.027$ from the SH$_{0}$ES collaboration~\cite{Riess2022a}.}
\end{itemize}

For cosmic shear, we use the measurements of $S_8$ rather than the full likelihood, because the latter requires an algorithm to compute nonlinearities, which is currently only available for the $\Lambda$CDM model. For the SH$_{0}$ES measurement, we use $M_b$ rather than the Hubble constant $H_0$ as we are combining with the Pantheon sample, see~\cite{Benevento:2020fev, Camarena:2021jlr, Efstathiou:2021ocp}, and correspondingly we assess tensions using $M_b$.

\subsection{Criteria to assess tensions}

We assess the impact of the dark sector on cosmological tensions by means of several criteria. First, we compare the minimum $\chi^2$ of the stepped dark radiation (SDR) model under consideration for a given dataset with the minimum $\chi^2$ of the $\Lambda$CDM model with the same dataset, $\Delta\chi^2\equiv \chi^2_{\text{sdr}}-\chi^2_{\Lambda \text{CDM}}$. Obviously, if $\Delta\chi^2>0$, then the dark sector model is disfavored compared to $\Lambda$CDM. Even when $\Delta\chi^2<0$, the evidence for the dark sector model is not necessarily relevant, because of the additional parameters. 

Additionally, we determine the tension between the posteriors $\mathcal{P}_{\MCMC}$ for $M_b$ (or $S_8$) in a given model and its measurement, also represented by a distribution $\mathcal{P}_{m}$, by integrating the cross-correlation of the two distributions (i.e. the probability of parameter differences between the distributions), as described in~\cite{Raveri:2021wfz}. This method is useful when posteriors are non-Gaussian, as is often the case in models that modify cosmology around recombination. More specifically, when the posterior  $\mathcal{P}_{m}$ is Gaussian (a good approximation for the SH$_0$ES measurement of $M_b$ and for the $S_8$ measurements by cosmic shear surveys), the probability of a difference between the two distributions is given by 
\beq
\Delta = \int^{\infty}_{-\infty} \mathcal{P}_{\MCMC}(\theta_1)\frac{1}{2}\left(1 \pm \mbox{erf}\left(\frac{\theta_1-\mu_{m}}{\sqrt{2}\sigma_{m}}\right)\right) d\theta_1
\eeq
where $\theta_1$ represents the parameter of interest. We have taken $\mathcal{P}_{m}$ to be a Gaussian with mean $\mu_{m}$ and variance $\sigma_{m}^2$. The $+$ sign ($-$ sign) then corresponds to $\mu_{m}<\mu_{\MCMC}$ ($\mu_{m}>\mu_\MCMC$)~\cite{Raveri:2021wfz}, where $\mu_{\MCMC}$ is the mean of the posterior from our search. A tension between the two distributions as a multiple $IT$ of standard deviations of a pure Gaussian is determined by solving
\beq
\Delta = \int_{-\infty}^{IT}  \frac{1}{\sqrt{2\pi}}e^{-\frac{1}{2}x^2}dx.
\eeq
We will refer to this measure of the tension as the ``integrated tension" (IT). This computation improves on the more commonly used ``Gaussian tension" (GT) defined by
\beq
GT = \frac{|\mu_m-\mu_\MCMC|}{\sqrt{\sigma_m^2+\sigma_\MCMC^2}},
\label{GT}
\eeq
since in general $\mathcal{P}_\MCMC$ is non-Gaussian. A simple intuition for the IT measure can be understood as follows: if the measurement were infinitely precise like a delta function $\mathcal{P}_{m}(\theta_1)=\delta(\theta_1-\theta_0)$, one wishes to determine the probability of the accuracy of this measurement $\theta_0$ given the distribution $\mathcal{P}_{\MCMC}$. Then, the value ``$IT\,\sigma$" simply denotes the placement of $\theta_0$ in the distribution $\mathcal{P}_{\MCMC}$ as a multiple of standard deviations away from the mean.

\begin{table*}
\begin{tabular} {| l | c| c|}
\hline\hline
 \multicolumn{1}{|c|}{ Parameter} &  \multicolumn{1}{|c|}{Baseline} &  \multicolumn{1}{|c|}{Baseline + FS}\\
\hline\hline
$H_0 \,[\textrm{km}/\textrm{s}/\textrm{Mpc}]$ & $67.66~(67.9)^{+0.41}_{-0.41}     $ & $67.82~(67.99)^{+0.39}_{-0.39}     $\\
$S_8                       $ & $0.825~(0.823)^{+0.010}_{-0.010}   $ & $0.8202~(0.8217)^{+0.0099}_{-0.0099}$\\
$M_b                       $ & $-19.419~(-19.413)^{+0.012}_{-0.011} $ & $-19.414~(-19.411)^{+0.011}_{-0.011} $\\
\hline
\hline
$Q_{\textrm{\tiny DMAP}}^{M_b}$ & $5.73\sigma $ & $5.52\sigma $\\
\hline
$M_b$ GT & $5.63\sigma $ & $5.52\sigma $\\
\hline
$M_b$ IT & $5.63\sigma $ & $5.52\sigma $\\
\hline
$Q_{\textrm{\tiny DMAP}}^{S_8}$ & $3.46\sigma $ & $3.01\sigma $\\
\hline
$S_8$ GT & $3.24\sigma $ & $3.02\sigma $\\
\hline
$S_8$ IT & $3.23\sigma $ & $3.0\sigma $\\
\hline
\end{tabular}
\caption{Measures of tension are given for the $\Lambda$CDM model, including $\QDMAP$, Gaussian tension (GT), and integrated tension (IT) for both $M_b$ and $S_8$.  Mean (best-fit) $\pm 1\sigma$ are also given for $H_0$, $S_8$, and $M_b$.}
\label{tab:Ltens}
\end{table*}

We compute both the IT and GT measures for $M_b$ and $S_8$ for different datasets. In the case of $S_8$, we will use a combined value of the two priors from KiDS-100 and DES-Y3 (as indicated in our dataset above). This combined value comes from approximating each of the two $S_8$ measurements as a Gaussian with mean $\mu_i$ and variance $\sigma_i^2$, and taking the product (joint probability) of those distributions. This gives a new Gaussian probability distribution with  mean $\mu = (\mu_1 \sigma_2^2 + \mu_2 \sigma_1^2)/(\mu_1^2+\mu_2^2)$ and variance $\sigma^2 = \sigma_1^2\sigma_2^2/(\sigma_1^2+\sigma_2^2)$. We use the positive error for each of the $S_8$ measures, since their means lie below the MCMC inferences. We therefore use $S_8=0.767$ as the mean value and $0.014$ as the upper $1\sigma$ error bar, for computing the tension. On the other hand, we use both $S_8$ measurements in our MCMC analysis, see above. For GTs, in the case of asymmetric error bars obtained from the posteriors of our MCMC, we use the upper (lower) $1\sigma$ error bar for $M_b$ ($S_8$) to compute the tension, as in~\cite{Schoneberg:2021qvd}. We notice that in~\cite{Joseph:2022jsf}, the $1\sigma$ error bar for computing GT is obtained by taking half of the $2\sigma$ range in the posterior instead. This latter method consistently gives a smaller GT than in our work or in~\cite{Schoneberg:2021qvd}.

In addition to the criteria above, we also report the values of: the {\it difference of the maximum a posteriori}
$\QDMAP^{M_b}=\sqrt{\chi^2_{\text{w/} M_b}-\chi^2_{\text{w/o}~ M_b}}$~\cite{Raveri:2018wln} (see also~\cite{Schoneberg:2021qvd}) between the minimum $\chi^2$'s obtained by fitting the same model to a dataset with and without the measurement of $M_b$ (or $\QDMAP^{S_8}$ for $S_8$); the Akaike Information Criterion~\cite{1100705} (see also~\cite{Liddle:2007fy}) $\AIC^{M_b}\equiv \Delta\chi^2 + 2\times($\# of added free param.s$)$ for a dataset which includes the measurement of $M_b$ (or $\AIC^{S_8}$ for $S_8$). With these criteria, we report tensions in the $\Lambda$CDM model in Table~\ref{tab:Ltens}. One can see that with the latest measurements of $M_b$ from SH$_0$ES, the tension with $\Lambda$CDM firmly exceeds $5 \sigma$.

In the following subsections, we present results of our searches for dark sectors with mass thresholds and also report updated results for the simpler scenario of interacting dark radiation without a mass threshold. To highlight difference with previous work, we analyze in steps the effects of: different prior choices, fixing some parameters of the model, and the inclusion of galaxy-clustering data. Our reference prior choices are indicated in the last column of Table~\ref{tab:WZdata} and in Table~\ref{tab:priorsidm}.

\subsection{Dark radiation}

We first focus on the pure stepped dark radiation (\SDR) model, setting to zero the interactions with the dark matter. Before giving the final result of our search, we discuss the implications of certain prior choices, as well as of the inclusion of datasets beyond our baseline.

\subsubsection*{Narrow vs broad priors on $\log_{10} z_t$}\label{sec:widepriors}

\begin{figure}
\begin{minipage}{0.48\textwidth}
         \resizebox{1.05\textwidth}{!}{
\begin{tabular} {| l | c| c|}
\hline\hline
 \multicolumn{1}{|c|}{ Parameter} &  \multicolumn{1}{|c|}{$\log_{10}z_t \in [4.0,4.6]$} &  \multicolumn{1}{|c|}{$\log_{10}z_t \in [3.0,5.0]$}\\
\hline\hline
$\Delta N^{\textrm{\scriptsize IR}}_{\textrm{\scriptsize eff}}$ & $ < 0.597~(0.551)$ & $ < 0.59~(0.551)$
\\
$\log_{10} z_t             $ & Unconstrained $(4.3)$ & Unconstrained $(4.3)$\\
\hline
$H_0 \,[\textrm{km}/\textrm{s}/\textrm{Mpc}]$ & $69.30~(70.68)^{+0.86}_{-1.3}      $ & $69.11~(70.68)^{+0.80}_{-1.3}      $\\
$S_8                       $ & $0.829~(0.837)^{+0.011}_{-0.011}   $ & $0.827~(0.837)^{+0.011}_{-0.011}   $\\
$M_b                       $ & $-19.369~(-19.325)^{+0.025}_{-0.038} $ & $-19.374~(-19.325)^{+0.024}_{-0.037} $\\
\hline
$\Delta\chi^2$ & $-0.41$ & $-0.41$\\
\hline
$Q_{\textrm{\tiny DMAP}}^{M_b}$ & $2.55\sigma $ & $2.55\sigma $\\
\hline
$M_b$ GT & $3.12\sigma $ & $3.38\sigma $\\
\hline
$M_b$ IT & $2.56\sigma $ & $2.7\sigma $\\
\hline
$\Delta	\textrm{AIC}^{M_b}$ & $-22.67$ & $-22.67$\\
\hline
\end{tabular}
}
\end{minipage}
     \hfill
\begin{minipage}{0.48 \textwidth}

\includegraphics[width=1.0\linewidth]{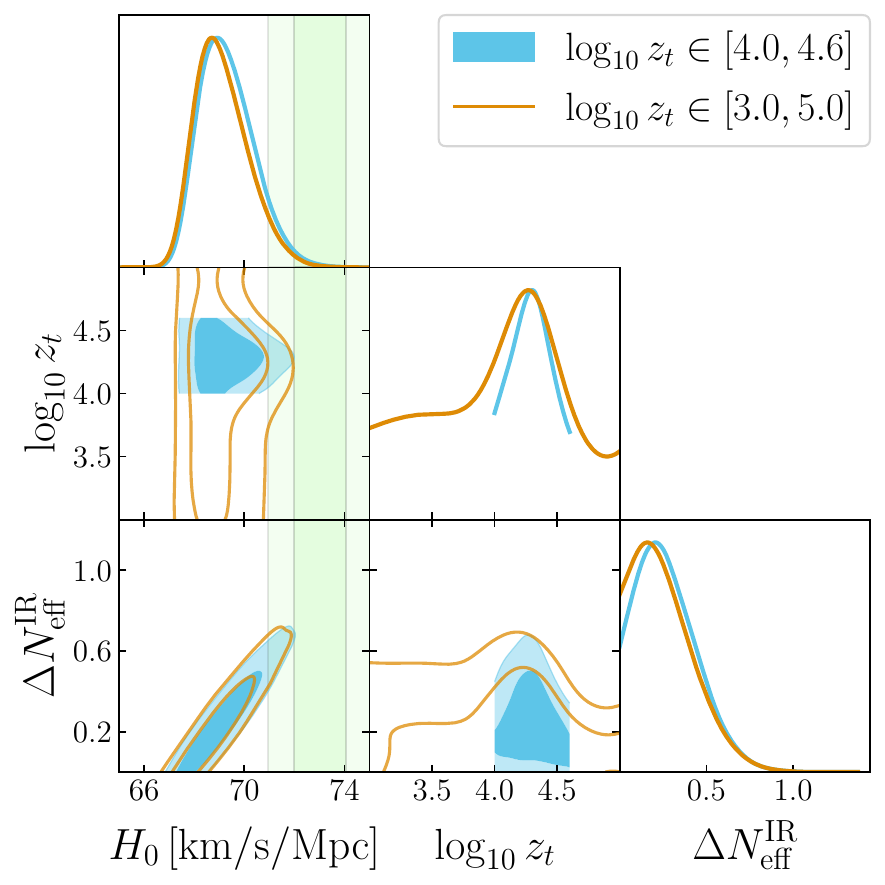}
\end{minipage}
        \caption{{\it Left}:  mean (best-fit) $\pm 1\sigma$ error of dark sector parameters obtained by fitting the two-parameter \SDR\, model (i.e. with $r_g$ fixed) to the baseline dataset P18+BAO+Pantheon, comparing two choices of prior on $\log_{10} \zt$. Upper bounds are presented at 95\% C.L., and parameters without constraints at 95\% C.L. within their prior boundaries are marked as unconstrained. Tension measures are reported with respect to the SH$_0$ES measurement of $M_b$. {\it Right}: one- and two-dimensional posterior distributions for dark sector parameters and $H_0$. The posterior of $H_0$ inferred by SH$_0$ES is shown by shaded vertical regions ($1$ and $2\sigma$ ranges).}
        \label{fig:widerpriors}
\end{figure}

We start by considering the effects of prior choices on the redshift $z_t$, which we sample logarithmically. We consider two choices: first, we set $\log_{10} z_t\in [4.0,4.6]$ as in~\cite{Aloni:2021eaq, Joseph:2022jsf}; second, we set slightly broader priors $\log_{10} z_t\in [3.0, 5.0]$.\footnote{Much wider prior boundaries have been considered in~\cite{Schoneberg:2022grr}, $\log_{10} z_t\in [0.0,10.0]$, which may however lead to strong volume effects.} In terms of the particle physics parameter, i.e. the mass threshold, we see from \eqref{eq:mass} that our choice roughly corresponds to scanning mass thresholds $0.1~\text{eV}\lesssim m\lesssim 10~\text{eV}$ for $\Delta N^{\text{IR}}_{\text{eff}}\sim 0.1$, whereas the choice of \cite{Aloni:2021eaq,Joseph:2022jsf} restricts the search to the very narrow range $\text{eV}\lesssim m\lesssim 4~\text{eV}$. In the absence of a particle physics model that predicts such a specific mass range, we find it more fair to adopt wider priors.

We fix the step size parameter $r_g=1.14$ in this analysis, motivated by a dark sector composed of one complex scalar and a Weyl fermion, as in~\cite{Aloni:2021eaq}. We use only our baseline dataset in this comparison.

Results are reported in Fig.~\ref{fig:widerpriors}, together with plots of posterior distributions. The following observations can be made. First, widening the priors does not affect the bestfit values of parameters, therefore bestfit-based tension measures (i.e. those based on minimum $\chi^2$) are similarly not altered. On the other hand, the GT is affected by the choice of priors (only a minor effect), since the $M_b$ posterior is now shifted to smaller values. For both choice of priors the GT is above $3\sigma$, slightly more so with the wider prior.

We also report a very minor improvement in $\chi^2$ compared to $\Lambda$CDM, i.e. $\Delta\chi^2\simeq -0.4$ with two extra free parameters. Our results in the left column of the table are overall in slight disagreement with those of~\cite{Aloni:2021eaq}, obtained with the same choices. In particular, we find a slightly larger $\Delta\chi^2$ (by one unit), i.e. less improvement of the fit compared to $\Lambda$CDM than in~\cite{Aloni:2021eaq, Joseph:2022jsf}. We also find a larger GT tension, due both to the new SH$_0$ES  measurement (with respect to~\cite{Aloni:2021eaq}) and the different prescription used to compute the GT (with respect to~\cite{Joseph:2022jsf}).

From now on, we fix our priors as $\log_{10} z_t\in [3.0,5.0]$, to (at least partially) account for the ``look elsewhere" effect related to the choice of very narrow priors.

\begin{figure}
     \begin{minipage}{0.48\textwidth}
         \centering
         \resizebox{1.1\textwidth}{!}{
\begin{tabular} {| l | c| c|}
\hline\hline
 \multicolumn{1}{|c|}{ Parameter} &  \multicolumn{1}{|c|}{$r_g$ fixed} &  \multicolumn{1}{|c|}{$r_g$ free}\\
\hline\hline
$\Delta N^{\textrm{\scriptsize IR}}_{\textrm{\scriptsize eff}}$ & $ < 0.59~(0.551)$ & $ < 0.546~(0.289)$\\
$\log_{10} z_t             $ & Unconstrained $(4.3)$ & Unconstrained $(4.29)$\\
$r_g                       $ & --- & Unconstrained $(4.0)$                         \\
\hline
$H_0 \,[\textrm{km}/\textrm{s}/\textrm{Mpc}]$ & $69.11~(70.68)^{+0.80}_{-1.3}      $ & $68.89~(69.34)^{+0.71}_{-1.1}      $\\
$S_8                       $ & $0.827~(0.837)^{+0.011}_{-0.011}   $ & $0.827~(0.834)^{+0.011}_{-0.011}   $\\
$M_b                       $ & $-19.374~(-19.325)^{+0.024}_{-0.037} $ & $-19.381~(-19.369)^{+0.021}_{-0.032} $\\
\hline
$\Delta\chi^2$ & $-0.41$ & $-1.4$\\
\hline
$Q_{\textrm{\tiny DMAP}}^{M_b}$ & $2.55\sigma $ & $2.74\sigma $\\
\hline
$M_b$ GT & $3.38\sigma $ & $3.74\sigma $\\
\hline
$M_b$ IT & $2.7\sigma $ & $3.03\sigma $\\
\hline
$\Delta \textrm{AIC}^{M_b}$ & $-22.67$ & $-20.67$\\
\hline
\end{tabular}
}
\vspace{1.5cm}
     \end{minipage}
     \hfill
     \begin{minipage}{0.5\textwidth}
         \centering
         \includegraphics[width=0.85\textwidth]{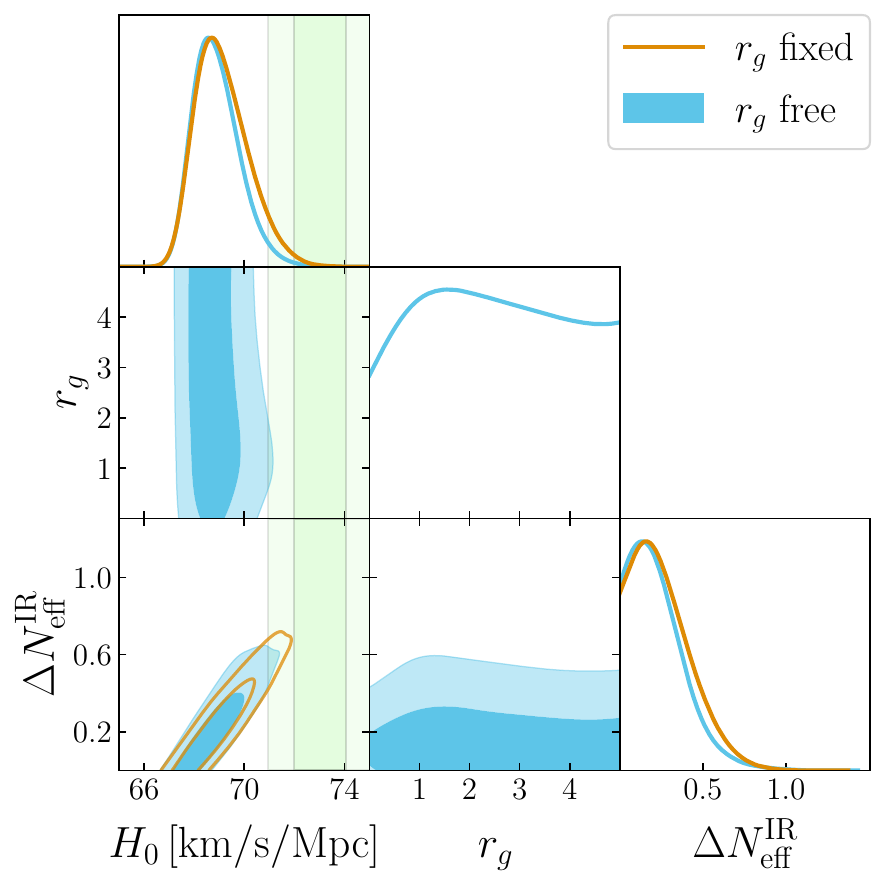}
     \end{minipage}
        \caption{{\it Left}:  mean (best-fit) $\pm 1\sigma$ error of dark sector parameters obtained by fitting the two- or three-parameter \SDR\, model (i.e. with $r_g$ fixed or free to vary) to the baseline dataset P18+BAO+Pantheon. Upper bounds are presented at 95\% C.L., and parameters without constraints at 95\% C.L. within their prior boundaries are marked as unconstrained. Tension measures are reported with respect to the SH$_0$ES measurement of $M_b$. {\it Right}: one- and two-dimensional posterior distributions for dark sector parameters and $H_0$. The posterior of $H_0$ inferred by SH$_0$ES is shown by shaded vertical regions ($1$ and $2\sigma$ ranges).}
        \label{fig:rgfree}
\end{figure}

\subsubsection*{Step size fixed vs free to vary}

We now study the implications of leaving the step size $r_g$ free to vary in our search. The motivation to do so is twofold: First, we currently do not have any compelling theory prediction for $r_g$, as there is not a specific mass spectrum for the dark sector which is better-motivated than any other one (for instance, there may be more than one complex scalar field and/or fermion in the models of~\cite{Aloni:2021eaq, Buen-Abad:2022kgf}). Second, to perform a fair comparison with other models for the $H_0$ tension, such as in particular early dark energy (where the three parameters $f_{\text{EDE}}, z_c$ and $\theta_i$ are kept free to vary). We thus vary $r_g\in [0, 5]$, where the choice of the upper prior is somewhat arbitrary.

The comparison with the previous results ($r_g$ fixed as in~\cite{Aloni:2021eaq, Joseph:2022jsf}) and posterior distributions are given in Fig.~\ref{fig:rgfree}. The following differences can be appreciated. First, the goodness-of-the-fit is slightly improved, as expected from the addition of one extra parameter, while the best-fit value of $r_g$ is twice as large as the previously fixed value. Second, all tension measures are now affected: in particular, the GT is now well above $3\sigma$ (and the integrated tension is at $3\sigma$), while the $Q_{\text{DMAP}}$ also approaches $3\sigma$.

\subsubsection*{Comparison to self-interacting dark radiation without mass threshold}

Before moving to the next step of our analysis, let us compare the SDR model to the self-interacting dark radiation (SIDR) model without a mass threshold, using only our baseline dataset, see Table~\ref{tab:SIDR}. SDR (i) refers to fixing the parameter $r_g$ and using the narrower prior $\log_{10} \zt \in [4.0,4.6]$ as in~\cite{Aloni:2021eaq, Joseph:2022jsf}. SDR (ii) refers to the strategy which we emphasize in this paper, namely leaving $r_g$ free to vary and using the wider prior $\log_{10} \zt \in [3.0,5.0]$.

One can see that for both strategies of analyzing SDR, the bound on $\Delta N_{\text{eff}}$ is significantly relaxed compared to SIDR. Additionally, for the implementation (i), the SDR model shows a relevant reduction in all tension measures with respect to the SH$_0$ES measurement of $M_b$.

However, in the analysis we highlight in this work, SDR (ii), one must make a more nuanced comparison. On the one hand, the Hubble tension (with respect to $M_b$) is no longer reduced compared to SIDR for specifically the tension measures which compare between the posterior distributions fit to the baseline dataset (GT and IT).
On the other hand, when comparing the goodness-of-fit measures of tension (namely those that depend on $\chi^2$), SIDR performs worse. While the $\Delta \chi^2$ for SIDR is only marginally worse than for SDR, the $\QDMAP$ tension measure, which compares the goodness-of-fit with and without the inclusion of the SH$_0$ES measurement, shows a significantly larger tension for SIDR. The $\AIC$ measure which compares the goodness-of-fit when including the SH$_0$ES measurement relative to $\Lambda$CDM, also shows a slightly worse tension for SIDR. The best-fit value of $H_0$ for the SIDR model is also significantly smaller than in the stepped scenario.

Overall, this comparison shows that the SDR model consistently alleviates the constraint on $\Delta N_{\text{eff}}$ compared to the SIDR model, while the impact on the $H_0$ tension is only slightly better in the SDR model than for SIDR, when sampling the parameter space more broadly. 

\begin{table*}
\begin{tabular} {| l | c| c| c|}
\hline\hline
 \multicolumn{1}{|c|}{ Parameter} &  \multicolumn{1}{|c|}{SIDR} &  \multicolumn{1}{|c|}{SDR (i)} &  \multicolumn{1}{|c|}{SDR (ii)}\\
\hline\hline
$\Delta N^{\textrm{\scriptsize IR}}_{\textrm{\scriptsize eff}}$ & $ < 0.456$ (95\% CL) &$ < 0.597~(0.551)$ & $ < 0.546~(0.289)$\\
\hline
$H_0 \,[\textrm{km}/\textrm{s}/\textrm{Mpc}]$ & $68.95~(68.38)^{+0.73}_{-1.2}      $ & $69.30~(70.68)^{+0.86}_{-1.3}      $ & $68.89~(69.34)^{+0.71}_{-1.1}      $\\
$S_8                       $ & $0.823~(0.818)^{+0.011}_{-0.011}   $ & $0.829~(0.837)^{+0.011}_{-0.011}   $ & $0.827~(0.834)^{+0.011}_{-0.011}   $\\
$M_b                       $ & $-19.380~(-19.397)^{+0.022}_{-0.034} $ & $-19.369~(-19.325)^{+0.025}_{-0.038} $ & $-19.381~(-19.369)^{+0.021}_{-0.032} $\\
\hline
$\Delta\chi^2$ & $-0.22$ & $-0.41$ & $-1.4$\\
\hline
$Q_{\textrm{\tiny DMAP}}^{M_b}$ & $3.48\sigma $ & $2.55\sigma $ & $2.74\sigma $\\
\hline
$M_b$ GT & $3.66\sigma $ & $3.12\sigma $ & $3.74\sigma $\\
\hline
$M_b$ IT & $2.95\sigma $ & $2.56\sigma $ & $3.03\sigma $\\
\hline
$\Delta\textrm{AIC}^{M_b}$ & $-18.88$ & $-22.67$ & $-20.67$\\
\hline
\end{tabular}
\caption{Mean (best-fit) $\pm 1\sigma$ error of $H_0$, $S_8$, and $M_b$ are given along with a 95\% C.L. upper bound on $\Delta N_{\text{eff}}$ for the SIDR model obtained by fitting to the baseline dataset P18+BAO+Pantheon. For comparison, the SDR model is shown with: (i) $r_g$ fixed and $\log_{10} \zt \in [4.0,4.6]$ and (ii) $r_g$ free and $\log_{10} \zt \in [3.0,5.0]$. Tension measures are reported with respect to the SH$_0$ES measurement of $M_b$.}
\label{tab:SIDR}
\end{table*}

\subsubsection*{Adding full-shape data}

\begin{table*}

\begin{tabular} {| l | c| c| c| c|}
\hline\hline
 \multicolumn{1}{|c|}{ Parameter} &  \multicolumn{1}{|c|}{Baseline} &  \multicolumn{1}{|c|}{Baseline $+$ FS} &  \multicolumn{1}{|c|}{Baseline $+$ FS $+M_b$} &  \multicolumn{1}{|c|}{Priors}\\
\hline\hline
$\Delta N^{\textrm{\scriptsize IR}}_{\textrm{\scriptsize eff}}$ &$ < 0.546~(0.289)$ & $ < 0.55~(0.08)$ & $0.69~(0.63)^{+0.14}_{-0.23}      $ & $[0,\infty)$\\
$\log_{10} z_t             $ & Unconstrained $(4.29)$ & Unconstrained $(4.97)$ & Unconstrained $(4.22)$ & [3,5]\\
$r_g                       $ & Unconstrained $(4.0)$  & Unconstrained $(2.34)$  & Unconstrained $(1.14)$ & [0,5]\\
\hline
$H_0 \,[\textrm{km}/\textrm{s}/\textrm{Mpc}]$ & $68.89~(69.34)^{+0.71}_{-1.1}      $ & $69.01~(68.37)^{+0.66}_{-1.1}      $ & $71.71~(72.17)^{+0.83}_{-0.80}     $ & ---\\
$S_8                       $ & $0.827~(0.834)^{+0.011}_{-0.011}   $ & $0.821~(0.824)^{+0.010}_{-0.010}   $ & $0.816~(0.82)^{+0.010}_{-0.012}   $& ---\\
$M_b                       $ & $-19.381~(-19.369)^{+0.021}_{-0.032} $ & $-19.378~(-19.4)^{+0.019}_{-0.032} $ & $-19.300~(-19.285)^{+0.024}_{-0.024} $& ---\\
\hline
$\Delta\chi^2$ & $-1.4$ & $-1.62$ & $-24.69$& ---\\
\hline
$Q_{\textrm{\tiny DMAP}}^{M_b}$ & $2.74\sigma $ & $2.72\sigma $ & ---& ---\\
\hline
$M_b$ GT & $3.74\sigma $ & $3.77\sigma $ & $1.29\sigma $& ---\\
\hline
$M_b$ IT & $3.03\sigma $ & $2.94\sigma $ & $1.29\sigma $& ---\\
\hline
$\Delta \textrm{AIC}^{M_b}$ & $-20.67$ & $-18.68$ & ---& ---\\
\hline
\end{tabular}

\caption{Mean (best-fit) $\pm 1\sigma$ error of dark sector parameters obtained by fitting the three-parameter \SDR\, model to three datasets: the baseline dataset P18+BAO+Pantheon, the baseline + FS, and the baseline + FS + $M_b$. Upper bounds are presented at 95\% C.L., and parameters without constraints at 95\% C.L. within their prior boundaries are marked as unconstrained. Tension measures are reported with respect to the S$H_0$ES measurement of $M_b$. Priors for the \SDR\, parameters are given in the last column.}
\label{tab:WZdata}
\end{table*}

\begin{figure}
\centering
\includegraphics[width=0.6\textwidth]{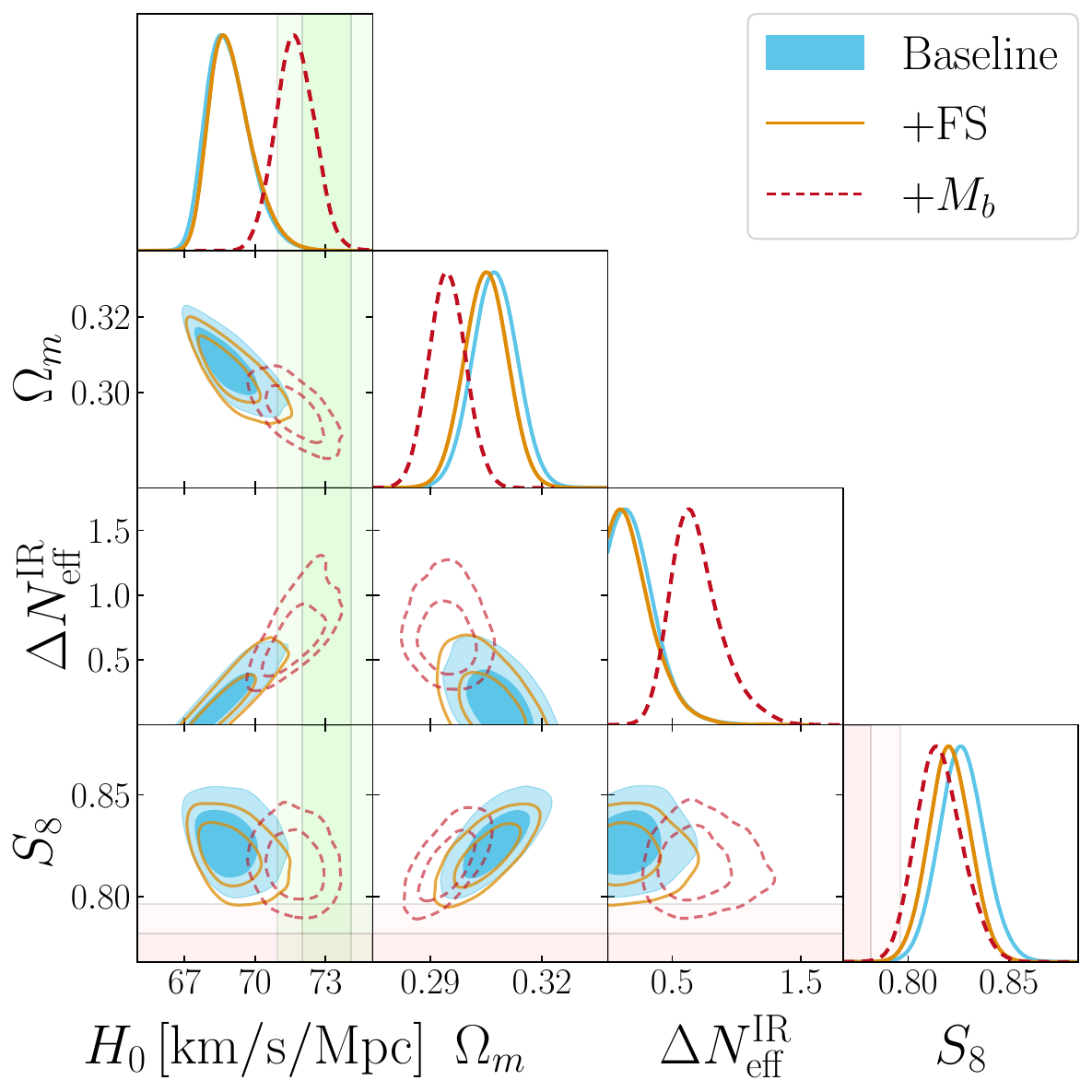}
\caption{One- and two-dimensional posterior distributions for selected parameters fit to three different datasets: the baseline dataset P18+BAO+Pantheon, the baseline + FS, and the baseline +FS+$M_b$. The light green (lighter green) vertical bars show the 1-$\sigma$ (2-$\sigma$) bounds of the SH$_0$ES measurement of $H_0$, and the light pink (lighter pink) horizontal bars show the 1-$\sigma$ (2-$\sigma$) bounds of the combined $S_8$ measurement from KiDS-1000 and DES-Y3. For more posteriors, see Appendix~\ref{app:data}.}
\label{fig:WZdata}
\end{figure}

Finally, we consider the addition of BOSS data on the full-shape (FS) of the power spectrum of galaxies, which has been shown to impact other proposals to address the Hubble tension (see e.g.~\cite{Hill:2020osr, Ivanov:2020ril, DAmico:2020ods} and the recent reassessment~\cite{Simon:2022adh} for the EDE scenario).

Results are reported in Table~\ref{tab:WZdata} and Fig.~\ref{fig:WZdata}. Overall, we observe the effects of FS on SDR models with our prior choices to be very mild. In particular, both the constraint on $\NIR$ and the $\Delta\chi^2$ are not significantly affected, although the best-fit values of $M_b$ and $H_0$ are indeed significantly smaller than their values without FS data. The $\Delta \text{AIC}$ is slightly increased, signaling that FS data, while not imposing strong constraints, also do not prefer values of $H_0$ as large as those required to fully alleviate the $H_0$ tension. In fact, when adding a prior on $M_b$ from the SH$_0$ES measurement, we find a residual $1.3\sigma$ tension. 

We find that the FS likelihood constrains SDR models similarly to EDE models (see Table III of~\cite{Simon:2022adh}), although the EDE model has $\Delta\chi^2\simeq -5$ with the same number of parameters as the SDR model.

\begin{figure}
    \centering
    \includegraphics[width=0.4\textwidth]{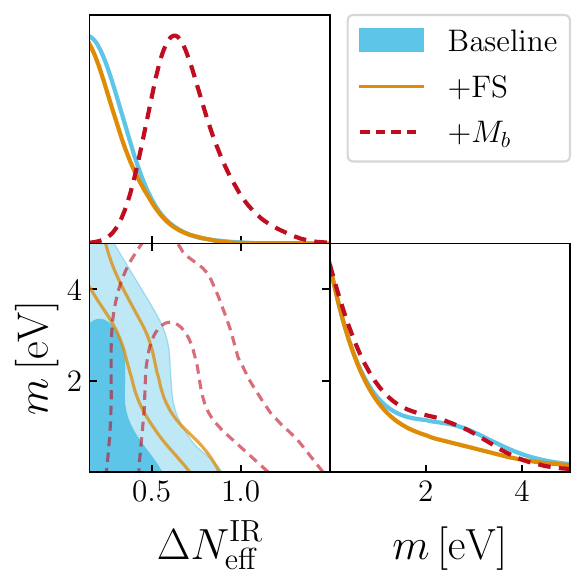}
    \caption{One- and two-dimensional posterior distributions for $m$ and $\Delta N_{\text{eff}}^{\text{IR}}$ fit to three different datasets: the baseline dataset P18+BAO+Pantheon, the baseline + FS, and the baseline +FS+$M_b$. The constraint on $m$ becomes more limiting for larger $N_{\text{eff}}^{\text{IR}}$, while for $N_{\text{eff}}^{\text{IR}}\to 0 $, one should no longer trust this constraint since $m\to0$ as well, see~\eqref{eq:mass}.}
    \label{fig:mpost}
\end{figure}

We show also in Fig.~\ref{fig:mpost} the posterior distribution for the mass scale $m$ of the threshold compared to $\Delta N_{\text{eff}}^{\text{IR}}$, obtained as a derived parameter by means of~\eqref{eq:mass}. One can see that with increasing $\Delta N_{\text{eff}}^{\text{IR}}$, the value of $m$ allowed by data at 95\% confidence decreases. On the other hand, as $\Delta N_{\text{eff}}^{\text{IR}}\to 0$, one can see from~\eqref{eq:mass} that $m\to 0$ as well, and thus the data no longer constrain $m$ in this limit, which can be seen by the fact that the 2-dimensional posterior continues to rise at small $\Delta N_{\text{eff}}^{\text{IR}}$. One should only trust the one-dimensional posterior for $m$ for sufficiently large $\Delta N_{\text{eff}}^{\text{IR}}$, keeping in mind that no such upper-bound on $m$ is possible with $\Delta N_{\text{eff}}^{\text{IR}}=0$.

We conclude that dark radiation models with mass thresholds around the epoch of recombination lead to a significant relaxation of the constraint on $\Delta N_{\text{eff}}$, obviously in particular with respect to the free-streaming case, but also with respect to the self-interacting (SIDR) model without a mass threshold. A comparison of the SDR and SIDR models including the FS dataset is given in Table~\ref{tab:sidrlss}, where one can see the relaxation of the bound on $\Delta N_{\text{eff}}$ is still present, but the improvement of the $H_0$ tension is largely gone, while the $
\chi^2$ is only minimally improved considering that the SDR model has two extra parameters compared to the SIDR model.

\begin{table*}

\begin{tabular} {| l | c| c|}
\hline\hline
 \multicolumn{1}{|c|}{ Parameter} &  \multicolumn{1}{|c|}{SIDR} &  \multicolumn{1}{|c|}{SDR}\\
\hline\hline
$\Delta N^{\textrm{\scriptsize IR}}_{\textrm{\scriptsize eff}}$ & $ < 0.457~(0.157)$ & $ < 0.55~(0.08)$\\
\hline
$H_0 \,[\textrm{km}/\textrm{s}/\textrm{Mpc}]$ & $69.13~(69.11)^{+0.73}_{-1.1}      $ & $69.01~(68.37)^{+0.66}_{-1.1}      $\\
$S_8                       $ & $0.819~(0.809)^{+0.010}_{-0.010}   $ & $0.821~(0.824)^{+0.010}_{-0.010}   $\\
$M_b                       $ & $-19.375~(-19.376)^{+0.022}_{-0.034} $ & $-19.379~(-19.402)^{+0.020}_{-0.032} $\\
\hline
$\Delta\chi^2$ & $-0.23$ & $-1.62$\\
\hline
$M_b$ GT & $3.53\sigma $ & $3.77\sigma $\\
\hline
$M_b$ IT & $2.83\sigma $ & $2.94\sigma $\\
\hline
\end{tabular}
\caption{Mean (best-fit) $\pm 1\sigma$ error of dark sector parameters obtained by fitting the SIDR and SDR models to the baseline + FS dataset. Upper bounds are presented at 95\% C.L., and parameters without constraints at 95\% C.L. within their prior boundaries are marked as unconstrained. Tension measures are reported with respect to the S$H_0$ES measurement of $M_b$. Priors for the \SDR\, parameters are given in the last column.}
\label{tab:sidrlss}
\end{table*}

While in the SDR model the Hubble tension is alleviated from $\gtrsim 5$ to $\sim 3\sigma$, the minimal improvement in $\chi^2$ over $\Lambda$CDM despite three additional parameters, as well as the significant residual tension, suggest that these models struggle to provide a convincing framework to address the discrepancy in the determinations of $H_0$. Unsurprisingly, the inclusion of SH$_0$ES measurement of $M_b$ leads to a much more significant improvement over the $\Lambda$CDM model, with $\Delta\text{AIC}\simeq -19$. However, we stress that caution should be used when interpreting this result, as it is obtained combining datasets that are in significant $\sim 3\sigma$ tension among them.

Finally, let us comment on the $S_8$ tension in this model, before we consider interactions with the dark matter. We notice a minor impact of the SDR model on $S_8$, as compared to $\Lambda$CDM, $S_8=0.827\pm 0.011$ for P18+BAO+Pantheon, $S_8=0.821\pm 0.01$ with the addition of FS, although the best-fit values are somewhat larger than for $\Lambda$CDM. Indeed, when including $S_8$ measurements to the baseline dataset with SH$_0$ES, we do find a significant increase in $\Delta\text{AIC}^{M_b}$ (as usual compared to $\Lambda$CDM with the same dataset) of approximately six units (see Table~\ref{tab:appdata7} in Appendix~\ref{app:data}), signaling that $S_8$ measurements do indeed penalize the SDR model more than $\Lambda$CDM. The interested reader can find further results in Appendix~\ref{app:data}.

\subsection{Interactions with dark matter}

We now include dark radiation-dark matter interactions, modeled as described in Sec.~\ref{sec:SDR}. Our results are the first reported in the literature for the SIDM model. For the WIDM model, we perform a more comprehensive analysis than in~\cite{Joseph:2022jsf}, including different prior choices and the BOSS FS dataset discussed above. 

Given that the addition of interactions is strongly motivated by the $S_8$ tension, it is especially important to understand the prior dependency of the $S_8$ posteriors in these models. To this aim, we consider a logarithmic prior on the interaction strength parameter $\Gamma_0$, rather than the linear prior adopted in~\cite{Joseph:2022jsf}. This choice turns out to have an important impact on $S_8$, as we outline below. As can be appreciated in Fig.~\ref{fig:suppression}, the SIDM and WIDM models give similar suppressions of the matter power spectrum for small values of $\Gamma_0$. Therefore, we restrict our analysis of the SIDM model only to large values of the interaction strength originally considered in~\cite{Buen-Abad:2022kgf}, while we analyze the WIDM model only for small values of $\Gamma_0$, as proposed by~\cite{Joseph:2022jsf}. Additionally, we let the interacting dark matter fraction $f_\text{DM}$ free to vary. We use different prior choices for this parameter in the two models, due to different region of interest (small $f_\text{DM}$ for the SIDM model, large $f_\text{DM}$ for the WIDM model). Overall, both models are thus characterized by five parameters in addition to the standard six $\Lambda$CDM parameters.\footnote{For the SIDM model, we fix the parameter $b\equiv 1/\log[\pi/(g_{\psi}\alpha^3)]$ to a well-motivated value, $b\simeq 0.04$, which is obtained for $\alpha=10^{-4}$ and for minimal fermion content $g_\psi=7/2$, as in~\cite{Buen-Abad:2022kgf}. We notice that $b$ depends only logarithmically on fundamental parameters, and thus would anyway not change dramatically as the parameter space is explored.} We summarize our prior choices for the two models in Table~\ref{tab:priorsidm}. In this section we also report tension measures with respect to $S_8$.

\begin{table*}
\begin{tabular} {| l | c| c|}
\hline\hline
 \multicolumn{1}{|c|}{ Parameter} &  \multicolumn{2}{|c|}{Priors} \\
 \hline
 \multicolumn{1}{|c|}{ } &  \multicolumn{1}{|c|}{WIDM} &  \multicolumn{1}{|c|}{SIDM}\\
\hline\hline
$\Delta N^{\textrm{\scriptsize IR}}_{\textrm{\scriptsize eff}}$ & $[0.01,\infty)$ & $[0.01,\infty)$\\
$\log_{10} \Gamma_0        $ & $[-9,-5]$ & $[-2,6]$\\
$f_{\text{DM}}$           & $[0.1,1]$ &  -\\
$\log_{10} f_{\text{DM}}$           & - &  $[-4,0]$\\
\hline
\end{tabular}
\caption{Choices of priors for dark matter-dark radiation interaction models, WIDM and SIDM. The priors on $r_g$ and $\log_{10} \zt$ are the same as given in Table~\ref{tab:WZdata}.}
\label{tab:priorsidm}
\end{table*}

\subsubsection*{Baseline dataset plus full-shape}

\begin{figure}
     \centering
     \begin{minipage}{0.5\textwidth}
         \centering
         \resizebox{1.1\textwidth}{!}{
\begin{tabular} {| l | c| c|}
\hline\hline
 \multicolumn{1}{|c|}{ Parameter} &  \multicolumn{1}{|c|}{WIDM} &  \multicolumn{1}{|c|}{SIDM}\\
\hline\hline
$\Delta N^{\textrm{\scriptsize IR}}_{\textrm{\scriptsize eff}}$ & $ < 0.531~(0.092)$ & $ < 0.519~(0.268)$\\
$\log_{10} z_t             $ & Unconstrained $(4.18)$ & Unconstrained $(4.38)$\\
$r_g                       $ & Unconstrained $(4.87)$ & Unconstrained $(4.39)$\\
$\log_{10} \Gamma_0        $ & $ < -6.156~(-8.231)$ & $ < 4.259~(3.723)$\\
$\log_{10} \fdm          $ & Unconstrained $(-0.806)$ & $ < -2.031~(-3.903)$\\
\hline
$H_0 \,[\textrm{km}/\textrm{s}/\textrm{Mpc}]$ & $68.97~(68.37)^{+0.65}_{-1.1}      $ & $68.96~(69.45)^{+0.67}_{-1.1}      $\\
$S_8                       $ & $0.818~(0.826)^{+0.011}_{-0.011}   $ & $0.820~(0.828)^{+0.011}_{-0.011}   $\\
$M_b                       $ & $-19.379~(-19.396)^{+0.019}_{-0.031} $ & $-19.380~(-19.364)^{+0.020}_{-0.032} $\\
\hline
$\Delta\chi^2$ & $-0.63$ & $-1.04$\\
\hline
$Q_{\textrm{\tiny DMAP}}^{S_8}$ & $2.75\sigma $ & $2.55\sigma $\\
\hline
$S_8$ GT & $2.82\sigma $ & $2.89\sigma $\\
\hline
$S_8$ IT & $2.63\sigma $ & $2.65\sigma $\\
\hline
$\Delta\textrm{AIC}^{S_8}$ & $7.9$ & $6.43$\\
\hline
\end{tabular}
}
\vspace{0.5cm}
     \end{minipage}
     \hfill
     \begin{minipage}{0.48\textwidth}
         \centering
         \includegraphics[width=0.85\textwidth]{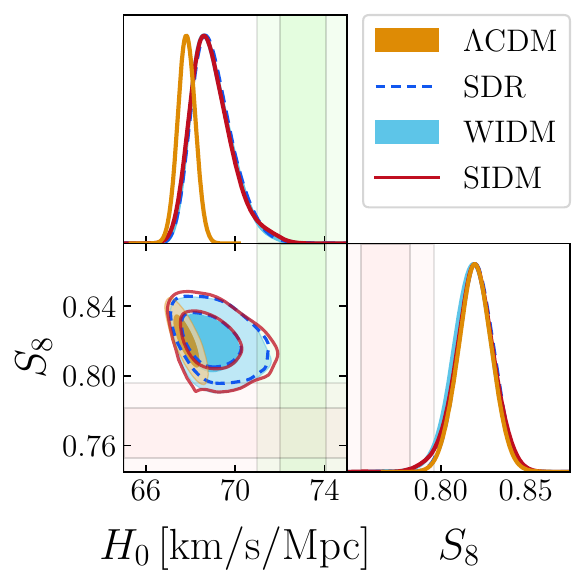}
     \end{minipage}
        \caption{{\it Left}: mean (best-fit) $\pm 1\sigma$ error of WIDM and SIDM parameters obtained by fitting to the baseline + FS dataset. Upper bounds are presented at 95\% C.L., and parameters without constraints at 95\% C.L. within their prior boundaries are marked as unconstrained. Tension measures are reported with respect to the combined $S_8$ measurement from KiDS-1000 and DES-Y3. {\it Right}: one- and two-dimensional posterior distributions are given for $H_0$ and $S_8$ obtained by fitting to the baseline + FS dataset for each model: $\Lambda$CDM, SDR, WIDM, and SIDM. The light green (lighter green) vertical bars show the 1-$\sigma$ (2-$\sigma$) bounds of the S$H_0$ES measurement of $H_0$, and the light pink (lighter pink) horizontal bars show the 1-$\sigma$ (2-$\sigma$) bounds of the combined $S_8$ measurement.}
        \label{fig:idmfs}
\end{figure}

We start by searching for the WIDM and SIDM models in our baseline + FS dataset (Planck18+BAO+Pantheon +FS). Results are reported in the left panel of Fig.~\ref{fig:idmfs}. Posteriors for $H_0$ and $S_8$ are plotted in the right panel of Fig.~\ref{fig:idmfs}, together with their posteriors obtained in the pure $\Lambda$CDM model as well as in the previously considered dark sector scenario without dark matter-dark radiation interactions.

The first and most important consideration concerns the $S_8$ parameter. As can be appreciated in the one-dimensional posterior shown in Fig.~\ref{fig:idmfs}, there is little-to-no significant difference among any of the models. Not surprisingly, the $S_8$
tension is only mildly lowered in the SIDM and WIDM models as compared to the $\Lambda$CDM model and the SDR model (e.g. the IT reported for $\Lambda$CDM is $3.0\sigma$, while for WIDM and SIDM, it is $2.63\sigma$ and $2.65\sigma$, respectively). This conclusion differs significantly from the claim in~\cite{Joseph:2022jsf}, whose $S_8$ posteriors are shifted toward significantly smaller values (with the corresponding tension below $2\sigma$). We have checked that this discrepancy is due to our choice of a logarithmic prior on the interaction strength, rather than to the addition of the FS dataset with respect to~\cite{Joseph:2022jsf}, see also Appendix~\ref{app:data} for further details. In this respect, we further notice that the best-fit value for the linearly sampled $\Gamma_0$ in~\cite{Joseph:2022jsf} is almost two orders of magnitude smaller ($\Gamma_0=5\cdot 10^{-9}~\text{km/s/Mpc}$) than the mean value of the posterior distribution ($\Gamma_0=2.95\cdot 10^{-7}~\text{km/s/Mpc}$), thereby questioning the use of a linear prior and justifying our choice. Moreover, we are able to place $95~\%$ C.L. upper limits on the interaction strength parameter $\Gamma_0$ in both models, with the prior choices on $f_{\text{DM}}$ reported in Table~\ref{tab:priorsidm} (for the WIDM model, a smaller lower prior boundary causes convergence problems).

While no MCMC results were reported for the SIDM model by the authors of~\cite{Buen-Abad:2022kgf}, we do not find compelling support for their claim that the model can simultaneously address the $H_0$ and $S_8$ tensions. In addition to the considerations on the $S_8$ posterior above, we indeed find that all measures of tensions with the weak lensing measurements of $S_8$ remain $\gtrsim 2.5~\sigma$. Moreover, the Akaike information criterion (computed with respect to $\Lambda$CDM model, including the $S_8$ priors) is positive, signaling that the $\Lambda$CDM model is actually preferred over both the SIDM and WIDM models once $S_8$ measurements are included. Furthermore, we find that the value of the interacting dark matter fraction suggested in~\cite{Buen-Abad:2022kgf}, i.e. $f_{\text{DM}}\simeq 1-5~\%$ is actually in tension with our $95~\%$ C.L. bound reported in Fig.~\ref{fig:idmfs}.

Finally, the Hubble tension remains alleviated in both models, at the same level of the \SDR\, model without interactions, and the relaxation of the bounds on $\Delta N_{\text{eff}}$ also remain qualitatively similar.

\subsubsection*{Adding $S_8$ and $M_b$ priors}

\begin{figure}
\centering
\includegraphics[width=0.6\textwidth]{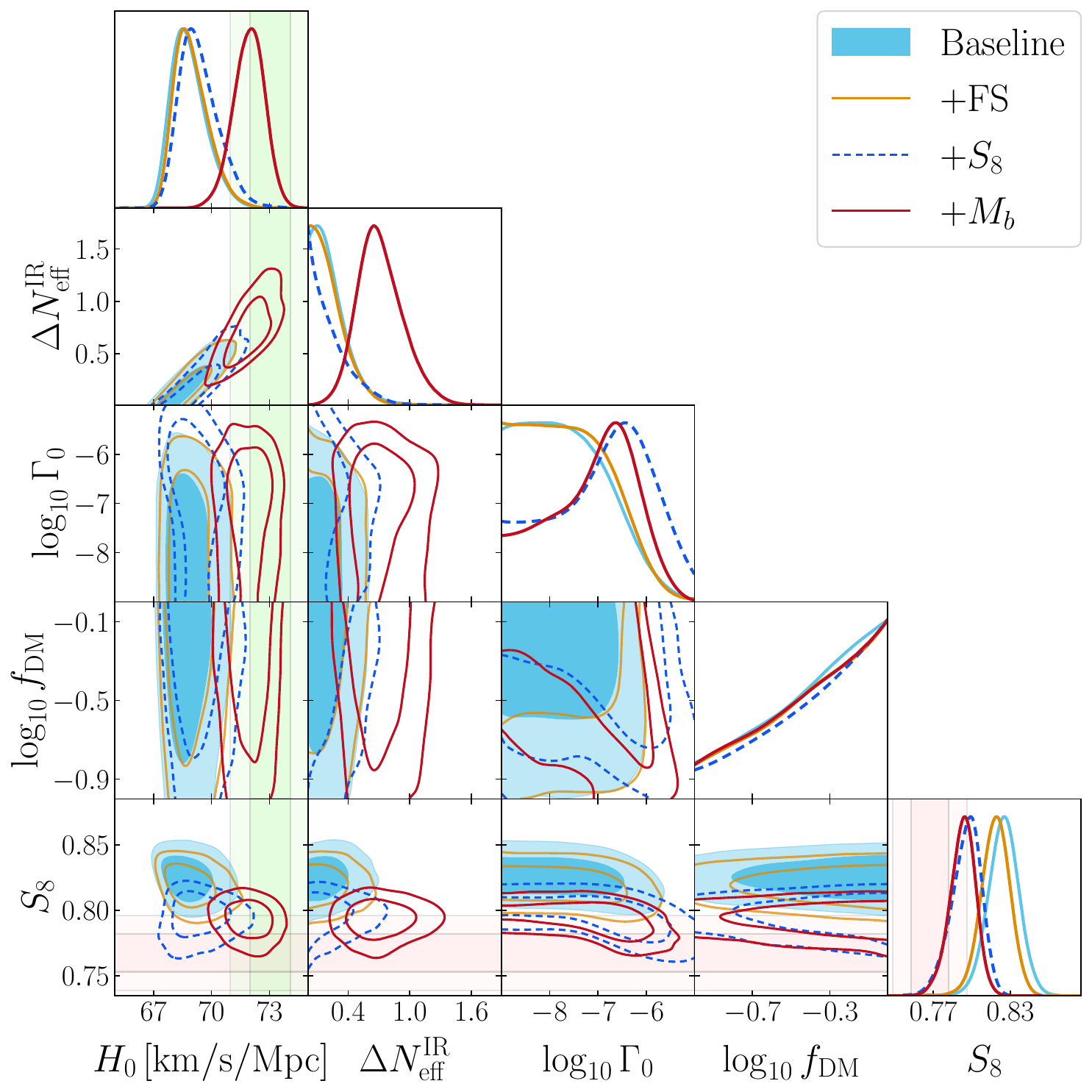}
\caption{One- and two-dimensional posterior distributions are given for selected model parameters, as well as $H_0$ and $S_8$, obtained by fitting the WIDM model to four datasets: the baseline, baseline + FS, baseline + FS + $S_8$, and baseline + FS + $S_8+M_b$. The light green (lighter green) vertical bars show the 1-$\sigma$ (2-$\sigma$) bounds of the S$H_0$ES measurement of $H_0$, and the light pink (lighter pink) horizontal bars show the 1-$\sigma$ (2-$\sigma$) bounds of the combined $S_8$ measurement from KiDS-1000 and DES-Y3.}
\label{fig:widmdata}
\end{figure}

\begin{figure}
\centering
\includegraphics[width=0.6\textwidth]{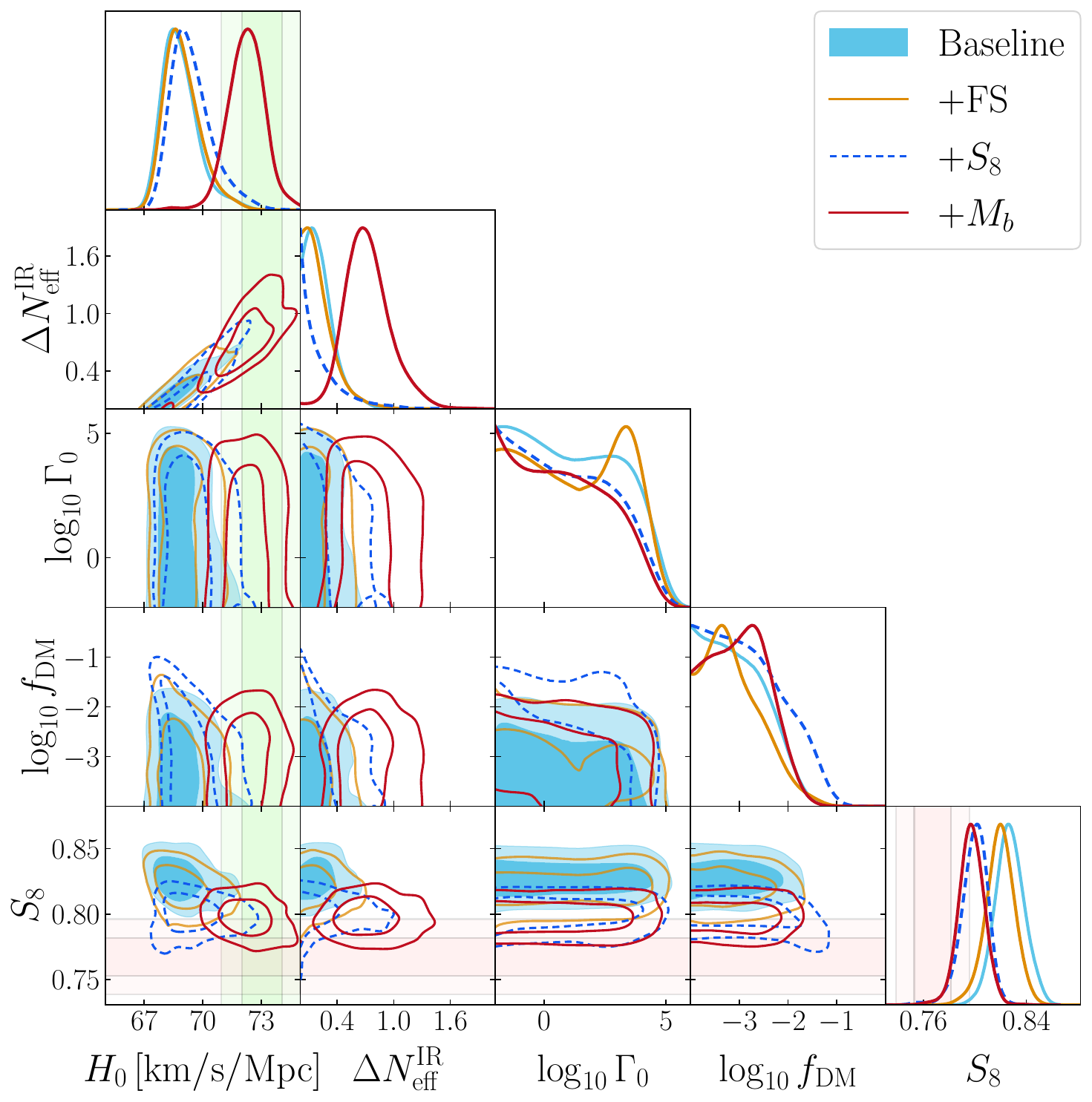}
\caption{One- and two-dimensional posterior distributions are given for selected model parameters, as well as $H_0$ and $S_8$, obtained by fitting the SIDM model to four datasets: the baseline, baseline + FS, baseline + FS + $S_8$, and baseline + FS + $S_8+M_b$. The light green (lighter green) vertical bars show the 1-$\sigma$ (2-$\sigma$) bounds of the S$H_0$ES measurement of $H_0$, and the light pink (lighter pink) horizontal bars show the 1-$\sigma$ (2-$\sigma$) bounds of the combined $S_8$ measurement from KiDS-1000 and DES-Y3.}
\label{fig:sidmdata}
\end{figure}

For completeness, we report results including priors on the $S_8$ and $M_b$ parameters in Table~\ref{tab:idmS8Mb}. Our MCMC chains for the SIDM model with these priors have somewhat larger Gelman-Rubin parameter $R-1<0.06$ than in previous runs. Figures~\ref{fig:widmdata} and~\ref{fig:sidmdata} show the posterior distributions for selected model parameters as well as $H_0$ and $S_8$ for the WIDM and SIDM models, respectively. The inclusion of the $S_8$ prior unsurprisingly demonstrates an alleviation of the $S_8$ tension with respect to $\Lambda$CDM, although with significant residual tensions $\gtrsim 1.6\sigma$ (for a detailed comparison, see Appendix~\ref{app:data}, in particular Tables~\ref{tab:appdata4} and~\ref{tab:appdata5}). Moreover, the improvement in $\chi^2$ is very small given the number of additional parameters, which implies the above mentioned positive large $\Delta$AIC. Furthermore, with the inclusion of both the $S_8$ and $M_b$ priors, one can see that although the Hubble tension is significantly reduced, the $S_8$ tension is found to be the same for SIDM and WIDM as in $\Lambda$CDM (see Appendix~\ref{app:data} for more results, in particular Table~\ref{tab:appdata5}). We therefore find that the inclusion of the dark matter interactions on top of the SDR component is not favored by the data.

\begin{table*}
\begin{tabular} {| l | c| c| c| c|}
\hline\hline
\multicolumn{1}{|c|}{} &  \multicolumn{2}{|c|}{WIDM} &  \multicolumn{2}{|c|}{SIDM}\\
\hline
 \multicolumn{1}{|c|}{ Parameter} &  \multicolumn{1}{|c|}{Basline + FS + $S_8$} &  \multicolumn{1}{|c|}{Baseline + FS + $S_8$ + $M_b$} &  \multicolumn{1}{|c|}{Basline + FS + $S_8$} &  \multicolumn{1}{|c|}{Baseline + FS + $S_8$ + $M_b$}\\
\hline\hline
$\Delta N^{\textrm{\scriptsize IR}}_{\textrm{\scriptsize eff}}$ & $ < 0.616~(0.177)$ & $0.71~(0.65)^{+0.18}_{-0.25}      $ & $ < 0.718~(0.011)$ & $0.72~(0.59)^{+0.19}_{-0.26}      $\\
$\log_{10} z_t             $ & Unconstrained $(4.53)$ & $ < 4.432~(4.12)$ & Unconstrained $(3.45)$    & $ < 4.555~(3.74)$\\
$r_g                       $ & Unconstrained $(0.19)$ & Unconstrained $(0.53)$ & Unconstrained $(4.29)$ & $ < 4.098~(0.27)$\\
$\log_{10} \Gamma_0        $ & Unconstrained $(-6.288)$ & $ < -5.837~(-5.746)$ & $ < 4.167~(2.379)$ & $ < 3.983~(-0.853)$\\
$\log_{10} f_{\textrm{DM}} $ & Unconstrained $(-0.139)$ & Unconstrained $(-0.824)$ & $ < -1.496~(-1.348)$ & $ < -2.017~(-3.616)$\\
\hline
$H_0 \,[\textrm{km}/\textrm{s}/\textrm{Mpc}]$ & $69.35~(68.79)^{+0.75}_{-1.2}      $ & $71.94~(72.13)^{+0.86}_{-0.75}     $ & $69.42~(67.73)^{+0.70}_{-1.2}      $ & $72.24~(72.48)^{+0.93}_{-0.89}     $\\
$S_8                       $ & $0.795~(0.793)^{+0.013}_{-0.0097}  $ & $0.793~(0.782)^{+0.011}_{-0.0089}  $ & $0.800~(0.773)^{+0.011}_{-0.0084}  $ & $0.7974~(0.7901)^{+0.0096}_{-0.0093}$\\
$M_b                       $ & $-19.369~(-19.384)^{+0.022}_{-0.035} $ & $-19.294~(-19.283)^{+0.025}_{-0.022} $ & $-19.368~(-19.417)^{+0.020}_{-0.037} $ & $-19.286~(-19.279)^{+0.027}_{-0.025} $\\
\hline
$\Delta\chi^2$ & $-2.1$ & $-20.57$ & $-3.57$ & $-18.08$\\
\hline
$S_8$ GT & $1.62\sigma $ & $1.51\sigma $ & $1.94\sigma $ & $1.75\sigma $\\
\hline
$S_8$ IT & $1.48\sigma $ & $1.43\sigma $ & $1.76\sigma $ & $1.69\sigma $\\
\hline
\end{tabular}
\caption{Mean (best-fit) $\pm 1\sigma$ error of WIDM and SIDM parameters obtained by fitting to two datasets: the baseline + FS + $S_8$ and baseline + FS + $S_8+M_b$. Upper bounds are presented at 95\% C.L., and parameters without constraints at 95\% C.L. within their prior boundaries are marked as unconstrained. Tension measures are reported with respect to the combined $S_8$ measurement from KiDS-1000 and DES-Y3.}
\label{tab:idmS8Mb}
\end{table*}

\section{Conclusions}\label{sec:conclusions}

Interacting dark sectors with mass thresholds are an interesting BSM possibility, the cosmological evolution of which can resemble that of the SM bath. When their mass scale is $\mathcal{O}(0.1-10)~\text{eV}$, the heavy degrees of freedom transfer entropy to the remaining light particles around and slightly before the epoch of recombination. The resulting step-like increase in the dark radiation abundance compared to that of neutrinos allows for larger values of $\Delta N_{\text{eff}}$ and, as a consequence, for larger values of $H_0$ than in other dark radiation models.

In this work, we have carefully assessed the constraint on $\Delta N_{\text{eff}}$ in this stepped dark radiation (SDR) model, as well as the possibility to alleviate the $H_0$ tension, by means of a combination of CMB, BAO, LSS, and Pantheon supernovae datasets. With respect to previous work~\cite{Aloni:2021eaq, Joseph:2022jsf}, we have allowed for wider prior boundaries on the redshift and the size of the step-like feature (but made a more restrictive choice than~\cite{Schoneberg:2022grr}, to avoid volume effects), reflecting the lack of a well-motivated narrow theoretical prediction for those parameters, and included full-shape information on the BOSS DR12 galaxy clustering power spectrum. Firstly, we found $\Delta N_{\text{eff}}\leq 0.55$ at $95\%$ C.L. with our prior choices and using the full Planck18 likelihood combined with BAO and Pantheon data, significantly relaxing the bound $\Delta N_{\text{eff}}\leq 0.46$ at $95\%$ C.L. for the interacting dark radiation scenario without a mass threshold. While the model succeeds in raising the Hubble constant, we assessed the tension with the SH$_0$ES measurement to be around the $3\sigma$ level,  independently of which prescription is used to compute it among several proposed in the literature. 

Overall, our results are more limiting than those presented in~\cite{Aloni:2021eaq, Joseph:2022jsf}, where the model is claimed to perform significantly better than the simpler self-interacting dark radiation (SIDR) scenario without a mass threshold. In contrast, our analysis suggests that the value of the Hubble constant remains in tension with SH$_0$ES in both the SDR and SIDR models at similar levels. These differences are to be attributed to: the tight prior ranges used in~\cite{Aloni:2021eaq, Joseph:2022jsf}, as well as the choice to keep the step size fixed in their MCMC analysis. The discrepancies found in this work are similar to those reported for Early Dark Energy (EDE) models (see e.g.~\cite{Smith:2020rxx}), when fixing certain parameters or choosing narrow prior ranges. 

Adding LSS data does not significantly affect the constraint on $\Delta N_{\text{eff}}$ nor the tension with SH$_0$ES. Nonetheless, the improvement in the fit with respect to the $\Lambda$CDM model is minimal both with and without LSS data, $\Delta \chi^2\simeq -1.5$ with three extra parameters. This is similar to the result presented in~\cite{Aloni:2021eaq, Joseph:2022jsf}, which is however obtained with only two free parameters. Additionally, let us compare with one of the most investigated competitor models to address the $H_0$ tension, EDE, for which Ref.~\cite{Simon:2022adh} reports (see Table VIII of~\cite{Simon:2022adh}) $\Delta \chi^2\simeq -4.7$ and $Q_{\text{DMAP}}\simeq -2.1$ with the same number of extra parameters and the same dataset including FS data. The EDE model thus performs significantly better than the SDR model, although it importantly does not have a simple particle physics realization. 

We also investigated two extensions of the SDR model~\cite{Joseph:2022jsf, Buen-Abad:2022kgf} that include interactions with dark matter to suppress matter fluctuations at late times and alleviate the $S_8$ tension. The two models differ in the way that interactions are turned-off below the mass threshold, due to different types of microphysical interactions (designed to capture either weak or strong interactions). In practice, these extensions add two more extra parameters: the strength of the interaction and the fraction of the dark matter that is interacting. We keep both parameters free to vary (in contrast with~\cite{Joseph:2022jsf}, where the DM fraction is fixed to one). We use logarithmic priors to sample the interaction strength and find the $S_8$ tension to remain close to the $3\sigma$ level, with only a minor improvement compared to the $\Lambda$CDM model. Our result differs significantly from that of~\cite{Joseph:2022jsf} for the weakly interacting model, which claims a reduction of the tension to $1.7~\sigma$. This should be attributed mostly to the choice of prior on the interaction strength, which is sampled linearly in~\cite{Joseph:2022jsf}. Our choice instead captures more fairly the ``look elsewhere" effect. Considering also the five additional parameters, we find that the $\Lambda$CDM model is significantly preferred over the extended SDR sector, even when adding a prior on $S_8$ from cosmic shear measurements. Concerning the model with strong interactions proposed in~\cite{Buen-Abad:2022kgf}, while the authors did not test their model against data, we find no evidence that this model can convincingly alleviate the $S_8$ tension. In fact, we obtain an upper bound on the interacting DM fraction of less than $1\%$ at $95\%$ C.L. for large momentum transfer rates $\Gamma_0\geq 10^{-2}~\text{Mpc}^{-1}$, thereby constraining the range $(1-5)\%$ suggested in~\cite{Buen-Abad:2022kgf}.
Furthermore, all of this is under the assumption that the extra relativistic species arise after BBN; if this were not the case, the constraints would be even stronger.

Despite their arguably not-so decisive impact on cosmological tensions, dark sectors with mass thresholds $(0.1-10)~\text{eV}$ are an interesting particle physics scenario, which can be significantly probed with current datasets and certainly more so with upcoming CMB and LSS surveys~\cite{SimonsObservatory:2018koc, CMB-S4:2022ght, Amendola:2016saw}. In this respect, it may be interesting to improve the modeling of the mass threshold transition, which currently relies on an effective fluid description that may not fully capture the implications of a transient significant fraction of massive particles in the dark sector bath for CMB and LSS perturbations. For now, our work provides up-to-date constraints on interacting dark radiation scenarios, that should prove useful for model-builders as well as cosmologists.

\section*{Acknowledgements}

We thank G. D'Amico and P. Zhang for help with the latest version of {\tt PyBird}, as well as M. Simonovic for useful discussions. We acknowledge use of the Tufts HPC research cluster. The work of I.J.A. is supported by the John F. Burlingame Graduate Fellowship in Physics at Tufts. The work of F.R.~is partly supported by the grant No. RYC2021-031105-I from the Ministerio de Ciencia e Innovación (Spain). M.P.H. is supported in part by National Science Foundation grant No. PHY-2013953.
F.R. thanks the Galileo Galilei Institute in Florence (Italy) for kind hospitality during the completion of this work.

\bibliography{biblio}

\appendix
\section{Time Evolution of Dark Radiation with Thresholds}\label{app:SDR}

We describe below the time-evolution of the background quantities for the stepped dark radiation fluid. For further details, see~\cite{Aloni:2021eaq,Joseph:2022jsf}.

The dark radiation fluid describes a sector with $g_*^{\textrm{UV}}$ effective relativistic degrees of freedom well before the step, and $g_*^{\textrm{IR}}$ after the step. The evolution of the energy density and pressure of the fluid as functions of the dark sector temperature $T_d$ are given by

\beq \rho(T_d)=g_*^{\textrm{IR}} \frac{T_d^4 \pi^2}{30} (1+r_g\hat{\rho}(x) )
\eeq
\beq p(T_d) = g_*^{\textrm{IR}} \frac{T_d^4 \pi^2}{90} (1+r_g \hat{p}(x) )
\eeq
\beq r_g \equiv \frac{g_*^{\textrm{UV}}-g_*^{\textrm{IR}}}{g_*^{\textrm{IR}}}
\eeq
where $x$ is the ratio of the mass scale to the dark sector temperature $x\equiv m/T_d$. Approximating the distribution functions of species in the dark sector by their Maxwell-Boltzmann distributions, as in~\cite{Aloni:2021eaq}, the functions $\hat{\rho}(x)$ and $\hat{p}(x)$ are given by 
\beq \hat{\rho}(x)\equiv \frac{x^2}{2}K_2(x) + \frac{x^3}{6}K_1(x)
\eeq
\beq \hat{p}(x) \equiv \frac{x^2}{2}K_2(x)
\eeq
where $K_n(x)$ is the $n$th-order Bessel function of the second kind.
The parameter $r_g$ determines the effective size of the step, relating the effective number of neutrino species at early times $\NUV$ and late times $\NIR$ 
\beq \frac{\NIR}{\NUV} = (1+\rg)^{1/3}
\eeq

The step occurs at a redshift $z_t = 1/a_t -1$, where the scale factor of the transition is defined to be $a_t \equiv T_{d0}/m$ with $T_{d0}$ the dark sector temperature today. This definition gives the following relation between the dark sector temperature and the phenomenological parameters $r_g$ and $a_t$
\beq \left(\frac{x a_t}{a}\right)^3 = 1+\frac{r_g}{4}(3\hat{\rho}(x)+\hat{p}(x))
\eeq
Based on this relation, the scale factor dependence of all parameters can be determined through $x(a)$. Far away from the step, there is a simple relationship for $x(a)$; at $z\gg\zt$, $x a_t = (1+r_g)^{1/3} a$, while at $z\ll\zt$, $x a_t = a$. With these relations, the full background evolution of the dark radiation fluid is determined by the three parameters $\NIR$, $\zt$ (or $a_t$), and $\rg$.

The perturbation equations~\eqref{deltaeq} and~\eqref{thetaeq} can be computed in terms of the equation of state $w$ and sound speed $c_s^2$
\beq w(x) = \frac{1}{3}-\frac{\rg}{3}\frac{\hat{\rho}(x)-\hat{p}(x)}{1+\rg \hat{p}(x)}
\eeq
\beq c_s^2(x) = \frac{1}{3} - \frac{\rg}{36}\frac{x^2 \hat{p}(x)}{1+\rg(\frac{3}{4}\hat{\rho}(x)+(\frac{1}{4}+\frac{x^2}{12})\hat{p}(x))}
\eeq

\section{Momentum Transfer Between Dark Matter and Dark Radiation}\label{app:SDRDM}

We parameterize the momentum transfer rate between dark matter and stepped dark radiation as in \eqref{gammaform}, repeated below.
\begin{equation}
\Gamma = \Gamma_0\left(\frac{1+z_t}{x}\right)^2 \left[1+b h_1(x)\right] h_2(x)
\end{equation}

For the WIDM model, the time evolution of the momentum transfer rate is given by~\cite{Joseph:2022jsf}

\begin{minipage}{0.5\textwidth} 
\beq h_1(x) = 0
\eeq
\end{minipage}
\begin{minipage}{0.5\textwidth}
\beq h_2(x) = (1- 0.05 x^{1/2}+0.131 x)^{-4}
\eeq
\end{minipage}
\\

For the SIDM model, the time evolution is given by~\cite{Buen-Abad:2022kgf}

\begin{minipage}{0.5\textwidth} 
\beq h_1(x) = \ln\left(\frac{K_2(x)}{2(x K_0(x)+K_1(x))^2}\right)
\eeq
\end{minipage}
\begin{minipage}{0.5\textwidth}
\beq h_2(x) = \frac{e^{-x}}{2} (2+x(2+x))
\eeq
\end{minipage}
\\

In this case, since $h_1(x)\neq 0$, there is an extra free parameter $b$. In terms of the fundamental parameters of the model in~\cite{Buen-Abad:2022kgf}, this extra parameter is given by 
\beq 
b \equiv \left[ \ln \left(\frac{\pi}{g^\psi_{*} \alpha^3}\right)\right]^{-1}
\eeq
where $g^\psi_{*}$ corresponds to the degrees of freedom in the SDR component that becomes massive at $\zt$ (the fermion $\psi$ in the reference model of~\cite{Buen-Abad:2022kgf}) and $\alpha$ is the coupling strength of the interaction. This parameter is only logarithmically dependent on these fundamental parameters, and thus its value does not vary much. In this work, we have kept this parameter fixed (for the sake of comparing to WIDM), using the values suggested in~\cite{Buen-Abad:2022kgf}: $\alpha=10^{-4}$ and $g^\psi_{*}=4\times 7/8$ such that $b\approx 0.04$.

\section{Detailed MCMC Results}
\label{app:data}

We present below plots and tables containing the details of our analysis. This includes posterior distribution plots and tables reflecting statistics for a larger set of ($\Lambda$CDM and model-specific) parameters, as well as tables of $\chi^2$ values for each dataset derived from several different fits of each model. For all our runs, we used a jumping factor of $2.0$ and produced more than eight chains. We used the BBN table {\tt BBN\_2017\_marcucci.dat} to relate $Y_{\text{He}}$ to $\omega_b$ and $N_{\text{eff}}=3.046$ at BBN. As mentioned in the main text, we did not include the contribution of self-interacting radiation at BBN, accounting for the possibility that it may be produced after BBN.

\subsection*{SDR narrow vs broad priors}

In Table~\ref{tab:appztpriors}, we show the complete set of free parameters for the SDR model (with the step size fixed as done in \cite{Aloni:2021eaq}), comparing two fits to the baseline data set with different choices of the prior on $\log_{10} \zt$. The wider choice of the prior $\log_{10} \zt \in [3.0,5.0]$ can be thought of as a more conservative choice, and since the MCMC analysis does not obtain reliable constraints on the value $\log_{10} z_t$, we use this wider prior for the rest of our analysis. 

Except for a mild fluctuation in the tensions (i.e. a slight increase in the $M_b$ GT and IT, along with a slight decrease in the $S_8$ GT and IT), the widening of the prior does not strongly affect the inferred statistics. For instance, the conclusion that the presence of the step in $\Delta N_{\textrm{eff}}$ relaxes the constraint on $\Delta N_{\textrm{eff}}$ remains unchanged. In addition, the Markov chains generated with the wider prior to not obtain a better fit to the data, and thus the $\chi^2$ (and therefore $\Delta_\chi^2$ and $\Delta$AIC) do not change.


\begin{table*}
\begin{tabular} {| l | c| c|}
\hline\hline
 \multicolumn{1}{|c|}{ Parameter} &  \multicolumn{1}{|c|}{$\log_{10}z_t \in [4.0,4.6]$} &  \multicolumn{1}{|c|}{$\log_{10}z_t \in [3.0,5.0]$}\\
\hline\hline
$100 \omega_b              $ & $2.253~(2.257)^{+0.015}_{-0.017}   $ & $2.251~(2.257)^{+0.015}_{-0.016}   $\\
$\omega_{cdm }             $ & $0.1240~(0.1289)^{+0.0023}_{-0.0036}$ & $0.1234~(0.1289)^{+0.0020}_{-0.0035}$\\
$\ln 10^{10}A_s            $ & $3.050~(3.049)^{+0.013}_{-0.015}   $ & $3.048~(3.049)^{+0.014}_{-0.014}   $\\
$n_{s }                    $ & $0.9715~(0.9773)^{+0.0046}_{-0.0055}$ & $0.9691~(0.9773)^{+0.0042}_{-0.0056}$\\
$\tau_{reio }              $ & $0.0574~(0.0573)^{+0.0067}_{-0.0076}$ & $0.0575~(0.0573)^{+0.0068}_{-0.0075}$\\
$\Delta N^{\textrm{\scriptsize IR}}_{\textrm{\scriptsize eff}}$ & $ < 0.597~(0.551)$& $ < 0.59~(0.551)$\\
$\log_{10} z_t             $ & Unconstrained $(4.3)$ & Unconstrained $(4.3)$\\
\hline
$H_0 \,[\textrm{km}/\textrm{s}/\textrm{Mpc}]$ & $69.30~(70.68)^{+0.86}_{-1.3}      $ & $69.11~(70.68)^{+0.80}_{-1.3}      $\\
$S_8                       $ & $0.829~(0.837)^{+0.011}_{-0.011}   $ & $0.827~(0.837)^{+0.011}_{-0.011}   $\\
$M_b                       $ & $-19.369~(-19.325)^{+0.025}_{-0.038} $ & $-19.374~(-19.325)^{+0.024}_{-0.037} $\\
\hline
$\Delta\chi^2$ & $-0.41$ & $-0.41$\\
\hline
$Q_{\textrm{\tiny DMAP}}^{M_b}$ & $2.55\sigma $ & $2.55\sigma $\\
\hline
$M_b$ GT & $3.12\sigma $ & $3.38\sigma $\\
\hline
$M_b$ IT & $2.56\sigma $ & $2.7\sigma $\\
\hline
$S_8$ GT & $3.41\sigma $ & $3.32\sigma $\\
\hline
$S_8$ IT & $3.43\sigma $ & $3.31\sigma $\\
\hline
$\Delta \textrm{AIC}^{M_b}$ & $-22.67$ & $-22.67$\\
\hline
\end{tabular}
\caption{Mean (best-fit) $\pm 1\sigma$ error of all free parameters, and $S_8$ and $M_b$, obtained by fitting the two-parameter \SDR\, model (i.e. with $r_g$ fixed) to the baseline dataset P18+BAO+Pantheon, comparing two choices of prior on $\log_{10} \zt$. Upper bounds are presented at 95\% C.L., and parameters without constraints at 95\% C.L. within their prior boundaries are marked as unconstrained. Tension measures are reported with respect to the S$H_0$ES measurement of $M_b$ and with respect to the combined $S_8$ measurement from KiDS-1000 and DES-Y3.}
\label{tab:appztpriors}
\end{table*}

\subsection*{SDR step size fixed vs free}

In Table~\ref{tab:apprg}, we show the complete set of free parameters for the SDR model, comparing the cases of having the step size as a fixed or free parameter. The specific choice of the step size reflects a specified particle physics model. However, many such particle physics models may be able to produce the phenomenology of the dark radiation fluid with a step, and therefore one can think of the step size as a free parameter. 

This choice allows the MCMC to find a set of parameters which fit the data better than when the step size is fixed, which is reflected in the improvement in the $\Delta\chi^2$. On the other hand, when fitting to the dataset baseline + $M_b$, the same set of best-ft parameters is found in both the step-size-fixed and -free cases. Thus, the $\Delta$AIC is worsened (i.e. made more positive) with the step size free because of the addition of a new free parameter. However, since the $\Delta$AIC is already quite negative, this is not significant evidence against the model. Overall, the addition of a new free parameter, the step size, leads to a better fit to data.

On the other hand, the tension with $M_b$ is increased when the step size is made free (i.e. $\QDMAP^{M_b}$, $M_b$ GT, and $M_b$ IT are all increased), but not significantly beyond the $\sim 3 \sigma$ level overall. 

\begin{table*}
\begin{tabular} {| l | c| c|}
\hline\hline
 \multicolumn{1}{|c|}{ Parameter} &  \multicolumn{1}{|c|}{$r_g$ fixed} &  \multicolumn{1}{|c|}{$r_g$ free}\\
\hline\hline
$100 \omega_b              $ & $2.251~(2.257)^{+0.015}_{-0.016}   $ & $2.248~(2.244)^{+0.015}_{-0.017}   $\\
$\omega_{cdm }             $ & $0.1234~(0.1289)^{+0.0020}_{-0.0035}$ & $0.1227~(0.1234)^{+0.0018}_{-0.0030}$\\
$\ln 10^{10}A_s            $ & $3.048~(3.049)^{+0.014}_{-0.014}   $ & $3.050~(3.066)^{+0.014}_{-0.015}   $\\
$n_{s }                    $ & $0.9691~(0.9773)^{+0.0042}_{-0.0056}$ & $0.9695~(0.9772)^{+0.0042}_{-0.0063}$\\
$\tau_{reio }              $ & $0.0575~(0.0573)^{+0.0068}_{-0.0075}$ & $0.0575~(0.0616)^{+0.0067}_{-0.0076}$\\
$\Delta N^{\textrm{\scriptsize IR}}_{\textrm{\scriptsize eff}}$ &$ < 0.59~(0.551)$ & $ < 0.546~(0.289)$\\
$\log_{10} z_t             $ & Unconstrained $(4.3)$ & Unconstrained $(4.29)$\\
$r_g                       $ & --- & Unconstrained $(4.0)$                         \\
\hline
$H_0 \,[\textrm{km}/\textrm{s}/\textrm{Mpc}]$ & $69.11~(70.68)^{+0.80}_{-1.3}      $ & $68.89~(69.34)^{+0.71}_{-1.1}      $\\
$S_8                       $ & $0.827~(0.837)^{+0.011}_{-0.011}   $ & $0.827~(0.834)^{+0.011}_{-0.011}   $\\
$M_b                       $ & $-19.374~(-19.325)^{+0.024}_{-0.037} $ & $-19.381~(-19.369)^{+0.021}_{-0.032} $\\
\hline
$\Delta\chi^2$ & $-0.41$ & $-1.4$\\
\hline
$Q_{\textrm{\tiny DMAP}}^{M_b}$ & $2.55\sigma $ & $2.74\sigma $\\
\hline
$M_b$ GT & $3.38\sigma $ & $3.74\sigma $\\
\hline
$M_b$ IT & $2.7\sigma $ & $3.03\sigma $\\
\hline
$S_8$ GT & $3.32\sigma $ & $3.27\sigma $\\
\hline
$S_8$ IT & $3.31\sigma $ & $3.28\sigma $\\
\hline
$\Delta \textrm{AIC}^{M_b}$ & $-22.67$ & $-20.67$\\
\hline
\end{tabular}
\caption{Mean (best-fit) $\pm 1\sigma$ error of all free parameters, and $S_8$ and $M_b$, obtained by fitting the two- or three-parameter \SDR\, model (i.e. with $r_g$ fixed or free to vary) to the baseline dataset P18+BAO+Pantheon. Upper bounds are presented at 95\% C.L., and parameters without constraints at 95\% C.L. within their prior boundaries are marked as unconstrained. Tension measures are reported with respect to the S$H_0$ES measurement of $M_b$ and with respect to the combined $S_8$ measurement from KiDS-1000 and DES-Y3.}
\label{tab:apprg}
\end{table*}

\subsection*{SDR compared to SIDR}

In Table~\ref{tab:appsidr}, we show the complete set of free parameters for the SIDR model and compare to two implementations of the SDR model: SDR (i) refers to fixing the parameter $r_g$ and using the narrower prior $\log_{10} \zt \in [4.0,4.6]$ as done in~\cite{Aloni:2021eaq}, and SDR (ii) refers to leaving $r_g$ free to vary and using the wider prior $\log_{10} \zt \in [3.0,5.0]$. We see here that the ability of the SDR model to improve upon SIDR with regard to the $H_0$ tension is absent in the implementation (ii) compared to implementation (i) where this improvement is successful. 

In Table~\ref{tab:appsidrlss}, we show a comparison of SIDR and SDR (ii) fit with the baseline + FS dataset, showing similar results to the baseline dataset.

We note here also that SIDR shows a mildly smaller tension with $S_8$ compared to SDR (both (i) and (ii)). Meanwhile, the WIDM model reduces the tension marginally lower than SIDR [see Table~\ref{tab:appdata1}].

\begin{table*}
\begin{tabular} {| l | c| c| c|}
\hline\hline
 \multicolumn{1}{|c|}{ Parameter} &  \multicolumn{1}{|c|}{SIDR} &  \multicolumn{1}{|c|}{SDR (i)} &  \multicolumn{1}{|c|}{SDR (ii)}\\
\hline\hline
$100 \omega_b              $ & $2.255~(2.252)^{+0.017}_{-0.017}   $ & $2.253~(2.257)^{+0.015}_{-0.017}   $ & $2.248~(2.244)^{+0.015}_{-0.017}   $\\
$\omega_{cdm }             $ & $0.1226~(0.12)^{+0.0018}_{-0.0030}$ & $0.1240~(0.1289)^{+0.0023}_{-0.0036}$ & $0.1227~(0.1234)^{+0.0018}_{-0.0030}$\\
$\ln 10^{10}A_s            $ & $3.046~(3.048)^{+0.014}_{-0.015}   $ & $3.050~(3.049)^{+0.013}_{-0.015}   $ & $3.050~(3.066)^{+0.014}_{-0.015}   $\\
$n_{s }                    $ & $0.9667~(0.9678)^{+0.0038}_{-0.0037}$ & $0.9715~(0.9773)^{+0.0046}_{-0.0055}$ & $0.9695~(0.9772)^{+0.0042}_{-0.0063}$\\
$\tau_{reio }              $ & $0.0578~(0.0588)^{+0.0066}_{-0.0077}$ & $0.0574~(0.0573)^{+0.0067}_{-0.0076}$ & $0.0575~(0.0616)^{+0.0067}_{-0.0076}$\\
$\Delta N^{\textrm{\scriptsize IR}}_{\textrm{\scriptsize eff}}$ & $ < 0.456~(0.072)$ & $ < 0.597~(0.551)$ & $ < 0.546~(0.289)$\\
$\log_{10} z_t             $ & --- &  Unconstrained $(4.3)$ & Unconstrained $(4.29)$\\
$r_g                       $ & --- & --- & Unconstrained $(4.0)$ \\
\hline
$H_0 \,[\textrm{km}/\textrm{s}/\textrm{Mpc}]$ & $68.95~(68.38)^{+0.73}_{-1.2}      $ & $69.30~(70.68)^{+0.86}_{-1.3}      $ & $68.89~(69.34)^{+0.71}_{-1.1}      $\\
$S_8                       $ & $0.823~(0.818)^{+0.011}_{-0.011}   $ & $0.829~(0.837)^{+0.011}_{-0.011}   $ & $0.827~(0.834)^{+0.011}_{-0.011}   $\\
$M_b                       $ & $-19.380~(-19.397)^{+0.022}_{-0.034} $ & $-19.369~(-19.325)^{+0.025}_{-0.038} $ & $-19.381~(-19.369)^{+0.021}_{-0.032} $\\
\hline
$\Delta\chi^2$ & $-0.22$ & $-0.41$ & $-1.4$\\
\hline
$Q_{\textrm{\tiny DMAP}}^{M_b}$ & $3.48\sigma $ & $2.55\sigma $ & $2.74\sigma $\\
\hline
$M_b$ GT & $3.66\sigma $ & $3.12\sigma $ & $3.74\sigma $\\
\hline
$M_b$ IT & $2.95\sigma $ & $2.56\sigma $ & $3.03\sigma $\\
\hline
$S_8$ GT & $3.14\sigma $ & $3.41\sigma $ & $3.27\sigma $\\
\hline
$S_8$ IT & $3.15\sigma $ & $3.43\sigma $ & $3.28\sigma $\\
\hline
$\Delta\textrm{AIC}^{M_b}$ & $-18.88$ & $-22.67$ & $-20.67$\\
\hline
\end{tabular}
\caption{Mean (best-fit) $\pm 1\sigma$ error of all free parameters, and $S_8$ and $M_b$, obtained by fitting the SIDR model to the baseline dataset P18+BAO+Pantheon. For comparison, the SDR model is shown with: (i) $r_g$ fixed and $\log_{10} \zt \in [4.0,4.6]$ and (ii) $r_g$ free and $\log_{10} \zt \in [3.0,5.0]$. Upper bounds are presented at 95\% C.L., and parameters without constraints at 95\% C.L. within their prior boundaries are marked as unconstrained. Tension measures are reported with respect to the S$H_0$ES measurement of $M_b$ and with respect to the combined $S_8$ measurement from KiDS-1000 and DES-Y3.}
\label{tab:appsidr}
\end{table*}

\begin{table*}
\begin{tabular} {| l | c| c|}
\hline\hline
 \multicolumn{1}{|c|}{ Parameter} &  \multicolumn{1}{|c|}{SIDR} &  \multicolumn{1}{|c|}{SDR}\\
\hline\hline
$100 \omega_b              $ & $2.258~(2.266)^{+0.015}_{-0.018}   $ & $2.251~(2.247)^{+0.015}_{-0.016}   $\\
$\omega_{cdm }             $ & $0.1223~(0.1212)^{+0.0018}_{-0.0030}$ & $0.1220~(0.1202)^{+0.0016}_{-0.0028}$\\
$\ln 10^{10}A_s            $ & $3.045~(3.035)^{+0.015}_{-0.014}   $ & $3.048~(3.057)^{+0.014}_{-0.015}   $\\
$n_{s }                    $ & $0.9673~(0.968)^{+0.0038}_{-0.0037}$ & $0.9691~(0.9696)^{+0.0039}_{-0.0053}$\\
$\tau_{reio }              $ & $0.0578~(0.0534)^{+0.0068}_{-0.0075}$ & $0.0576~(0.0617)^{+0.0071}_{-0.0071}$\\
$\Delta N^{\textrm{\scriptsize IR}}_{\textrm{\scriptsize eff}}$ & $ < 0.457~(0.157)$ & $ < 0.55~(0.08)$\\
$\log_{10} z_t             $ & --- & Unconstrained $(4.97)$\\
$r_g                       $ & --- & Unconstrained $(2.34)$   \\
\hline
$H_0 \,[\textrm{km}/\textrm{s}/\textrm{Mpc}]$ & $69.13~(69.11)^{+0.73}_{-1.1}      $ & $69.01~(68.37)^{+0.66}_{-1.1}      $\\
$S_8                       $ & $0.819~(0.809)^{+0.010}_{-0.010}   $ & $0.821~(0.824)^{+0.010}_{-0.010}   $\\
$M_b                       $ & $-19.375~(-19.376)^{+0.022}_{-0.034} $ & $-19.379~(-19.402)^{+0.020}_{-0.032} $\\
\hline
$\Delta\chi^2$ & $-0.23$ & $-1.62$\\
\hline
$M_b$ GT & $3.53\sigma $ & $3.77\sigma $\\
\hline
$M_b$ IT & $2.83\sigma $ & $2.94\sigma $\\
\hline
$S_8$ GT & $2.9\sigma $ & $3.01\sigma $\\
\hline
$S_8$ IT & $2.92\sigma $ & $3.01\sigma $\\
\hline
\end{tabular}
\caption{Mean (best-fit) $\pm 1\sigma$ error of all free parameters, and $S_8$ and $M_b$, obtained by fitting the SIDR model to the baseline + FS dataset. For comparison, the SDR model is shown with (ii) $r_g$ free and $\log_{10} \zt \in [3.0,5.0]$. Upper bounds are presented at 95\% C.L., and parameters without constraints at 95\% C.L. within their prior boundaries are marked as unconstrained. Tension measures are reported with respect to the S$H_0$ES measurement of $M_b$ and with respect to the combined $S_8$ measurement from KiDS-1000 and DES-Y3.}
\label{tab:appsidrlss}
\end{table*}

\subsection*{Detailed posteriors}

\begin{figure}
\centering
\includegraphics[width=0.85\textwidth]{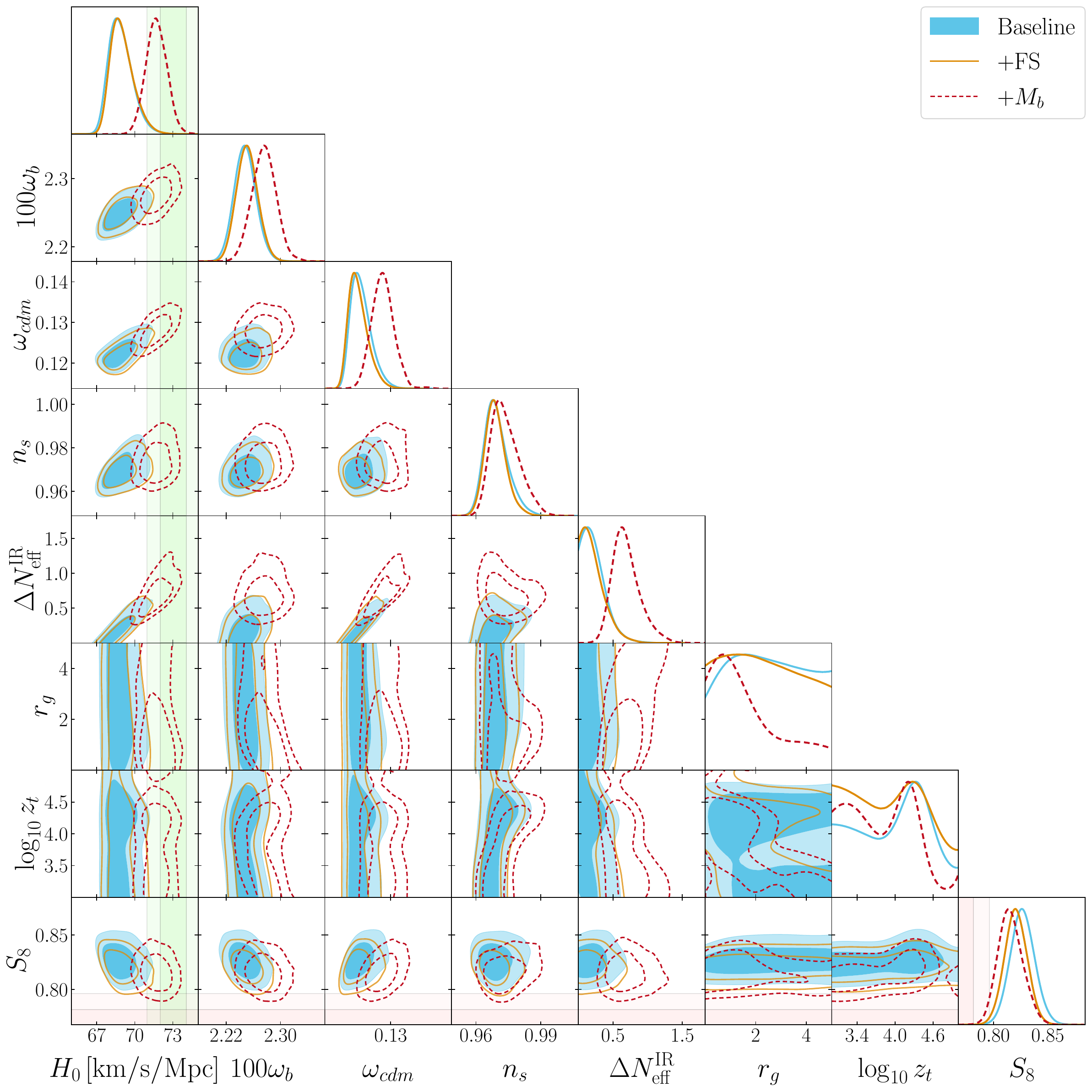}
\caption{One- and two-dimensional posterior distributions for selected parameters fit to three different datasets: the baseline dataset P18+BAO+Pantheon, the baseline + FS, and the baseline +FS+$M_b$. The light green (lighter green) vertical bars show the 1-$\sigma$ (2-$\sigma$) bounds of the S$H_0$ES measurement of $H_0$, and the light pink (lighter pink) horizontal bars show the 1-$\sigma$ (2-$\sigma$) bounds of the combined $S_8$ measurement from KiDS-1000 and DES-Y3.}
\label{fig:appSDR}
\end{figure}

In Fig.~\ref{fig:appSDR}, we present posterior distributions for several model parameters in the SDR model. We compare the posteriors obtained by fitting to three datasets: baseline, baseline + FS, and baseline + FS + $M_b$. These posteriors demonstrate clearly the capacity for the SDR model to relax the constraints on $\Delta N_{\textrm{eff}}$ as well as its ability to broaden the posterior distribution of $H_0$ and thus alleviate the Hubble tension. 

It can also be seen in the posterior distributions for $r_g$ and $\log_{10} \zt$ that the data do not clearly prefer any value (or even range of values) of these parameters.

\begin{figure}
\centering
\includegraphics[width=0.85\textwidth]{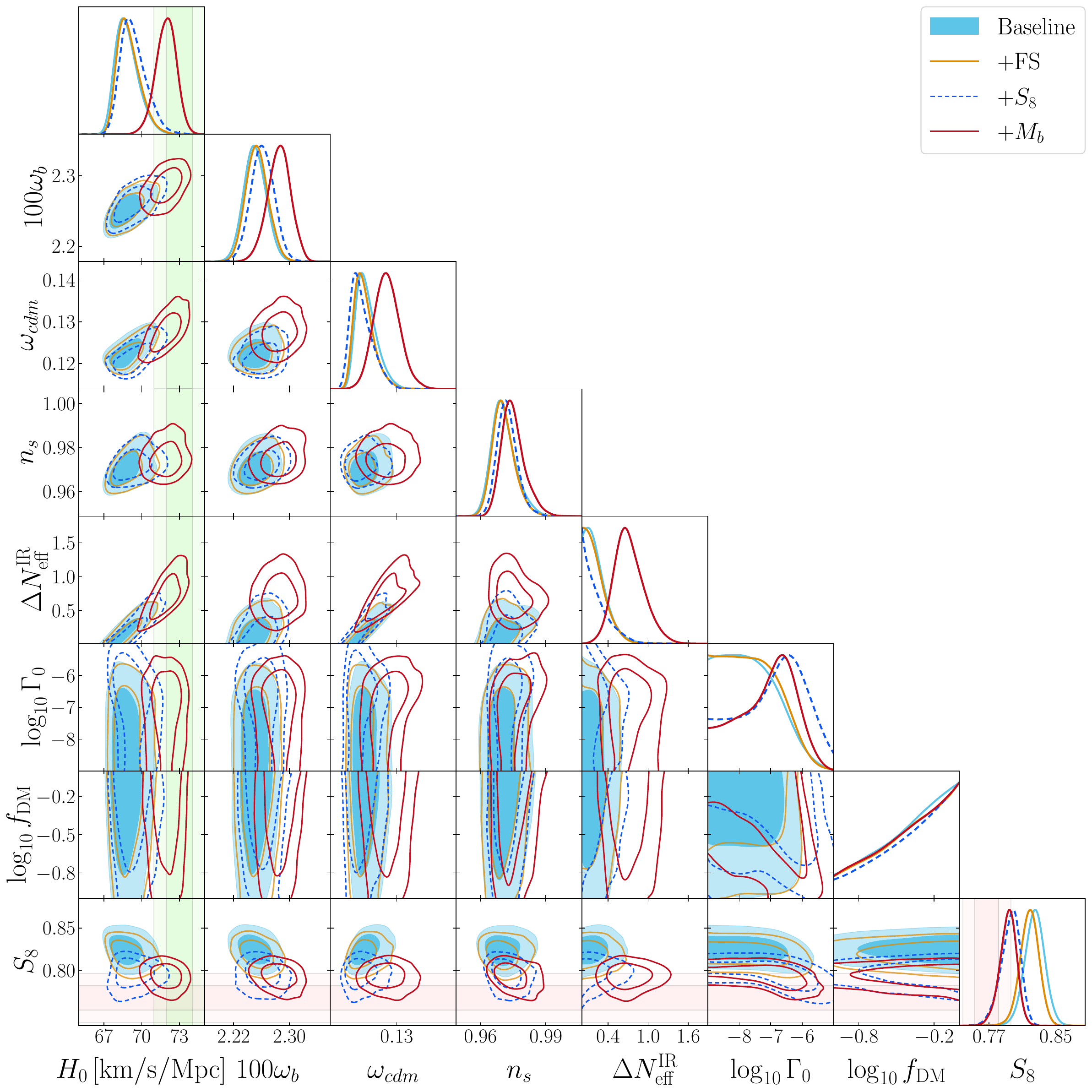}
\caption{One- and two-dimensional posterior distributions are given for selected model parameters, as well as $H_0$ and $S_8$, obtained by fitting the WIDM model to four datasets: the baseline, baseline + FS, baseline + FS + $S_8$, and baseline + FS + $S_8+M_b$. The light green (lighter green) vertical bars show the 1-$\sigma$ (2-$\sigma$) bounds of the S$H_0$ES measurement of $H_0$, and the light pink (lighter pink) horizontal bars show the 1-$\sigma$ (2-$\sigma$) bounds of the combined $S_8$ measurement from KiDS-1000 and DES-Y3.}
\label{fig:appWIDM}
\end{figure}

\begin{figure}
\centering
\includegraphics[width=0.85\textwidth]{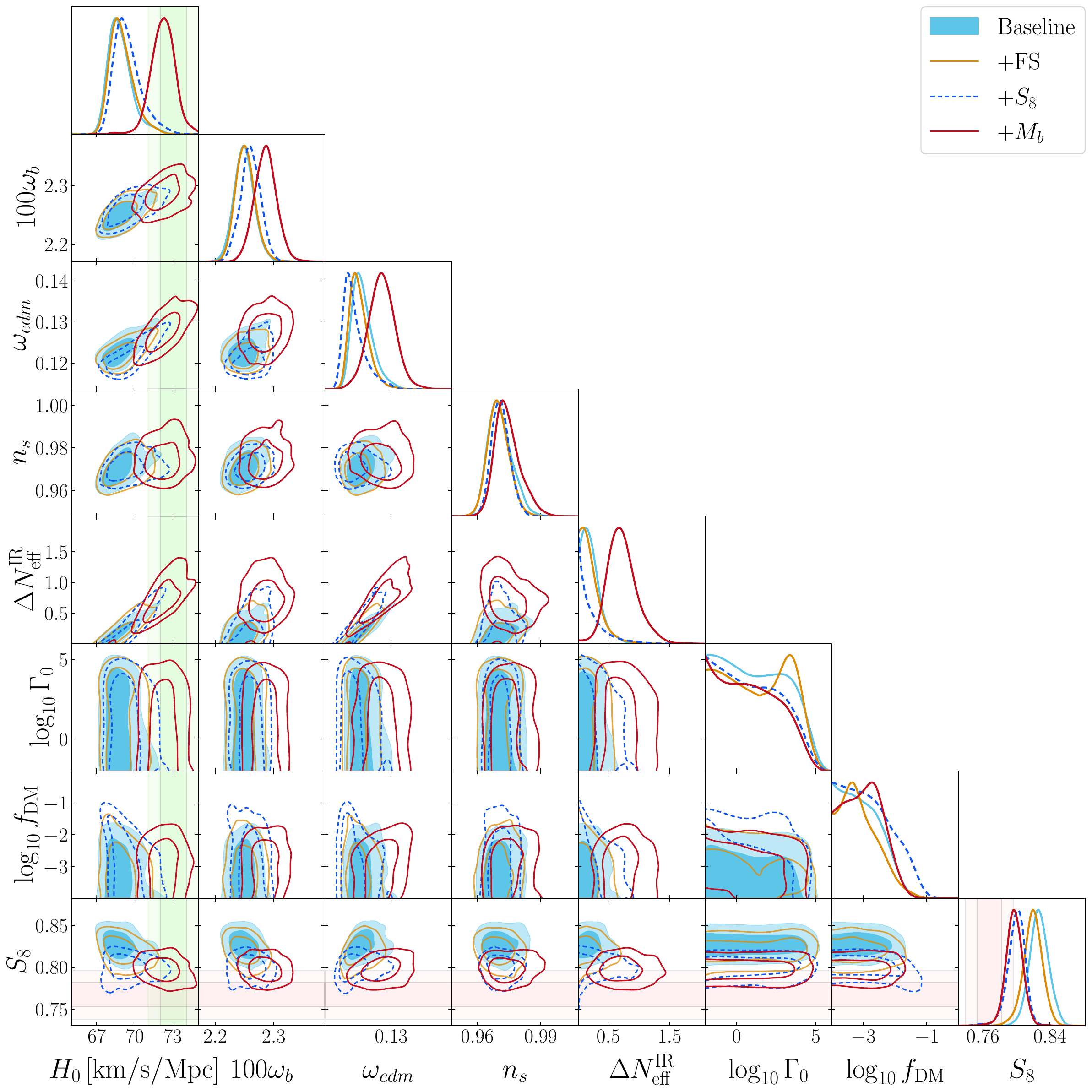}
\caption{One- and two-dimensional posterior distributions are given for selected model parameters, as well as $H_0$ and $S_8$, obtained by fitting the SIDM model to four datasets: the baseline, baseline + FS, baseline + FS + $S_8$, and baseline + FS + $S_8+M_b$. The light green (lighter green) vertical bars show the 1-$\sigma$ (2-$\sigma$) bounds of the S$H_0$ES measurement of $H_0$, and the light pink (lighter pink) horizontal bars show the 1-$\sigma$ (2-$\sigma$) bounds of the combined $S_8$ measurement from KiDS-1000 and DES-Y3.}
\label{fig:appSIDM}
\end{figure}

In Fig.s~\ref{fig:appWIDM} and~\ref{fig:appSIDM}, we show the posterior distributions for several model parameters for the WIDM and SIDM models, respectively.

\subsection*{Model comparison}

\begin{figure}
\centering
\includegraphics[width=0.75\textwidth]{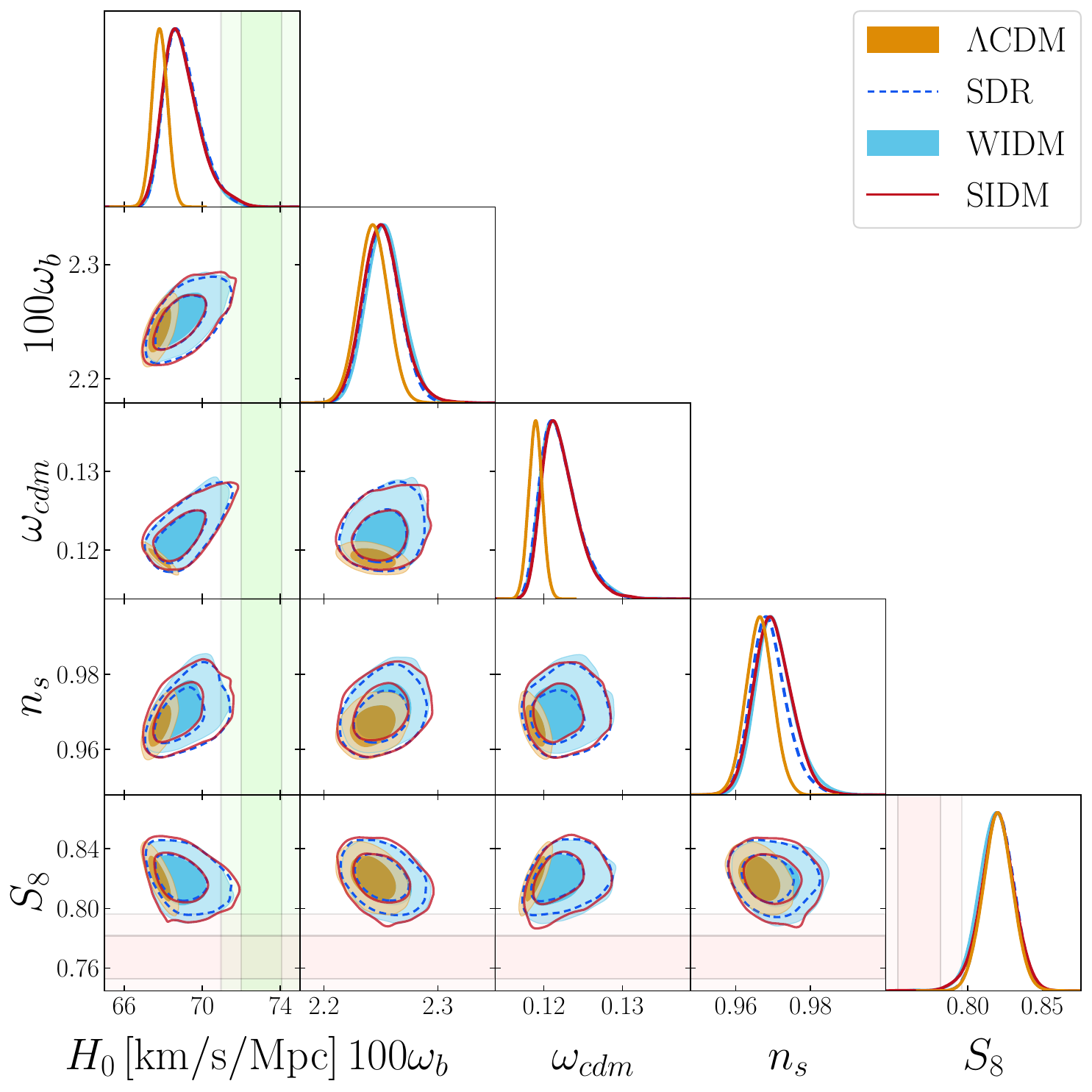}
\caption{One- and two-dimensional posterior distributions are given for selected parameters obtained by fitting to the baseline + FS dataset for each model: $\Lambda$CDM, SDR, WIDM, and SIDM. The light green (lighter green) vertical bars show the 1-$\sigma$ (2-$\sigma$) bounds of the S$H_0$ES measurement of $H_0$, and the light pink (lighter pink) horizontal bars show the 1-$\sigma$ (2-$\sigma$) bounds of the combined $S_8$ measurement from KiDS-1000 and DES-Y3.}
\label{fig:appModels}
\end{figure}

In Fig.~\ref{fig:appModels}, we plot posterior distributions of selected parameters for fits to the baseline + FS datset of each model discussed in this work: $\Lambda$CDM, SDR, WIDM, and SIDM. A key feature of this plot is that the inferred statistics for each parameter remain largely unchanged by the inclusion of dark matter interactions (i.e. the posteriors for WIDM and SIDM closely resemble the posteriors for SDR). This is especially significant when considering $S_8$, since this in turn indicates that the $S_8$ tension is not lessened by the addition of dark matter interactions.

Tables ~\ref{tab:appdata1}-~\ref{tab:appchi25} show the detailed statistics of parameters derived from the MCMC analyses performed in this work. Table~\ref{tab:appdata1} shows means, best fits, and errors reflecting fitting to the baseline dataset; Table~\ref{tab:appdata2} shows the same for the dataset baseline + FS; Table~\ref{tab:appdata3} shows baseline + FS + $M_b$; Table~\ref{tab:appdata7} shows baseline + $S_8$; Table~\ref{tab:appdata6} shows baseline + $S_8+M_b$; Table~\ref{tab:appdata4} shows baseline + FS + $S_8$; and Table~\ref{tab:appdata5} shows baseline + FS +$S_8$ + $M_b$. The $\chi^2$ values are given for each fit in Table~\ref{tab:appchi21}, Table~\ref{tab:appchi22}, Table~\ref{tab:appchi23}, Table~\ref{tab:appchi27}, Table~\ref{tab:appchi26}, Table~\ref{tab:appchi24}, and  Table~\ref{tab:appchi25}, respectively.

\begin{table*}
\begin{tabular} {| l | c| c| c| c|}
\hline\hline
 \multicolumn{1}{|c|}{ Parameter} &  \multicolumn{1}{|c|}{$\Lambda$CDM} &  \multicolumn{1}{|c|}{SDR} &  \multicolumn{1}{|c|}{WIDM} &  \multicolumn{1}{|c|}{SIDM}\\
\hline\hline
$100 \omega_b              $ & $2.240~(2.243)^{+0.013}_{-0.013}   $ & $2.248~(2.244)^{+0.015}_{-0.017}   $ & $2.251~(2.245)^{+0.015}_{-0.016}   $ & $2.249~(2.257)^{+0.016}_{-0.018}   $\\
$\omega_{cdm }             $ & $0.11930~(0.11902)^{+0.00091}_{-0.00091}$ & $0.1227~(0.1234)^{+0.0018}_{-0.0030}$ & $0.1227~(0.1231)^{+0.0017}_{-0.0030}$ & $0.1230~(0.1216)^{+0.0017}_{-0.0030}$\\
$\ln 10^{10}A_s            $ & $3.049~(3.054)^{+0.014}_{-0.014}   $ & $3.050~(3.066)^{+0.014}_{-0.015}   $ & $3.051~(3.06)^{+0.014}_{-0.015}   $ & $3.051~(3.055)^{+0.014}_{-0.016}   $\\
$n_{s }                    $ & $0.9658~(0.9683)^{+0.0037}_{-0.0037}$ & $0.9695~(0.9772)^{+0.0042}_{-0.0063}$ & $0.9707~(0.9743)^{+0.0043}_{-0.0063}$ & $0.9708~(0.9722)^{+0.0045}_{-0.0066}$\\
$\tau_{reio }              $ & $0.0574~(0.0596)^{+0.0071}_{-0.0072}$ & $0.0575~(0.0616)^{+0.0067}_{-0.0076}$ & $0.0578~(0.0591)^{+0.0069}_{-0.0076}$ & $0.0575~(0.0568)^{+0.0070}_{-0.0077}$\\
$\Delta N^{\textrm{\scriptsize IR}}_{\textrm{\scriptsize eff}}$ & --- & $ < 0.546~(0.289)$ & $ < 0.546~(0.179)$ & $ < 0.511~(0.105)$\\
$\log_{10} z_t             $ & --- & Unconstrained $(4.29)$ & Unconstrained $(4.42)$ & Unconstrained $(4.58)$\\
$r_g                       $ & --- & Unconstrained $(4.0)$ & Unconstrained $(3.74)$  & Unconstrained $(1.74)$ \\
$\log_{10} \Gamma_0        $ & --- &           ---     & $ < -6.156~(-6.324)$ & $ < 4.404~(1.416)$\\
$\log_{10} \fdm          $ & --- & --- & Unconstrained $(-0.354)$ & $ < -2.015~(-2.551)$\\
\hline
$H_0 \,[\textrm{km}/\textrm{s}/\textrm{Mpc}]$ & $67.66~(67.9)^{+0.41}_{-0.41}     $ & $68.89~(69.34)^{+0.71}_{-1.1}      $ & $68.84~(68.4)^{+0.67}_{-1.1}      $ & $68.85~(68.37)^{+0.65}_{-1.1}      $\\
$S_8                       $ & $0.825~(0.823)^{+0.010}_{-0.010}   $ & $0.827~(0.834)^{+0.011}_{-0.011}   $ & $0.825~(0.832)^{+0.012}_{-0.012}   $ & $0.827~(0.83)^{+0.011}_{-0.011}   $\\
$M_b                       $ & $-19.419~(-19.413)^{+0.012}_{-0.011} $ & $-19.381~(-19.369)^{+0.021}_{-0.032} $ & $-19.382~(-19.392)^{+0.020}_{-0.032} $ & $-19.382~(-19.395)^{+0.019}_{-0.032} $\\
\hline
$\Delta\chi^2$ & --- & $-1.4$ & $-1.83$ & $0.08$\\
\hline
$Q_{\textrm{\tiny DMAP}}^{M_b}$ & $5.73\sigma $ & $2.74\sigma $ & $2.9\sigma $ & $2.68\sigma $\\
\hline
$Q_{\textrm{\tiny DMAP}}^{S_8}$ & $3.46\sigma $ & --- & $2.93\sigma $ & $3.52\sigma $\\
\hline
$M_b$ GT & $5.63\sigma $ & $3.74\sigma $ & $3.86\sigma $ & $3.91\sigma $\\
\hline
$M_b$ IT & $5.63\sigma $ & $3.03\sigma $ & $3.1\sigma $ & $2.88\sigma $\\
\hline
$S_8$ GT & $3.24\sigma $ & $3.27\sigma $ & $3.08\sigma $ & $3.3\sigma $\\
\hline
$S_8$ IT & $3.23\sigma $ & $3.28\sigma $ & $2.9\sigma $ & $3.24\sigma $\\
\hline
$\Delta\textrm{AIC}^{M_b}$ & --- & $-20.67$ & $-16.21$ & $-15.51$\\
\hline
$\Delta\textrm{AIC}^{S_8}$ & --- & --- & $4.78$ & $10.56$\\
\hline
\end{tabular}
\caption{Mean (best-fit) $\pm 1\sigma$ error of all free parameters, and $S_8$ and $M_b$, obtained by fitting $\Lambda$CDM, SDR, WIDM, and SIDM  to the baseline dataset P18+BAO+Pantheon. Upper bounds are presented at 95\% C.L., and parameters without constraints at 95\% C.L. within their prior boundaries are marked as unconstrained. Tension measures are reported with respect to the S$H_0$ES measurement of $M_b$ and with respect to the combined $S_8$ measurement from KiDS-1000 and DES-Y3.}
\label{tab:appdata1}
\end{table*}

\begin{table*}
\begin{tabular} {| l | c| c| c| c|}
\hline\hline
 Dataset & $\Lambda$CDM & SDR & WIDM & SIDM\\
\hline\hline
Planck\_highl\_TTTEEE & $2350.41$ & $2350.47$ & $2349.25$ & $2352.79$\\
Planck\_lowl\_EE & $398.07$ & $397.96$ & $397.19$ & $396.47$\\
Planck\_lowl\_TT & $23.07$ & $21.92$ & $22.24$ & $22.29$\\
Planck\_lensing & $8.66$ & $9.09$ & $9.05$ & $8.84$\\
Pantheon & $1025.93$ & $1025.73$ & $1026.31$ & $1025.93$\\
bao\_boss\_dr12 & $4.3$ & $3.44$ & $4.67$ & $4.17$\\
bao\_smallz\_2014 & $1.25$ & $1.67$ & $1.14$ & $1.28$\\
\hline
$\chi^2_{\textrm{\scriptsize total}}$ & $3811.69$ & $3810.29$ & $3809.86$ & $3811.77$\\
\hline
\end{tabular}
        \caption{The $\chi^2$ value for each likelihood in the baseline dataset is given for each model, along with the total $\chi^2$.}
        \label{tab:appchi21}
\end{table*}

\begin{table*}
\begin{tabular} {| l | c| c| c| c|}
\hline\hline
 \multicolumn{1}{|c|}{ Parameter} &  \multicolumn{1}{|c|}{$\Lambda$CDM} &  \multicolumn{1}{|c|}{SDR} &  \multicolumn{1}{|c|}{WIDM} &  \multicolumn{1}{|c|}{SIDM}\\
\hline\hline
$100 \omega_b              $ & $2.243~(2.248)^{+0.013}_{-0.013}   $ & $2.251~(2.247)^{+0.015}_{-0.016}   $ & $2.254~(2.258)^{+0.015}_{-0.017}   $ & $2.251~(2.249)^{+0.015}_{-0.018}   $\\
$\omega_{cdm }             $ & $0.11893~(0.11861)^{+0.00086}_{-0.00087}$ & $0.1220~(0.1202)^{+0.0016}_{-0.0028}$ & $0.1221~(0.1205)^{+0.0016}_{-0.0028}$ & $0.1221~(0.1228)^{+0.0016}_{-0.0027}$\\
$\ln 10^{10}A_s            $ & $3.048~(3.061)^{+0.014}_{-0.014}   $ & $3.048~(3.057)^{+0.014}_{-0.015}   $ & $3.050~(3.057)^{+0.014}_{-0.015}   $ & $3.048~(3.06)^{+0.014}_{-0.015}   $\\
$n_{s }                    $ & $0.9664~(0.9679)^{+0.0036}_{-0.0036}$ & $0.9691~(0.9696)^{+0.0039}_{-0.0053}$ & $0.9706~(0.969)^{+0.0040}_{-0.0057}$ & $0.9700~(0.9808)^{+0.0045}_{-0.0055}$\\
$\tau_{reio }              $ & $0.0574~(0.0621)^{+0.0068}_{-0.0075}$ & $0.0576~(0.0617)^{+0.0071}_{-0.0071}$ & $0.0579~(0.0594)^{+0.0069}_{-0.0078}$ & $0.0576~(0.0617)^{+0.0071}_{-0.0071}$\\
$\Delta N^{\textrm{\scriptsize IR}}_{\textrm{\scriptsize eff}}$ & --- & $ < 0.55~(0.08)$ & $ < 0.531~(0.092)$ & $ < 0.544~(0.268)$\\
$\log_{10} z_t             $ & --- & Unconstrained $(4.97)$ & Unconstrained $(4.18)$ & Unconstrained $(4.38)$\\
$r_g                       $ & --- & Unconstrained $(2.34)$ & Unconstrained $(4.87)$  & Unconstrained $(4.39)$\\
$\log_{10} \Gamma_0        $ & --- &  --- & $ < -6.156~(-8.231)$ & $ < 4.286~(3.723)$\\
$\log_{10} f_{DM}          $ & --- & --- & Unconstrained $(-0.806)$ & $ < -2.022~(-3.903)$\\
\hline
$H_0 \,[\textrm{km}/\textrm{s}/\textrm{Mpc}]$ & $67.82~(67.99)^{+0.39}_{-0.39}     $ & $69.01~(68.37)^{+0.66}_{-1.1}      $ & $68.97~(68.37)^{+0.65}_{-1.1}      $ & $68.96~(69.45)^{+0.67}_{-1.1}      $\\
$S_8                       $ & $0.8202~(0.8217)^{+0.0099}_{-0.0099}$ & $0.821~(0.824)^{+0.010}_{-0.010}   $ & $0.818~(0.826)^{+0.011}_{-0.011}   $ & $0.820~(0.828)^{+0.011}_{-0.011}   $\\
$M_b                       $ & $-19.414~(-19.411)^{+0.011}_{-0.011} $ & $-19.379~(-19.402)^{+0.020}_{-0.032} $ & $-19.379~(-19.396)^{+0.019}_{-0.031} $ & $-19.380~(-19.364)^{+0.020}_{-0.032} $\\
\hline
$\Delta\chi^2$ & --- & $-1.62$ & $-0.63$ & $-1.04$\\
\hline
$Q_{\textrm{\tiny DMAP}}^{M_b}$ & $5.52\sigma $ & $2.72\sigma $ & --- & ---\\
\hline
$Q_{\textrm{\tiny DMAP}}^{S_8}$ & $3.01\sigma $ & --- & $2.75\sigma $ & $2.55\sigma $\\
\hline
$M_b$ GT & $5.52\sigma $ & $3.77\sigma $ & $3.81\sigma $ & $3.8\sigma $\\
\hline
$M_b$ IT & $5.52\sigma $ & $2.94\sigma $ & $2.99\sigma $ & $2.92\sigma $\\
\hline
$S_8$ GT & $3.02\sigma $ & $3.01\sigma $ & $2.82\sigma $ & $2.89\sigma $\\
\hline
$S_8$ IT & $3.0\sigma $ & $3.01\sigma $ & $2.63\sigma $ & $2.65\sigma $\\
\hline
$\Delta\textrm{AIC}^{M_b}$ & --- & $-18.69$ & --- & ---\\
\hline
$\Delta\textrm{AIC}^{S_8}$ & --- & --- & $7.9$ & $6.43$\\
\hline
\end{tabular}
\caption{Mean (best-fit) $\pm 1\sigma$ error of all free parameters, and $S_8$ and $M_b$, obtained by fitting $\Lambda$CDM, SDR, WIDM, and SIDM  to the baseline + FS dataset. Upper bounds are presented at 95\% C.L., and parameters without constraints at 95\% C.L. within their prior boundaries are marked as unconstrained. Tension measures are reported with respect to the S$H_0$ES measurement of $M_b$ and with respect to the combined $S_8$ measurement from KiDS-1000 and DES-Y3.}
\label{tab:appdata2}
\end{table*}

\begin{table*}
\begin{tabular} {| l | c| c| c| c|}
\hline\hline
 Dataset & $\Lambda$CDM & SDR & WIDM & SIDM\\
\hline\hline
Planck\_highl\_TTTEEE & $2354.04$ & $2352.28$ & $2355.64$ & $2353.02$\\
Planck\_lowl\_EE & $398.21$ & $398.04$ & $397.2$ & $397.8$\\
Planck\_lowl\_TT & $23.1$ & $22.64$ & $21.78$ & $21.31$\\
Planck\_lensing & $8.67$ & $8.79$ & $9.66$ & $9.41$\\
eft\_boss\_cmass\_bao & $85.17$ & $85.05$ & $83.93$ & $84.5$\\
eft\_boss\_lowz\_bao & $67.88$ & $67.94$ & $68.01$ & $69.75$\\
Pantheon & $1025.84$ & $1026.55$ & $1025.75$ & $1025.71$\\
bao\_smallz\_2014 & $1.48$ & $1.49$ & $1.82$ & $1.85$\\
\hline
$\chi^2_{\textrm{\scriptsize total}}$ & $3964.4$ & $3962.78$ & $3963.77$ & $3963.35$\\
\hline
\end{tabular}
        \caption{ The $\chi^2$ value for each likelihood in the baseline + FS dataset is given for each model, along with the total $\chi^2$.}
        \label{tab:appchi22}
\end{table*}

\begin{table*}
\begin{tabular} {| l | c| c|}
\hline\hline
 \multicolumn{1}{|c|}{ Parameter} &  \multicolumn{1}{|c|}{$\Lambda$CDM} &  \multicolumn{1}{|c|}{SDR}\\
\hline\hline
$100 \omega_b              $ & $2.259~(2.256)^{+0.013}_{-0.014}   $ & $2.276~(2.287)^{+0.017}_{-0.018}   $\\
$\omega_{cdm }             $ & $0.11744~(0.118)^{+0.00085}_{-0.00085}$ & $0.1281~(0.1277)^{+0.0025}_{-0.0029}$\\
$\ln 10^{10}A_s            $ & $3.053~(3.038)^{+0.014}_{-0.016}   $ & $3.049~(3.069)^{+0.015}_{-0.016}   $\\
$n_{s }                    $ & $0.9702~(0.9672)^{+0.0037}_{-0.0037}$ & $0.9736~(0.982)^{+0.0049}_{-0.0077}$\\
$\tau_{reio }              $ & $0.0610~(0.0554)^{+0.0068}_{-0.0080}$ & $0.0603~(0.0668)^{+0.0070}_{-0.0082}$\\
$\Delta N^{\textrm{\scriptsize IR}}_{\textrm{\scriptsize eff}}$ & --- & $0.69~(0.63)^{+0.14}_{-0.23}      $\\
$\log_{10} z_t             $ & --- & Unconstrained $(4.22)$\\
$r_g                       $ & --- & Unconstrained $(1.14)$\\
\hline
$H_0 \,[\textrm{km}/\textrm{s}/\textrm{Mpc}]$ & $68.57~(68.27)^{+0.39}_{-0.39}     $ & $71.71~(72.17)^{+0.83}_{-0.80}     $\\
$S_8                       $ & $0.8049~(0.805)^{+0.0099}_{-0.0099}$ & $0.816~(0.82)^{+0.010}_{-0.012}   $\\
$M_b                       $ & $-19.395~(-19.399)^{+0.011}_{-0.011} $ & $-19.300~(-19.285)^{+0.024}_{-0.024} $\\
\hline
$\Delta\chi^2$ & --- & $-24.69$\\
\hline
$Q_{\textrm{\tiny DMAP}}^{S_8}$ & $1.84\sigma $ & ---\\
\hline
$M_b$ GT & $4.87\sigma $ & $1.29\sigma $\\
\hline
$M_b$ IT & $4.57\sigma $ & $1.29\sigma $\\
\hline
$S_8$ GT & $2.16\sigma $ & $2.54\sigma $\\
\hline
$S_8$ IT & $2.15\sigma $ & $2.68\sigma $\\
\hline
\end{tabular}
\caption{Mean (best-fit) $\pm 1\sigma$ error of all free parameters, and $S_8$ and $M_b$, obtained by fitting $\Lambda$CDM and SDR to the baseline + FS + $M_b$ dataset. Upper bounds are presented at 95\% C.L., and parameters without constraints at 95\% C.L. within their prior boundaries are marked as unconstrained. Tension measures are reported with respect to the S$H_0$ES measurement of $M_b$ and with respect to the combined $S_8$ measurement from KiDS-1000 and DES-Y3.}
\label{tab:appdata3}
\end{table*}

\begin{table*}
\begin{tabular} {| l | c| c|}
\hline\hline
 Dataset & $\Lambda$CDM & SDR\\
\hline\hline
Planck\_highl\_TTTEEE & $2360.69$ & $2354.32$\\
Planck\_lowl\_EE & $397.24$ & $399.78$\\
Planck\_lowl\_TT & $22.45$ & $21.32$\\
Planck\_lensing & $9.97$ & $9.33$\\
eft\_boss\_cmass\_bao & $83.11$ & $84.03$\\
eft\_boss\_lowz\_bao & $69.38$ & $70.78$\\
Pantheon & $1025.66$ & $1026.03$\\
bao\_smallz\_2014 & $2.54$ & $3.15$\\
shoesMB & $23.83$ & $1.44$\\
\hline
$\chi^2_{\textrm{\scriptsize total}}$ & $3994.88$ & $3970.19$\\
\hline
\end{tabular}
        \caption{The $\chi^2$ value for each likelihood in the baseline + FS + $M_b$ dataset is given for each model, along with the total $\chi^2$.}
        \label{tab:appchi23}
\end{table*}

\begin{table*}
\begin{tabular} {| l | c|}
\hline\hline
 \multicolumn{1}{|c|}{ Parameter} &  \multicolumn{1}{|c|}{SDR}\\
\hline\hline
$100 \omega_b              $ & $2.259~(2.261)^{+0.014}_{-0.016}   $\\
$\omega_{cdm }             $ & $0.1212~(0.1214)^{+0.0017}_{-0.0030}$\\
$\ln 10^{10}A_s            $ & $3.038~(3.046)^{+0.014}_{-0.014}   $\\
$n_{s }                    $ & $0.9704~(0.9725)^{+0.0037}_{-0.0048}$\\
$\tau_{reio }              $ & $0.0547~(0.0584)^{+0.0070}_{-0.0070}$\\
$\Delta N^{\textrm{\scriptsize IR}}_{\textrm{\scriptsize eff}}$ & $ < 0.621~(0.316)$\\
$\log_{10} z_t             $ & Unconstrained $(3.61)$\\
$r_g                       $ & Unconstrained $(8.95)$ \\
\hline
$H_0 \,[\textrm{km}/\textrm{s}/\textrm{Mpc}]$ & $69.54~(69.89)^{+0.69}_{-1.1}      $\\
$S_8                       $ & $0.8052~(0.8039)^{+0.0085}_{-0.0085}$\\
$M_b                       $ & $-19.365~(-19.356)^{+0.020}_{-0.033} $\\
\hline
$\Delta\chi^2$ & $-0.97$\\
\hline
$Q_{\textrm{\tiny DMAP}}^{M_b}$ & $2.74\sigma $\\
\hline
$M_b$ GT & $3.3\sigma $\\
\hline
$M_b$ IT & $2.65\sigma $\\
\hline
$S_8$ GT & $2.27\sigma $\\
\hline
$S_8$ IT & $2.27\sigma $\\
\hline
$\Delta\textrm{AIC}^{M_b}$ & $-14.48$\\
\hline
\end{tabular}
\caption{Mean (best-fit) $\pm 1\sigma$ error of all free parameters, and $S_8$ and $M_b$, obtained by fitting SDR to the baseline +  $S_8$ dataset. Upper bounds are presented at 95\% C.L., and parameters without constraints at 95\% C.L. within their prior boundaries are marked as unconstrained.  Tension measures are reported with respect to the S$H_0$ES measurement of $M_b$ and with respect to the combined $S_8$ measurement from KiDS-1000 and DES-Y3.}
\label{tab:appdata7}
\end{table*}

\begin{table*}
\begin{tabular} {| l | c|}
\hline\hline
 Dataset & SDR\\
\hline\hline
Planck\_highl\_TTTEEE & $2354.88$\\
Planck\_lowl\_EE & $396.78$\\
Planck\_lowl\_TT & $22.12$\\
Planck\_lensing & $10.36$\\
Pantheon & $1025.76$\\
bao\_boss\_dr12 & $3.68$\\
bao\_smallz\_2014 & $2.44$\\
S8DESY3 & $3.15$\\
S8kids & $3.51$\\
\hline
$\chi^2_{\textrm{\scriptsize total}}$ & $3822.66$\\
\hline
\end{tabular}
        \caption{The $\chi^2$ value for each likelihood in the baseline + $S_8$ dataset is given for each model, along with the total $\chi^2$.}
        \label{tab:appchi27}
\end{table*}

\begin{table*}
\begin{tabular} {| l | c| c|}
\hline\hline
 \multicolumn{1}{|c|}{ Parameter} &  \multicolumn{1}{|c|}{$\Lambda$CDM} &  \multicolumn{1}{|c|}{SDR}\\
\hline\hline
$100 \omega_b              $ & $2.265~(2.267)^{+0.013}_{-0.013}   $ & $2.282~(2.283)^{+0.016}_{-0.016}   $\\
$\omega_{cdm }             $ & $0.11667~(0.11666)^{+0.00078}_{-0.00078}$ & $0.1270~(0.1283)^{+0.0025}_{-0.0027}$\\
$\ln 10^{10}A_s            $ & $3.049~(3.052)^{+0.014}_{-0.016}   $ & $3.037~(3.045)^{+0.014}_{-0.015}   $\\
$n_{s }                    $ & $0.9721~(0.9732)^{+0.0037}_{-0.0037}$ & $0.9727~(0.9786)^{+0.0037}_{-0.0058}$\\
$\tau_{reio }              $ & $0.0598~(0.0629)^{+0.0071}_{-0.0082}$ & $0.0569~(0.0572)^{+0.0068}_{-0.0076}$\\
$\Delta N^{\textrm{\scriptsize IR}}_{\textrm{\scriptsize eff}}$ & --- & $0.72~(0.64)^{+0.18}_{-0.26}      $\\
$\log_{10} z_t             $ & --- & $ < 4.423~(4.13)$\\
$r_g                       $ & --- & Unconstrained $(9.45)$\\
\hline
$H_0 \,[\textrm{km}/\textrm{s}/\textrm{Mpc}]$ & $68.92~(68.95)^{+0.36}_{-0.36}     $ & $72.00~(72.18)^{+0.81}_{-0.80}     $\\
$S_8                       $ & $0.7946~(0.7961)^{+0.0083}_{-0.0081}$ & $0.8009~(0.8118)^{+0.0086}_{-0.0088}$\\
$M_b                       $ & $-19.386~(-19.384)^{+0.010}_{-0.010} $ & $-19.293~(-19.288)^{+0.023}_{-0.023} $\\
\hline
$\Delta\chi^2$ & --- & $-20.48$\\
\hline
$M_b$ GT & $4.59\sigma $ & $1.11\sigma $\\
\hline
$M_b$ IT & $4.55\sigma $ & $1.11\sigma $\\
\hline
$S_8$ GT & $1.65\sigma $ & $1.99\sigma $\\
\hline
$S_8$ IT & $1.65\sigma $ & $2.0\sigma $\\
\hline
\end{tabular}
\caption{Mean (best-fit) $\pm 1\sigma$ error of all free parameters, and $S_8$ and $M_b$, obtained by fitting $\Lambda$CDM and SDR to the baseline +  $S_8+M_b$ dataset. Upper bounds are presented at 95\% C.L., and parameters without constraints at 95\% C.L. within their prior boundaries are marked as unconstrained. Tension measures are reported with respect to the S$H_0$ES measurement of $M_b$ and with respect to the combined $S_8$ measurement from KiDS-1000 and DES-Y3.}
\label{tab:appdata6}
\end{table*}

\begin{table*}
\begin{tabular} {| l | c| c|}
\hline\hline
 Dataset & $\Lambda$CDM & SDR\\
\hline\hline
Planck\_highl\_TTTEEE & $2359.97$ & $2357.05$\\
Planck\_lowl\_EE & $398.09$ & $396.43$\\
Planck\_lowl\_TT & $22.21$ & $21.32$\\
Planck\_lensing & $10.5$ & $10.42$\\
Pantheon & $1025.75$ & $1025.84$\\
bao\_boss\_dr12 & $3.9$ & $4.64$\\
bao\_smallz\_2014 & $2.61$ & $3.05$\\
S8DESY3 & $1.8$ & $4.9$\\
S8kids & $2.4$ & $4.85$\\
shoesMB & $23.44$ & $1.7$\\
\hline
$\chi^2_{\textrm{\scriptsize total}}$ & $3850.68$ & $3830.2$\\
\hline
\end{tabular}
        \caption{The $\chi^2$ value for each likelihood in the baseline + $S_8+M_b$ dataset is given for each model, along with the total $\chi^2$.}
        \label{tab:appchi26}
\end{table*}

\begin{table*}
\begin{tabular} {| l | c| c| c|}
\hline\hline
 \multicolumn{1}{|c|}{ Parameter} &  \multicolumn{1}{|c|}{$\Lambda$CDM} &  \multicolumn{1}{|c|}{WIDM} &  \multicolumn{1}{|c|}{SIDM}\\
\hline\hline
$100 \omega_b              $ & $2.252~(2.245)^{+0.013}_{-0.013}   $ & $2.261~(2.268)^{+0.017}_{-0.016}   $ & $2.260~(2.231)^{+0.017}_{-0.016}   $\\
$\omega_{cdm }             $ & $0.11768~(0.11765)^{+0.00076}_{-0.00076}$ & $0.1214~(0.1227)^{+0.0017}_{-0.0032}$ & $0.1209~(0.1192)^{+0.0013}_{-0.0031}$\\
$\ln 10^{10}A_s            $ & $3.040~(3.04)^{+0.014}_{-0.014}   $ & $3.044~(3.038)^{+0.015}_{-0.015}   $ & $3.041~(3.072)^{+0.015}_{-0.016}   $\\
$n_{s }                    $ & $0.9688~(0.9699)^{+0.0036}_{-0.0036}$ & $0.9723~(0.9676)^{+0.0038}_{-0.0050}$ & $0.9710~(0.9673)^{+0.0038}_{-0.0047}$\\
$\tau_{reio }              $ & $0.0548~(0.0537)^{+0.0067}_{-0.0072}$ & $0.0561~(0.0548)^{+0.0071}_{-0.0072}$ & $0.0556~(0.0705)^{+0.0073}_{-0.0074}$\\
$\Delta N^{\textrm{\scriptsize IR}}_{\textrm{\scriptsize eff}}$ & --- & $ < 0.616~(0.177)$ & $ < 0.718~(0.011)$\\
$\log_{10} z_t             $ & --- & Unconstrained $(4.53)$ & Unconstrained $(3.45)$ \\
$r_g                       $ & --- & Unconstrained $(0.19)$ & Unconstrained $(4.29)$   \\
$\log_{10} \Gamma_0        $ & --- & Unconstrained $(-6.288)$ &$ < 4.167~(2.379)$\\
$\log_{10} f_{DM}          $ & --- & Unconstrained $(-0.139)$ & $ < -1.496~(-1.348)$\\
\hline
$H_0 \,[\textrm{km}/\textrm{s}/\textrm{Mpc}]$ & $68.38~(68.37)^{+0.35}_{-0.35}     $ & $69.35~(68.79)^{+0.75}_{-1.2}      $ & $69.42~(67.73)^{+0.70}_{-1.2}      $\\
$S_8                       $ & $0.8029~(0.8034)^{+0.0083}_{-0.0082}$ & $0.795~(0.793)^{+0.013}_{-0.0097}  $ & $0.800~(0.773)^{+0.011}_{-0.0084}  $\\
$M_b                       $ & $-19.3999~(-19.3999)^{+0.0099}_{-0.010}$ & $-19.369~(-19.384)^{+0.022}_{-0.035} $ & $-19.368~(-19.417)^{+0.020}_{-0.037} $\\
\hline
$\Delta\chi^2$ & --- & $-2.1$ & $-3.57$\\
\hline
$Q_{\textrm{\tiny DMAP}}^{M_b}$ & $4.98\sigma $ & $2.53\sigma $ & $3.21\sigma $\\
\hline
$M_b$ GT & $5.11\sigma $ & $3.36\sigma $ & $3.4\sigma $\\
\hline
$M_b$ IT & $4.98\sigma $ & $2.62\sigma $ & $2.43\sigma $\\
\hline
$S_8$ GT & $2.15\sigma $ & $1.62\sigma $ & $1.94\sigma $\\
\hline
$S_8$ IT & $2.15\sigma $ & $1.48\sigma $ & $1.76\sigma $\\
\hline
$\Delta\textrm{AIC}^{M_b}$ & --- & $-10.52$ & $-8.08$\\
\hline
\end{tabular}
\caption{Mean (best-fit) $\pm 1\sigma$ error of all free parameters, and $S_8$ and $M_b$, obtained by fitting $\Lambda$CDM, WIDM, and SIDM to the baseline + FS + $S_8$ dataset. Upper bounds are presented at 95\% C.L., and parameters without constraints at 95\% C.L. within their prior boundaries are marked as unconstrained. Tension measures are reported with respect to the S$H_0$ES measurement of $M_b$ and with respect to the combined $S_8$ measurement from KiDS-1000 and DES-Y3.}
\label{tab:appdata4}
\end{table*}

\begin{table*}
\begin{tabular} {| l | c| c| c|}
\hline\hline
 Dataset & $\Lambda$CDM & WIDM & SIDM\\
\hline\hline
Planck\_highl\_TTTEEE & $2356.99$ & $2358.37$ & $2352.09$\\
Planck\_lowl\_EE & $397.54$ & $396.07$ & $402.57$\\
Planck\_lowl\_TT & $22.03$ & $22.96$ & $23.42$\\
Planck\_lensing & $10.42$ & $9.97$ & $9.94$\\
eft\_boss\_cmass\_bao & $84.14$ & $85.06$ & $88.1$\\
eft\_boss\_lowz\_bao & $67.99$ & $68.35$ & $66.22$\\
Pantheon & $1025.9$ & $1025.83$ & $1025.88$\\
bao\_smallz\_2014 & $2.18$ & $1.33$ & $1.28$\\
S8DESY3 & $2.93$ & $1.38$ & $0.01$\\
S8kids & $3.33$ & $2.02$ & $0.36$\\
\hline
$\chi^2_{\textrm{\scriptsize total}}$ & $3973.45$ & $3971.35$ & $3969.88$\\
\hline
\end{tabular}
        \caption{The $\chi^2$ value for each likelihood in the baseline + FS + $S_8$ dataset is given for each model, along with the total $\chi^2$.}
        \label{tab:appchi24}
\end{table*}

\begin{table*}
\begin{tabular} {| l | c| c| c|}
\hline\hline
 \multicolumn{1}{|c|}{ Parameter} &  \multicolumn{1}{|c|}{$\Lambda$CDM} &  \multicolumn{1}{|c|}{WIDM} &  \multicolumn{1}{|c|}{SIDM}\\
\hline\hline
$100 \omega_b              $ & $2.267~(2.274)^{+0.013}_{-0.013}   $ & $2.285~(2.296)^{+0.017}_{-0.016}   $ & $2.286~(2.3)^{+0.019}_{-0.019}   $\\
$\omega_{cdm }             $ & $0.11659~(0.1169)^{+0.00074}_{-0.00074}$ & $0.1276~(0.1291)^{+0.0029}_{-0.0033}$ & $0.1277~(0.1262)^{+0.0031}_{-0.0031}$\\
$\ln 10^{10}A_s            $ & $3.047~(3.06)^{+0.014}_{-0.015}   $ & $3.042~(3.04)^{+0.014}_{-0.016}   $ & $3.035~(3.033)^{+0.017}_{-0.016}   $\\
$n_{s }                    $ & $0.9725~(0.973)^{+0.0036}_{-0.0035}$ & $0.9750~(0.9785)^{+0.0040}_{-0.0062}$ & $0.9743~(0.9734)^{+0.0044}_{-0.0069}$\\
$\tau_{reio }              $ & $0.0596~(0.0635)^{+0.0070}_{-0.0078}$ & $0.0580~(0.0556)^{+0.0069}_{-0.0074}$ & $0.0563~(0.0568)^{+0.0083}_{-0.0082}$\\
$\Delta N^{\textrm{\scriptsize IR}}_{\textrm{\scriptsize eff}}$ & --- & $0.71~(0.65)^{+0.18}_{-0.25}      $ & $0.72~(0.59)^{+0.19}_{-0.26}      $\\
$\log_{10} z_t             $ & --- & $ < 4.432~(4.12)$ & $ < 4.555~(3.74)$\\
$r_g                       $ & --- & Unconstrained $(0.53)$ & $ < 4.098~(0.27)$\\
$\log_{10} \Gamma_0        $ & --- & $ < -5.837~(-5.746)$ & $ < 3.983~(-0.853)$\\
$\log_{10} f_{DM}          $ & --- & Unconstrained $(-0.824)$ & $ < -2.017~(-3.616)$\\
\hline
$H_0 \,[\textrm{km}/\textrm{s}/\textrm{Mpc}]$ & $68.97~(68.96)^{+0.34}_{-0.34}     $ & $71.94~(72.13)^{+0.86}_{-0.75}     $ & $72.24~(72.48)^{+0.93}_{-0.89}     $\\
$S_8                       $ & $0.7930~(0.8003)^{+0.0079}_{-0.0079}$ & $0.793~(0.782)^{+0.011}_{-0.0089}  $ & $0.7974~(0.7901)^{+0.0096}_{-0.0093}$\\
$M_b                       $ & $-19.3844~(-19.3832)^{+0.0099}_{-0.0098}$ & $-19.294~(-19.283)^{+0.025}_{-0.022} $ & $-19.286~(-19.279)^{+0.027}_{-0.025} $\\
\hline
$\Delta\chi^2$ & --- & $-20.57$ & $-18.08$\\
\hline
$M_b$ GT & $4.57\sigma $ & $1.11\sigma $ & $0.85\sigma $\\
\hline
$M_b$ IT & $4.54\sigma $ & $1.14\sigma $ & $0.85\sigma $\\
\hline
$S_8$ GT & $1.57\sigma $ & $1.51\sigma $ & $1.75\sigma $\\
\hline
$S_8$ IT & $1.56\sigma $ & $1.43\sigma $ & $1.69\sigma $\\
\hline
\end{tabular}
\caption{Mean (best-fit) $\pm 1\sigma$ error of all free parameters, and $S_8$ and $M_b$, obtained by fitting $\Lambda$CDM, WIDM, and SIDM to the baseline + FS + $S_8+M_b$ dataset. Upper bounds are presented at 95\% C.L., and parameters without constraints at 95\% C.L. within their prior boundaries are marked as unconstrained. Tension measures are reported with respect to the S$H_0$ES measurement of $M_b$ and with respect to the combined $S_8$ measurement from KiDS-1000 and DES-Y3.}
\label{tab:appdata5}
\end{table*}

\begin{table*}
\begin{tabular} {| l | c| c| c|}
\hline\hline
 Dataset & $\Lambda$CDM & WIDM & SIDM\\
\hline\hline
Planck\_highl\_TTTEEE & $2359.11$ & $2364.24$ & $2363.08$\\
Planck\_lowl\_EE & $398.34$ & $396.12$ & $396.75$\\
Planck\_lowl\_TT & $22.36$ & $21.4$ & $21.68$\\
Planck\_lensing & $9.62$ & $11.91$ & $12.43$\\
eft\_boss\_cmass\_bao & $84.57$ & $83.93$ & $82.94$\\
eft\_boss\_lowz\_bao & $67.31$ & $68.13$ & $69.23$\\
Pantheon & $1025.71$ & $1026.81$ & $1025.81$\\
bao\_smallz\_2014 & $2.55$ & $2.72$ & $3.21$\\
S8DESY3 & $2.46$ & $0.3$ & $1.42$\\
S8kids & $2.96$ & $0.9$ & $2.06$\\
shoesMB & $23.26$ & $1.24$ & $1.58$\\
\hline
$\chi^2_{\textrm{\scriptsize total}}$ & $3998.27$ & $3977.7$ & $3980.19$\\
\hline
\end{tabular}
        \caption{The $\chi^2$ value for each likelihood in the baseline + FS + $S_8+M_b$ dataset is given for each model, along with the total $\chi^2$.}
        \label{tab:appchi25}
\end{table*}

\appendix

\end{document}